\documentstyle[12pt]{article}
\input epsf
\renewcommand{\theequation}{\thesection.\arabic{equation}}
\renewcommand{\title}[1]{\large\bf \mbox{}\\ \mbox{}\\ \mbox{}\\ \mbox{}\\
     #1\bigskip\medskip\\}
\renewcommand{\author}[1]{\large #1\\ \smallskip}
\newcommand{\address}[1]{{\narrower\normalsize\it #1\\}\bigskip}
\def\ade{$A$--$D$--$E$\space}
\def\weight#1#2#3#4#5{#1\!\!\left(\matrix{#5&#4\cr #2&#3\cr}\right)}
\def\wt#1#2#3#4#5#6{#1\!\!\mbox{
$\left(\matrix{#5&#4\cr#2&#3\cr}\biggm|\mbox{$#6$}\right)$}}
\def\cell#1#2#3#4#5#6{C\!\!\left(\matrix{#6&#5&#4\cr #1&#2&#3\cr}\right)}
\def\dddots{\mathinner{\mkern1mu\raise1pt\vbox{\kern1pt\hbox{.}}
                       \mkern2mu\raise4pt\hbox{.}
                       \mkern2mu\raise7pt\hbox{.}\mkern1mu}}

\def\mat {\pmatrix}
\def\smat#1{\mbox{\small $\mat{#1}$}}
\def\be {\begin{equation}}
\def\ee {\end{equation}}
\def\ba {\begin{array}}
\def\ea {\end{array}}
\def\bea {\begin{eqnarray}}
\def\eea {\end{eqnarray}}
\def\bra#1{\langle #1|}
\def\ket#1{|#1\rangle}

\def\disp{\displaystyle}
\def\aa{\overline{a}}
\def\bb{\overline{b}}
\def\cc{\overline{c}}
\def\tiny{\scriptsize}
\def\pb#1{\parbox{.75truein}{\rule[-.05in]{0in}{.2in}\hfil #1 \hfil}}
\def\fb#1{\framebox[.75truein][c]{\rule[-.05in]{0in}{.2in}{#1}}}
\long\def\twocol#1&#2\\{\par
\begin{minipage}[t]{2.9in}
#1
\end{minipage}
\begin{minipage}[t]{2.9in}
#2
\end{minipage}\smallskip\par}

\def\rawpicture #1 by #2 (#3){
  \vbox to #2{
    \hrule width #1 height 0pt depth 0pt
    \vfill
    \special{picture #3} % this is the low-level interface
    }
  }
\def\scaledpicture #1 by #2 (#3 scaled #4){{
  \dimen0=#1 \dimen1=#2
  \divide\dimen0 by 1000 \multiply\dimen0 by #4
  \divide\dimen1 by 1000 \multiply\dimen1 by #4
  \rawpicture \dimen0 by \dimen1 (#3 scaled #4)}
  }
\newcommand{\zr}[1]{\mbox{\hspace*{#1em}}}
\newcommand{\Z}{\mbox{\sf Z\zr{-0.45}Z}}

\def\punit#1{\hspace{#1\unitlength}}

\def\diagface#1#2#3#4#5#6#7{\rule[-3.8\unitlength]{0in}{7.6\unitlength}
\begin{picture}(9,4)(-#6,-#7)
\put(1.5,0.5){\vector(1,-1){3}}
\put(4.5,3.5){\vector(1,-1){3}}
\put(4.5,3.5){\vector(-1,-1){3}}
\put(7.5,0.5){\vector(-1,-1){3}}
\put(1.2,0.5){\makebox(0,0)[r]{\scriptsize \mbox{$#1$}}}
\put(4.5,-2.8){\makebox(0,0)[t]{\scriptsize \mbox{$#2$}}}
\put(7.8,0.5){\makebox(0,0)[l]{\scriptsize \mbox{$#3$}}}
\put(4.5,3.8){\makebox(0,0)[b]{\scriptsize \mbox{$#4$}}}
\put(4.5,0.5){\makebox(0,0){\scriptsize \mbox{$#5$}}}
\end{picture}}

\def\sqface#1#2#3#4#5#6#7{\rule[-2.8\unitlength]{0in}{5.6\unitlength}
\begin{picture}(4,4)(-#6,-#7)
\put(0,-2){\vector(1,0){4}}
\put(4,2){\vector(0,-1){4}}
\put(0,2){\vector(1,0){4}}
\put(0,2){\vector(0,-1){4}}
\put(0,-3.2){\makebox(0,0)[b]{\scriptsize \mbox{$#1$}}}
\put(4,-3.2){\makebox(0,0)[b]{\scriptsize \mbox{$#2$}}}
\put(4,2.5){\makebox(0,0)[b]{\scriptsize \mbox{$#3$}}}
\put(0,2.5){\makebox(0,0)[b]{\scriptsize \mbox{$#4$}}}
\put(2,0){\makebox(0,0){\scriptsize \mbox{$#5$}}}
\end{picture}}

\def\sqedges#1#2#3#4{
\begin{picture}(0,0)(0,-.3)
\put(2,-3.2){\makebox(0,0)[b]{\scriptsize \mbox{$#1$}}}
\put(4.4,0){\makebox(0,0)[l]{\scriptsize \mbox{$#2$}}}
\put(2,2.5){\makebox(0,0)[b]{\scriptsize \mbox{$#3$}}}
\put(-.4,0){\makebox(0,0)[r]{\scriptsize \mbox{$#4$}}}
\end{picture}}

\begin{document}
%\begin{titlepage}
\begin{center}
\title{INTERTWINERS AND \ade LATTICE MODELS}
\author{Paul A. Pearce and Yu-kui Zhou }
\address{Mathematics Department, University of Melbourne,\\
            Parkville, Victoria 3052, Australia }

\begin{abstract}
Intertwiners between \ade lattice models are presented and the general
theory developed.  The intertwiners
are discussed at three levels: at the level of the adjacency matrices, at
the level of the cell calculus
intertwining the face algebras and at the level of the row transfer
matrices. A convenient graphical
representation of the intertwining cells is introduced. The utility of the
intertwining relations in
studying the spectra of the \ade models is emphasized. In particular, it is
shown that the existence of an
intertwiner implies that many eigenvalues of the \ade row transfer matrices
are exactly in common for a
finite system and, consequently, that the corresponding central charges and
scaling
dimensions can be identified.
\end{abstract} %\vspace{0.5cm}
%{\small PACS numbers:}
\end{center}
%\end{titlepage}

\section{ Introduction }

The notion of intertwiners and cell calculus has been developed by many
authors
\cite{Ocneanu,PasqEtiol,Roch:90,FrZu:89p} and is potentially very useful in
studying two-dimensional
lattice models. Here we continue the general development of intertwiners
for critical \ade lattice models, including the Temperley-Lieb
\cite{Pasq:87,OwBa:87,Ginsparg:89?,Paul:90} and dilute
\cite{WaNiSe:92,Roch:92} models
and point out the implications of these intertwining relations for the
study of spectra.

The paper is
organized as follows. In this section we define the critical \ade lattice
models and indicate how
they are constructed from the adjacency matrices. In Section~2 we discuss
the elementary intertwiners
between the \ade adjacency matrices. In Section~3 we present intertwiners
at the level of the faces and
discuss the cell calculus. In particular, we introduce  cell graphs as a
convenient graphical
representation of the intertwining cells. In Section~4 we discuss
intertwiners at the level of
the row transfer matrices and indicate the consequences for the spectra of
\ade lattice models. Finally,
in the Appendices, we give a comprehensive listing of the \ade intertiners
and cells.

\subsection{\ade Lattice Models }

The \ade lattice models are interaction-round-a-face or IRF models
\cite{Baxter:82} that generalize
the restricted solid-on-solid (RSOS) models solved by Andrews, Baxter and
Forrester \cite{ABF:84} in
1984. At criticality, these models are given by solutions of the
Yang-Baxter
equation \cite{Yang:67,Baxter:82} based on the Temperley-Lieb algebra and
are associated with the
classical and affine \ade Dynkin diagrams \cite{Pasq:87,OwBa:87,Paul:90}
shown in Figure~1. States
at adjacent sites of the square lattice must be adjacent on the Dynkin
diagram. The face weights of
faces not satisfying this adjacency condition for each pair of adjacent
sites around a face vanish.

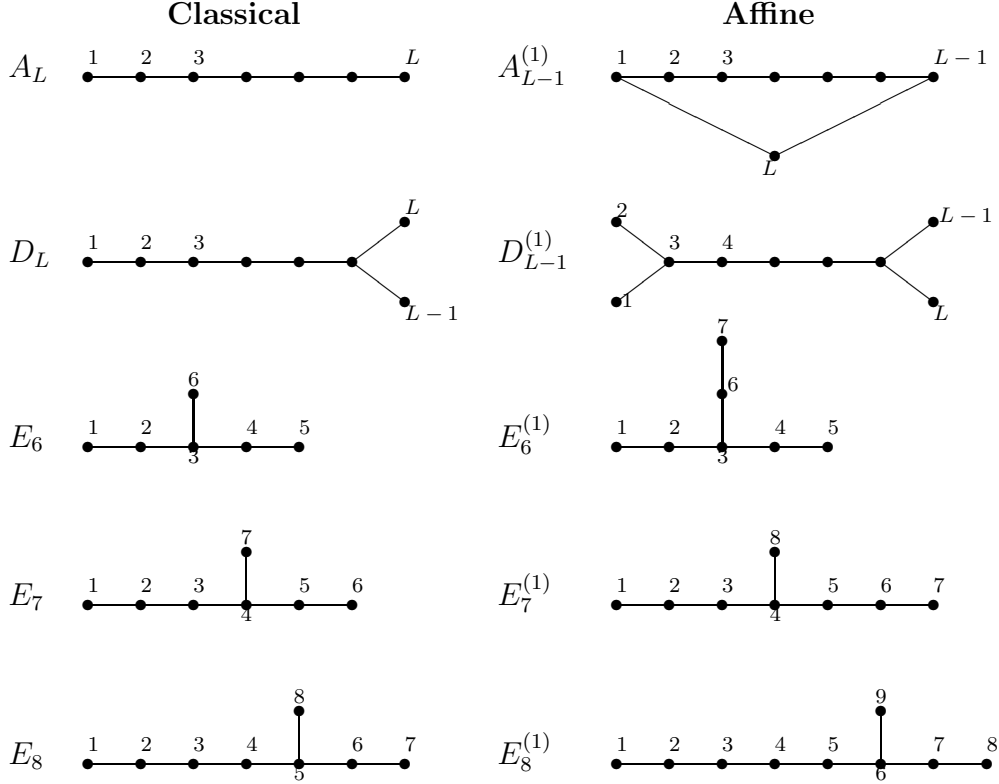
\begin{figure}[htb]
\begin{center}
%\framebox{
\begin{picture}(400,300)(-20,-180)
\put(50,100){\bf Classical } \put(260,100){\bf Affine }

\put(-10,80){$A_L$}
\multiput(20,80)(20,0){6}{\line(1,0){20}}
\multiput(20,80)(20,0){7}{\circle*{4}}
\put(20,85){\tiny 1} \put(40,85){\tiny 2} \put(60,85){\tiny 3}
\put(140,85){\tiny $L$}

\put(175,80){$A_{L-1}^{(1)}$}
\multiput(220,80)(20,0){6}{\line(1,0){20}}
\put(280,50){\line(2,1){60}}\put(280,50){\line(-2,1){60}}
\multiput(220,80)(20,0){7}{\circle*{4}} \put(280,50){\circle*{4}}
\put(220,85){\tiny 1} \put(240,85){\tiny 2} \put(260,85){\tiny 3}
\put(340,85){\tiny $L-1$}
\put(275,43){\tiny $L$}

\put(-10,10){$D_L$}
\multiput(20,10)(20,0){5}{\line(1,0){20}}
\put(120,10){\line(4,3){20}} \put(120,10){\line(4,-3){20}}
\multiput(20,10)(20,0){6}{\circle*{4}}
\multiput(140,25)(0,-30){2}{\circle*{4}}
\put(20,15){\tiny 1} \put(40,15){\tiny 2} \put(60,15){\tiny 3}
\put(140,28){\tiny $L$}\put(140,-12){\tiny $L-1$}

\put(175,10){$D_{L-1}^{(1)}$}
\multiput(240,10)(20,0){4}{\line(1,0){20}}
\put(240,10){\line(-4,3){20}} \put(240,10){\line(-4,-3){20}}
\put(320,10){\line(4,3){20}}  \put(320,10){\line(4,-3){20}}
\multiput(240,10)(20,0){5}{\circle*{4}}
\multiput(220,25)(0,-30){2}{\circle*{4}}
\multiput(340,25)(0,-30){2}{\circle*{4}}
\put(220,27){\tiny 2} \put(240,15){\tiny 3} \put(260,15){\tiny 4}
\put(342,25){\tiny $L-1$}
\put(340,-12){\tiny $L$} \put(222,-7){\tiny 1}

\put(-10,-60){$E_6$}
\multiput(20,-60)(20,0){4}{\line(1,0){20}}
\multiput(20,-60)(20,0){5}{\circle*{4}}
\put(60,-40){\line(0,-1){20}} \put(60,-40){\circle*{4}}
\put(20,-55){\tiny 1} \put(40,-55){\tiny 2} \put(80,-55){\tiny 4}
\put(100,-55){\tiny 5}
\put(58,-66){\tiny 3} \put(58,-37){\tiny 6}

\put(175,-60){$E_6^{(1)}$}
\multiput(220,-60)(20,0){4}{\line(1,0){20}}
\multiput(220,-60)(20,0){5}{\circle*{4}}
\multiput(260,-40)(0,20){2}{\line(0,-1){20}}
\multiput(260,-40)(0,20){2}{\circle*{4}}
\put(220,-55){\tiny 1} \put(240,-55){\tiny 2} \put(280,-55){\tiny 4}
\put(300,-55){\tiny 5}
\put(258,-66){\tiny 3} \put(262,-39){\tiny 6}  \put(258,-17){\tiny 7}

\put(-10,-120){$E_7$}
\multiput(20,-120)(20,0){5}{\line(1,0){20}}
\multiput(20,-120)(20,0){6}{\circle*{4}}
\put(80,-100){\line(0,-1){20}} \put(80,-100){\circle*{4}}
\put(20,-115){\tiny 1} \put(40,-115){\tiny 2} \put(60,-115){\tiny 3}
\put(100,-115){\tiny 5}
\put(120,-115){\tiny 6} \put(78,-126){\tiny 4} \put(78,-97){\tiny 7}

\put(175,-120){$E_7^{(1)}$}
\multiput(220,-120)(20,0){6}{\line(1,0){20}}
\multiput(220,-120)(20,0){7}{\circle*{4}}
\put(280,-100){\line(0,-1){20}} \put(280,-100){\circle*{4}}
\put(220,-115){\tiny 1} \put(240,-115){\tiny 2} \put(260,-115){\tiny 3}
\put(300,-115){\tiny 5}
\put(320,-115){\tiny 6} \put(340,-115){\tiny 7} \put(278,-126){\tiny 4}
\put(278,-97){\tiny 8}

\put(-10,-180){$E_8$}
\multiput(20,-180)(20,0){6}{\line(1,0){20}}
\multiput(20,-180)(20,0){7}{\circle*{4}}
\put(100,-160){\line(0,-1){20}} \put(100,-160){\circle*{4}}
\put(20,-175){\tiny 1} \put(40,-175){\tiny 2} \put(60,-175){\tiny 3}
\put(80,-175){\tiny 4}
\put(120,-175){\tiny 6} \put(140,-175){\tiny 7} \put(98,-186){\tiny 5}
\put(98,-157){\tiny 8}

\put(175,-180){$E_8^{(1)}$}
\multiput(220,-180)(20,0){7}{\line(1,0){20}}
\multiput(220,-180)(20,0){8}{\circle*{4}}
\put(320,-160){\line(0,-1){20}} \put(320,-160){\circle*{4}}
\put(220,-175){\tiny 1} \put(240,-175){\tiny 2} \put(260,-175){\tiny 3}
\put(280,-175){\tiny 4}
\put(300,-175){\tiny 5} \put(340,-175){\tiny 7} \put(360,-175){\tiny 8}
\put(318,-186){\tiny 6} \put(318,-157){\tiny 9}
\end{picture}
%}
\end{center}
\caption{Dynkin diagrams of the classical and affine \ade Lie algebras}
\end{figure}

The face weights
of the  classical \ade models at criticality are given by \cite{Pasq:87}
\begin{equation}
\wt Wabcdu=
 {\sin (\lambda -u) \over \sin \lambda }\ \delta_{a,c}A_{a,b}A_{a,d} +
       {\sin u \over \sin \lambda } \sqrt{{S_a S_c \over S_b S_d}}\
\delta_{b,d}A_{a,b}A_{b,c}
\label{eq:cADEface}
\end{equation}
where $u$ is the spectral parameter and $\lambda=\pi/h $ is the crossing
parameter. The
Coxeter exponent $h$ is given for each algebra in Table~1. The $S_a$ are
the positive components of the
Perron-Frobenius eigenvector $S$ of the adjacency matrix as given in
Table~2. The adjacency matrix $A$ has
elements
\begin{equation}
A_{a,b}=\left\{ \begin{array}{ll}  1,\qquad &\mbox{$(a,b)$ adjacent} \\
                                  0,\qquad &\mbox{otherwise.}
                 \end{array}  \right.
\end{equation}

\begin{table}[p]
\begin{center}\begin{tabular}{ccc} \hline  \\
%\multicolumn{3}{c}{{\bf Coxeter Exponents }  }\\ \\ \hline \\
Lie algebra    &   h    &             $ m_j$                          \\
\hline  \\
$A_L$          & $L+1$  & $ 1,2,3 ,\cdots ,L $             \\
$D_L$          &2$L-2$  & $ 1,3,5,\cdots ,2L-3,L-1  $    \\
$E_6$          & 12     & $ 1,4,5,7,8,11   $             \\
$E_7$          & 18     & $ 1,5,7,9,11,13,17,   $ \\
$E_8$          & 30     & $ 1,7,11,13,17,19,23,29  $ \\
$A_{L-1}^{(1)}$ & $L$    &$  0,2,4,\cdots ,2L-2  $ \\
$D_{L-1}^{(1)}$ &2($L-3$)&$  0,2,4,\cdots ,2(L-3),L-3,L-3  $ \\
$E_6^{(1)}$     & 6      &$  0,2,2,3,4,4,6  $ \\
$E_7^{(1)}$     &  12    &$  0,3,4,6,6,8,9,12  $ \\
$E_8^{(1)}$     &  30    & $ 0,6,10,12,15,18,20,24,30  $ \\ \\ \hline
\end{tabular}\end{center}
\caption{The Coxeter number $h$ and the Coxeter exponents $m_j$ for \ade
classical
and affine algebras.}
\end{table}
\begin{table}[p]\begin{center}
\begin{tabular}{cc} \hline \\
Lie algebra    &      Perron-Frobenius Eigenvector  \\   \hline  \\
$A_L$          &  ${ \left(  \sin {\pi \over L+1},\sin {2\pi \over
L+1},\cdots ,
           \sin {L\pi \over L+1} \right) }$  \\
$D_L$          & $ \left( 2\cos {(L-2)\pi \over 2L-2},\cdots ,2\cos {2\pi
\over 2L-2},
           2\cos {\pi \over 2L-2},1,1 \right) $    \\
$E_6$          & $ \left( \sin {\pi\over 12},\sin {\pi\over 6},\sin
{\pi\over 4},
       \sin {\pi\over 3}-{\sin {\pi\over 4}\over 2\cos {\pi\over 12}},
         \sin {5\pi\over 12}-\sin {\pi\over 4},{\sin {\pi\over 4}\over
2\cos {\pi\over 12}}\right)$ \\
$E_7$          & $ \left( \sin {\pi\over 18},\sin {\pi\over 9},\sin
{\pi\over 6},
         \sin {2\pi\over 9},\sin {5\pi\over 18}-{\sin {2\pi\over 9}\over
2\cos {\pi\over 18}},
            \sin {\pi\over 3}-\sin {2\pi\over 9},
             {\sin {2\pi\over 9}\over 2\cos {\pi\over 18}} \right) $ \\
$E_8$          & $ \left({ \sin {\pi\over 30},\sin {\pi\over 15},\sin
{\pi\over 10},
         \sin {2\pi\over 15},\sin {\pi\over 6},\sin {\pi\over 5}-{\sin
{\pi\over 6}\over
          2\cos {\pi\over 30}}, \sin {7\pi\over 30}-\sin {\pi\over 6},
               {\sin {\pi\over 6}\over 2\cos {\pi\over 30}} }\right) $ \\
$A_{L-1}^{(1)}$ & $ \left( 1,1,\cdots ,1 \right) $ \\
$D_{L-1}^{(1)}$ & $ \left( 1,1,2,2,\cdots ,2,2,1,1 \right) $ \\
$E_6^{(1)}$     &    $ \left( 1,2,3,2,1,2,1 \right) $ \\
$E_7^{(1)}$     &    $ \left( 1,2,3,4,3,2,1,2 \right) $ \\
$E_8^{(1)}$     &    $\left( 1,2,3,4,5,6,4,2,3 \right) $ \\ \\ \hline
\end{tabular}\end{center}
\caption{The Perron-Frobenius eigenvectors for simply-laced classical and
affine Lie algebras. The order of the
components is fixed by the labelling of sites in the Dynkin diagrams of
Figure~1. The corresponding
eigenvalue is $\Lambda_{max}=2$ for the affine algebras and
$\Lambda_{max}=2\cos(\pi/h)$ for the classical
algebras where the Coxeter number $h$ is listed in Table~1.}
\end{table}

The face weights of the affine $A_{L-1}^{(1)}$
and  $D_{L-1}^{(1)}$  models \cite{PeSe:88,KuYa:88} at criticality can be
written as
\begin{eqnarray}
& & \wt Wabcdu=
 \left[{\sin (\lambda -u) \over \sin \lambda }\ \delta_{a,c} +
       {\sin u \over \sin \lambda }\ \sqrt{{S_a S_c \over S_b S_d}}\
\delta_{b,d}\right.  \nonumber
\\ &+&\left.\left(1- {\sin (\lambda -u) \over \sin \lambda }-{\sin u \over
\sin \lambda }\right)
\sqrt{{S_a\over S_b}\left(3-{S_a\over S_b}-{S_b\over S_a}\right)}
\ \overline{\delta}_{a,c}  \overline{\delta}_{b,d}\right]
A_{a,b}A_{b,c}A_{c,d}A_{d,a} \label{eq:aAface}
\end{eqnarray}
where the crossing parameter $\lambda$ is arbitrary. For the affine $A$
model,
$\overline{\delta}$ is the usual Kronecker delta whereas, for the affine
$D$ model,
$\overline{\delta}$ is modified to
\begin{equation}
\overline{\delta}_{a,b} =\overline{\delta}_{b,a}=\left\{ \begin{array}{ll}
                 1 , &\mbox{if $a=b$} \\
                 1 , &\mbox{if $(a,b)=(1,2)$ or $(L-1,L)$} \\
                 0 ,           &\mbox{otherwise.}
                        \end{array}   \right.
\end{equation}

The classical and affine \ade models admit off-critical elliptic extensions
to the trigonometric
solutions of the Yang-Baxter equations given here. For the $A_{L-1}^{(1)}$
and $D_{L-1}^{(1)}$
models these occur for $\lambda=s\pi/L$ and $\lambda=\pi/(L-3)$
respectively. There is as yet no known
trigonometric solution to the Yang-Baxter equations for the IRF models
whose adjacency conditions are
given by the Dynkin diagrams of the exceptional affine Lie algebras
$E_6^{(1)}$, $E_7^{(1)}$ and
$E_8^{(1)}$. However, rational solutions are known~\cite{Ginsparg:89?} for
all the affine \ade
algebras
\begin{equation}
\wt Wabcdu=
 {\lambda -u \over  \lambda } \delta_{a,c} A_{a,b}A_{a,d} +
       {u \over \lambda } \sqrt{{S_a S_c \over S_b S_d}}\delta_{b,d}
A_{a,b}A_{b,c} \label{eq:aEface}
\end{equation}
where $\lambda$ is arbitrary.

The face weights of all the critical \ade models are invariant under the
symmetries of the Dynkin
diagrams and satisfy the crossing symmetry
\begin{equation}
\wt Wabcdu=
\sqrt{S_a S_c \over S_bS_d }\
\wt Wdabc{\lambda -u}.
\end{equation}

\subsection{ Dilute \ade Models}

Recently, dilute \ade models were constructed \cite{WaNiSe:92,Roch:92} by
extending the methods of
Pasquier \cite{Pasq:87} and Owczarek and Baxter \cite{OwBa:87}.
These models allow adjacent sites on the lattice to be either in the same
state or adjacent states on the
Dynkin diagram.
The face weights of the  dilute \ade models are \cite{WaNiSe:92,Roch:92}
\begin{eqnarray}
& &\wt Wabcdu
=\rho_1 (u)
  \delta_{a,b,c,d} +\rho_2 (u) \delta_{a,b,c} A_{a,d}+\rho_3 (u)
\delta_{a,c,d} A_{a,b} \nonumber \\
& &\quad \mbox{} +\sqrt{S_a \over S_b}\rho_4 (u) \delta_{b,c,d} A_{a,b}
     +\sqrt{S_c \over S_a}\rho_5 (u) \delta_{a,b,d} A_{a,c}
     +\rho_6 (u) \delta_{a,b} \delta_{c,d} A_{a,c}  \\
& &\quad \mbox{}  +\rho_7 (u) \delta_{a,d} \delta_{c,b} A_{a,b} +\rho_8 (u)
\delta_{a,c} A_{a,b} A_{a,d} +
       \sqrt{{S_a S_c \over S_b S_d}} \rho_9 (u) \delta_{b,d} A_{a,b}
A_{b,c} \nonumber
\end{eqnarray}
where the generalized Kronecker delta is $1$ if all its arguments take the
same value and is
zero if its arguments are not the same value. The $A_{a,b}$ are elements of
the adjacency
matrix for the corresponding classical or affine Dynkin diagram of
Figure~1. The weight functions are
\begin{eqnarray}
& & \rho_1 (u)={\sin (2\lambda ) \cos (3\lambda ) +\sin u \cos (u+3\lambda
) \over
                  \sin (2\lambda )\cos (3\lambda )},
       \;\;\; \rho_2 (u)=\rho_3 (u)={\cos (u+3\lambda )\over \cos (3\lambda
)} \nonumber \\
& & \rho_4 (u)=\rho_5 (u)={\sin u \over \cos (3\lambda )},
\;\;\;\;\;\;\;\;\;\;\;\;\;\;\;
         \;\;\;\;\;\;\;\;\;\;\;\;\;\;\;
       \;\; \rho_6 (u)=\rho_7 (u)={\sin u \cos (u+3\lambda ) \over \sin
(2\lambda )\cos (3\lambda )}\nonumber \\
& & \rho_8 (u)={\sin (u+2\lambda ) \cos (u+3\lambda ) \over \sin (2\lambda
)\cos (3\lambda )},\;\;\;\;\;\;\;\;\;\;\;\;\;
         \;\;\;\;\;\;
       \;\; \rho_9 (u)={\sin u \cos (u+\lambda ) \over \sin (2\lambda )\cos
(3\lambda )}
\end{eqnarray}
where the crossing parameter $\lambda = {\pi \over 4} (1\pm{1\over h})$.
The dilute models admit an off-critical elliptic solution to the
Yang-Baxter equations only for the classical $A$
models.

\section{Adjacency Matrix Intertwiners}
\setcounter{equation}{0}

Consider a connected graph such as one of the Dynkin diagrams in Figure~1.
Each node of the graph  is taken as
a possible spin state for spins placed on the square lattice. States at
adjacent sites of the square lattice
must be adjacent on the graph.  It turns out that there is a deep relation
between two lattice models
constructed in this way if an intertwiner can be constructed between them.
The first indication of the
existence of such an intertwiner is the existence of an intertwiner between
the adjacency matrices of the two
graphs. In this section we discuss intertwiners at the level of the
adjacency matrices focusing on the \ade
Dynkin diagrams. We will see that much of the theory of this section
carries over, more or less directly, to
the level of the row transfer matrices where the results and consequences
are not so trivial.

Consider the adjacency matrices of the \ade Dynkin diagrams shown in
Figure~1. It is straightforward to
calculate the eigenvalues and orthonormalized eigenvectors.  Remarkably,
the adjacency matrix eigenvalues can
all be written in the form $\Lambda_j=2\cos(m_j\pi/h)$ where the Coxeter
number $h$ and Coxeter exponents
$m_j$ are listed in Table~1. From the Perron-Frobenius theorem, it is known
that such irreducible
nonnegative matrices possess a real and non-degenerate largest eigenvalue
and that the corresponding
Perron-Frobenius eigenvector $S$ has all its components nonnegative. These
Perron-Frobenius vectors are listed in
Table~2 for the \ade adjacency matrices. The critical \ade models are
constructed from these eigenstates as indicated
in the previous section. The maximal eigenvalue $\Lambda_{max}$ is a
strictly increasing function of the
matrix elements. It follows that the only connected graphs for which
$\Lambda_{max} =2$ are the Dynkin diagrams of
the  affine \ade Lie algebras. The allowed connected graphs for which
$\Lambda_{max} < 2$ are the Dynkin
diagrams of the classical \ade Lie algebras.

\subsection{Properties of Intertwiners}

Let $A$ and $G$ be adjacency matrices. In general, these are arbitrary
square matrices with
nonnegative integer elements. The adjacency matrix $C$ is said to
intertwine $A$ and $G$ if
\begin{equation}
   A C =C G. \label{eq:adjinter}
\end{equation}
In general $C$ is a rectangular matrix with nonegative integer elements. It
is easy to show that this
intertwining relation
\begin{equation}
   A\stackrel{C}{\sim}G \label{rel:adjinter}
\end{equation}
is an equivalence relation acting on symmetric matrices, that is, it is
(i)~reflexive, (ii)~symmetric and
(iii)~transitive:
\begin{eqnarray}
&&\ \ \mbox{(i)}\quad A\stackrel{I}{\sim}A\nonumber\\
&&\ \mbox{(ii)}\quad A\stackrel{C}{\sim}G\quad\mbox{implies}\quad
G\stackrel{C^T}{\sim}A\\
&&\mbox{(iii)}\quad A\stackrel{C}{\sim}B\quad\mbox{and}\quad
B\stackrel{C'}{\sim}G\quad\mbox{implies}\quad
A\stackrel{CC'}{\sim}G.\nonumber
\end{eqnarray}
The existence of the intertwiner reflects a symmetry relating the two
graphs associated with $A$ and $G$. In
particular, observe that the intertwining relation implies that
\begin{equation}
   [C C^T,A]=[C^T C,G]=0
\end{equation}
so that the symmetry operators $C C^T$ and $C^T C$ commute with $A$ and $G$
respectively and their eigenvalues
can be used as quantum numbers labelling the eigenvectors of $A$ and $G$.

The existence of an
intertwiner also implies an overlap between the eigenvalues of $A$ and $G$.
Let us suppose that $x$ is a simultaneous
eigenvector of $G$ and $C^T C$, that is,
\begin{equation}
   Gx=\lambda x,\qquad C^T Cx=\mu x.
\end{equation}
Then it follows from the intertwining relation that
\begin{equation}
   A(Cx)=CGx=\lambda Cx
\end{equation}
so that $\lambda$ is an eigenvalue of $A$ with eigenvector $Cx$ provided
that $Cx\ne 0$, or equivalently,
$x^TC^TCx=\mu x^T x\ne 0$. Hence $Cx$ is an eigenvector of $A$ provided $x$
is not annihilated by the symmetry
operator $C^T C$. Similarly, by the symmetry of the intertwining relation,
it follows that if $y$ is a
simultaneous eigenvector of $A$ and $CC^T$ then $C^Ty$ is an eigenvector of
$G$ provided $y$ is not annihilated
by the symmetry operator $CC^T$.

Note that if $x=S$ is the Perron-Frobenius eigenvector of $G$ then, since
$C$ is
nonnegative, $Cx$ is a positive vector and is therefore the
Perron-Frobenius eigenvector of $A$. Thus, if an
intertwiner exists, the maximal eigenvalues of $A$ and $G$ are necessarily
in common.
Notice also that there is an intertwining relation between the symmetry
operators $CC^T$ and
$C^TC$, namely,
\begin{equation}
   CC^T\stackrel{C}{\sim}C^TC.
\end{equation}
It follows immediately from the previous considerations that all the
nonzero eigenvalues of the symmetry
operators $CC^T$ and $C^TC$ are in common. The properties of the symmetry
operators are best understood by
writing them in terms of fusion adjacency matrices. At first sight, this
connection between intertwiners and
fusion may seem surprising.

\subsection{Adjacency Fusion Rules}

The Temperley-Lieb \ade models are given by representations of the affine
algebra $su(2)$. The higher-spin
representations of this spin algebra are obtained by taking tensor products
of the fundamental representation.
The analog of this process for the \ade face models is fusion. Starting
with a fundamental $A$, $D$ or $E$
solution of the Yang-Baxter equations it is possible to obtain a hierarchy
of ``higher-spin" solutions by
fusing blocks of faces together. The fusion hierachy of the classical \ade
lattice models has been discussed
by a number of authors \cite{DJKMO:87,DJKMO:88,BaRe:89,KlPe:92,Paul:91p}.
At each fusion level $n$, a different
adjacency condition is imposed on the states of the model. The adjacency
matrices of these fused models are
given by simple fusion rules.

Let $A$ be the adjacency matrix of a classical or affine \ade Dynkin
diagram. Then a fusion hierarchy of
mutually commuting adjacency matrices $A^{(n)}$ is defined by the recursion
relations
\begin{eqnarray}
& & A^{(n)} A^{(1)}=A^{(n-1)} + A^{(n+1)} \;   \nonumber \\
& & A^{(0)}=I ,\; \;\;\; A^{(1)}=A       \label{eq:adjfusion}
\end{eqnarray}
where $I$ is the identity matrix and $n=1,2,\ldots $ For the classical
\ade models these fusion rules close with $A^{(h-1)}=0$ and
\be
A^{(h-2)}=\left\{ \begin{array}{ll}
                I , &\quad \mbox{for$\;D_L, \; E_7\;$and$\;E_8 $}  \\
                R , &\quad \mbox{for$\;A_L\;$and$\;E_6$}
                \end{array}   \right.
\ee
where $h$ is the Coxeter number and $R$ is the corresponding height
reflection operator of the model
defined as
\be
R_{a,b}=\delta_{a,h-b}.
\ee
Using the adjacency intertwining relation (\ref{eq:adjinter}) and the
fusion rules it follows that the
same intertwining relations hold between the fused adjacency matrices, that
is,
\begin{equation}
 A^{(n)} C=C G^{(n)}.
\end{equation}
If we define the Chebyshev polynomials of the second kind ${\cal
U}^{(n)}(z)$ by the same recursion
(\ref{eq:adjfusion}) with ${\cal U}^{(0)}(z)=1$ and ${\cal U}^{(1)}(z)=z$
then the eigenvalues of $A^{(n)}$
are given by
\be
\Lambda^{(n)}_j={\cal U}^{(n)}(2\cos m_j).
\ee

The row transfer matrices of the fused classical \ade models also satisfy a
fusion hierarchy \cite{Paul:91p}.
Indeed, the matrices  $A^{(n)}$ appearing in the above fusion rules are
precisely the adjacency matrices for the level $n$ fused models. In this
way it is possible to obtain the
adjacency matrices for all the fused \ade lattice models. Note that the
fusion rules are also valid for the
affine models but, in that case, the hierarchy is infinite with no closure.

\subsection{Adjacency Intertwiners}

The simplest example of an intertwiner relates the $A$ and $D$ Dynkin
diagrams \cite{Roch:90}.  The $A_L$
diagram posesses a $\Z_2$ symmetry corresponding to reflection about the
midpoint of the graph. For odd $L$,
this symmetry can be used to construct an intertwiner with $D_{(L+3)/2}$ by
an orbifold procedure
\cite{FendGins}. Notice that the Dynkin diagrams $A_L$ and $D_{(L+3)/2}$
have the same Coxeter number and
hence the same maximal eigenvalue as required for the existence of an
intertwiner. To obtain this
intertwiner by the orbifold procedure, the nodes $n$ and $L-n+1$ of $A$ are
mapped to the node $n'$ of $D$ and
the midpoint node ${L+1\over 2}$ of $A$ is mapped to the two nodes
$({L+1\over 2})'$ and $\overline
{({L+1\over 2})'}$ of $D$.  We can draw the graphs of $A$, $D$ and the
intertwiner $C$ as in Figure~2.
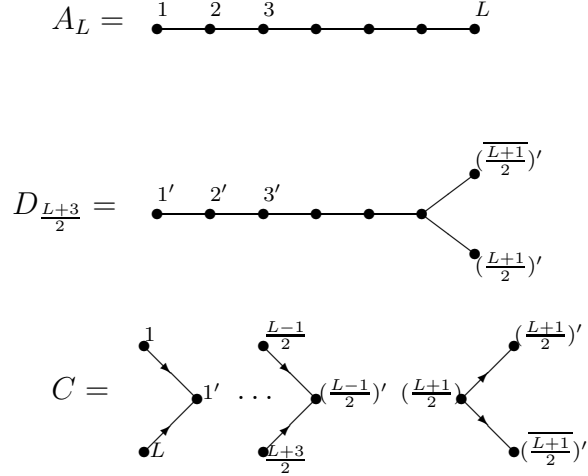
\begin{figure}[htbp]
\begin{center}
\begin{picture}(400,180)(-105,-90)

\put(-20,80){$A_L=\mbox{}$}
\multiput(20,80)(20,0){6}{\line(1,0){20}}
\multiput(20,80)(20,0){7}{\circle*{4}}
\put(20,85){\tiny 1} \put(40,85){\tiny 2} \put(60,85){\tiny 3}
\put(140,85){\tiny $L$}

\put(-35,10){$D_{L+3 \over 2}=\mbox{}$}
\multiput(20,10)(20,0){5}{\line(1,0){20}}
\put(120,10){\line(4,3){20}} \put(120,10){\line(4,-3){20}}
\multiput(20,10)(20,0){6}{\circle*{4}}
\multiput(140,25)(0,-30){2}{\circle*{4}}
\put(20,15){\tiny $1'$} \put(40,15){\tiny $2'$} \put(60,15){\tiny $3'$}
\put(140,28){\tiny $(\overline {L+1 \over 2})'$} \put(140,-12){\tiny $({L+1
\over 2})'$}

\put(-20,-60){$C=\mbox{}$}
\multiput(35,-60)(45,0){2}{\line(-1,1){10}}
\multiput(15,-40)(45,0){2}{\vector(1,-1){10}}
\multiput(35,-60)(45,0){2}{\line(-1,-1){10}}
\multiput(15,-80)(45,0){2}{\vector(1,1){10}}
\put(135,-60){\vector(1,1){10}}  \put(135,-60){\vector(1,-1){10}}
\put(155,-40){\line(-1,-1){10}}  \put(155,-80){\line(-1,1){10}}
\multiput(15,-40)(45,0){2}{\circle*{4}}
\multiput(15,-80)(45,0){2}{\circle*{4}}
\multiput(35,-60)(45,0){2}{\circle*{4}} \put(135,-60){\circle*{4}}
\multiput(155,-40)(0,-40){2}{\circle*{4}}

\put(37,-60){\tiny $1'$}  \put(81,-60){\tiny $({L-1\over 2})'$}
\put(15,-38){\tiny 1}   \put(60,-38){\tiny ${L-1\over 2}$}
\put(17,-82){\tiny $L$} \put(60,-85){\tiny ${L+3\over 2}$}
\put(112,-60){\tiny (${L+1\over 2}$)}  \put(157,-82){\tiny $(\overline {L+1
\over 2})'$}
\put(155,-38){\tiny $({L+1\over 2})'$}
\put(50,-60){$\ldots $}

\end{picture}
\caption{The graphs of $A_L$, $D_{(L+3)/2}$ and their
       intertwiner $C$. Notice that the edges of the graph of $C$ are
oriented as indicated by the arrows.}
\end{center}
\end{figure}
It is
easy to check that the adjacency matrix $C$ for the $A$--$D$ intertwiner
obtained from the graph satisfies the
intertwining relation $A C=C D$. The intertwining matrix $C$ is listed
explicitly in Appendix~A. By direct
matrix multiplication we see that the symmetry operators are given by
\begin{equation}
CC^T=(I+R_A),\qquad C^TC=(I+R_D)
\end{equation}
where the matrices $R_A$ and $R_D$ implement the $\Z_2$ symmetry on $A$ and
$D$ respectively:
\be
R_A=A^{(L-1)}=\smat{0&0&\ldots&0&1\cr
          0&0&\ldots&1&0\cr
          \vdots&\vdots&\ddots&\vdots&\vdots\cr
          0&1&\ldots&0&0\cr
          1&0&\ldots&0&0},\qquad
R_D=\smat{1&0&\ldots&0&0\cr
          0&1&\ldots&0&0\cr
          \vdots&\vdots&\ddots&\vdots&\vdots\cr
          0&0&\ldots&0&1\cr
          0&0&\ldots&1&0}
\ee

Since $R_A$ and $R_D$ are involutions, the eigenvalues of the symmetry
operators are $\mu=0,2$. Clearly, the
eigenvalues of $A$ and $D$ with even eigenvectors ($\mu=2$) under the
$\Z_2$ symmetry will be in common and the
eigenvalues with odd eigenvectors ($\mu=0$) will not be in common. The
eigenvalues of $R_A$ are given by
\be
r_A={\cal U}^{(L-1)}(2\cos m_j)=\cases{1,&$m_j$ odd\cr -1,& $m_j$ even\cr}
\ee
so the eigenvalues of $A$ with odd Coxeter exponents are in common. The
overlap of these eigenvalues and
the labelling by quantum numbers is shown pictorially in Figure~3.
\begin{figure}[htbp]
%\hspace{.7in}\mbox{}\hfil\scaledpicture 5.11in by 2.92in
%(adjeig scaled 500)\hfil\mbox{}
\[\epsfbox{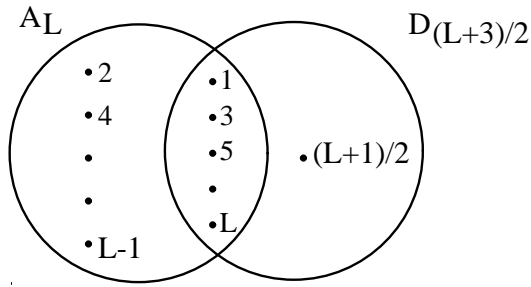}\]
\caption{The overlap of the eigenvalues of and $A_L$ and $D_{(L+3)/2}$
labelled by the Coxeter exponents
$m_j$. The common eigenvalues are even $(r_A=r_D=1)$ and the others are odd
$(r_A, r_D=-1)$ under the $\Z_2$
symmetry.}
\end{figure}

Intertwiners for all the adjacency matrices of the classical and affine
\ade algebras are listed in
Appendies~A and B. The classical adjacency intertwiners were obtained
previously by di Francesco and Zuber
\cite{FrZu:89p}. For the exceptional classical algebras we find
\be
C C^T=\cases{
I+A^{(6)},& for $A_{11}$--$E_6$\cr
I+A^{(8)}+A^{(16)},& for $A_{17}$--$E_7$\cr
I+A^{(10)}+A^{(18)}+A^{(28)},& for $A_{29}$--$E_8$\cr}
\ee
\be
C^T C=\cases{
I+E^{(6)},& for $A_{11}$--$E_6$\cr
I+E^{(8)}+E^{(16)},& for $A_{17}$--$E_7$\cr
I+E^{(10)}+E^{(18)}+E^{(28)},& for $A_{29}$--$E_8$\cr}
\ee
The eigenvalues of the symmetry operators can thus be written in terms of
the polynomials
${\cal U}^{(n)}(2\cos m_j)$. In each case they vanish except when $m_j$ is
a Coxeter exponent of $E_{6,7,8}$.

\subsection{Intertwiners and Projectors}

Loosely speaking, the action of intertwiners is to project onto the
eigenspaces of the common eigenvalues.
This suggests using projectors to construct the intertwiners. This is
indeed possible. Let $|G|$ denote the
number of vertices of $G$. The intertwining matrix $C$ between $A$ and $G$
is then a
$|A|\times |G|$ matrix. Let $|\psi^{(j)}\rangle$ denote the orthonormalized
eigenvectors of $A$ and
$|\phi^{(j)}\rangle$ denote those of $G$. Then the combination of
projectors
\be
C=\sum_j c_j|\psi^{(j)}\rangle\langle\phi^{(j)}|
\ee
where $c_j$ are coefficients and the summation is over the common
eigenvalues (or equivalently the common
Coxeter exponents) is a natural candidate for an  intertwiner. In
particular, it automatically satisfies the
intertwining relation (\ref{eq:adjinter}). But we require in addition that
the matrix $C$ has nonnegative
integer elements. So care is needed to choose the coefficients $c_j$
properly. A suitable choice
\cite{FrZu:89p} for the classical $A$--$D$ and $A$--$E$ intertwiners is
\be
C=\sum_j {\phi_1^{(j)}\over
\psi_1^{(j)}}|\psi^{(j)}\rangle\langle\phi^{(j)}|.
\ee
This choice gives the intertwiners listed in Appendix~A.

The affine case is complicated by the occurence of
degenerate eigenvalues. Let us use subscripts to distinguish the vectors
with the same Coxeter exponent in the
case of degeneracy.  The $A_{L-1}^{(1)}$--$D_{{L\over 2}+2}^{(1)}$
intertwiner can be obtained by the orbifold
procedure and this gives the intertwiner
\be
C=\cases{\disp{\sqrt{2}\sum_{m=0}^{L/2}\ket{\psi^{(2m)}}\bra{\phi^{(2m)}}},
&$L/2$ odd\cr
\disp{\sqrt{2}\sum_{m=1}^{L/2}(1-\delta_{m,L/4})\ket{\psi^{(2m)}}
\bra{\phi^{(2m)}}+\sqrt{2}\ket{\psi^{(L/2)}}\bra{\phi_1^{(L/2)}}},&$L/2$
even\cr}
\ee
We have also found the coefficients $c_j$ to obtain the adjacency matrix
intertwiner for the affine $A$--$E$
cases. The adjacency matrix interwiners for $A_5^{(1)}$--$E_6^{(1)}$,
$A_7^{(1)}$--$E_7^{(1)}$
and $A_9^{(1)}$--$E_8^{(1)}$ are given respectively by
\bea
C&=&2\ket{\psi^{(0)}}\bra{\phi^{(0)}}+
      \sqrt{2} \ket{\psi^{(2)}}\bra{\phi^{(2)}_1}+
      \sqrt{2} \ket{\psi^{(4)}}\bra{\phi^{(4)}_1}+
  2\ket{\psi^{(6)}}\bra{\phi^{(6)}}\\
C&=&\sqrt{6} \ket{\psi^{(0)}}\bra{\phi^{(0)}} +\sqrt{2}
\ket{\psi^{(2)}}\bra{\phi^{(3)}}\sqrt{2}
     +\sqrt{2} \ket{\psi^{(4)}}\bra{\phi^{(6)}_1}  \nonumber \\
& &\;\;\; + \ket{\psi^{(6)}}\bra{\phi^{(9)}}  +  \sqrt{6}
\ket{\psi^{(8)}}\bra{\phi^{(12)}}\\
C&=&\sqrt{12} \ket{\psi^{(0)}}\bra{\phi^{(0)}} +\sqrt{2}
\ket{\psi^{(2)}}\bra{\phi^{(6)}}
     +\sqrt{2} \ket{\psi^{(4)}}\bra{\phi^{(12)}}  \nonumber \\
& &\;\;\; +\sqrt{2} \ket{\psi^{(6)}}\bra{\phi^{(18)}}+\sqrt{2}
\ket{\psi^{(8)}}\bra{\phi^{(24)}}
   +\sqrt{12} \ket{\psi^{(10)}}\bra{\phi^{(30)}}.
\eea

Notice that the coefficients for the affine $A$--$G$ intertwiners are all
$\sqrt{2}$ except the first and last
which are given by
\be
c_{0}=\sqrt{{\sum_i (S^G_i)^2}\over h^A}.
\ee
The Coxeter exponents $m_j$ of the common eigenvalues can be simply read
off from
these formulas.  In this way we obtain the
adjacency matrices for the affine $A$--$D$ and $A$--$E$ intertwiners given
in Appendix~B.

%%%%%%%%%%%%%%%%%%%%%%%%%%%%%%%%%%%%%%%%%%%%%%%%

\section{Face Algebra and Cell Calculus}
\setcounter{equation}{0}

The face algebra of an \ade model is generated by the face transfer operators
$X_j(u)$. Let $a=\{a_1,a_2,\ldots,a_N\}$ and $a'=\{a'_1,a'_1,\ldots,a'_N\}$
be the configurations of two  consecutive rows of spins of the lattice and
apply periodic boundary conditions so that $a_{N+1}=a_1$ and $a'_{N+1}=a'_1$.
Then the elements of the face transfer matrices $X_j(u)$ are given by
\be
\bra a X_j(u)\ket{a'}=\wt
W{a_{j}}{a_{j+1}}{a'_{j}}{a_{j-1}}u
    \prod_{k\ne j}\delta(a_{k},a'_{k})
=
\setlength{\unitlength}{.0943in}
\diagface{a_j}{a_{j+1}}{a'_j}{a_{j-1}}{W}{0}{0}
\setlength{\unitlength}{.1in}
\label{FaceTM}
\ee
Notice that we have oriented the edges of the face with arrows. These are
used for reference and must match the orientation of the bonds on the square
lattice with all arrows pointing to the right or down. In terms of
face transfer operators the Yang-Baxter equation takes the form
\be
X_j(u)X_{j+1}(u+v)X_j(v)=X_{j+1}(v)X_j(u+v)X_{j+1}(u)\label{YBE}
\ee
or graphically
\be
\setlength{\unitlength}{.1in}
\rule[-8\unitlength]{0in}{16\unitlength}
\raisebox{1.5\unitlength}{$
\diagface{a_j}{a_{j+1}\ \ \ }{}{a_{j-1}}{u}{0}{0}\punit{-9}
\diagface{}{a_{j+2}}{}{}{u+v}{3}{-3}\punit{-9}
\diagface{}{\ \ \ a'_{j+1}}{a'_j}{a_{j-1}}{v}{6}{0}
$}\punit7=\punit2
\raisebox{-1.5\unitlength}{$
\diagface{a_{j+1}}{a_{j+2}}{}{a_j\ \ \ }{v}{0}{0}\punit{-9}
\diagface{}{}{}{a_{j-1}}{u+v}{3}{3}\punit{-9}
\diagface{}{a_{j+2}}{a'_{j+1}}{\ \ \ a'_j}{u}{6}{0}
$}
\ee
and the inversion or unitarity condition is
\be
X_j(u)X_j(-u)=\rho(u)\rho(-u)I.
\ee
In pictures the unitarity condition takes the form
\be
\begin{picture}(280,25)(50,30)

\put(100,30){$\disp{\sum_g}$}
\put(150,50){\vector(1,-1){10}} \put(150,50){\vector(-1,-1){10}}
\put(130,30){\line(1,1){10}}    \put(170,30){\line(-1,1){10}}

\put(130,30){\vector(1,-1){10}} \put(170,30){\vector(-1,-1){10}}
\put(150,10){\line(1,1){10}}    \put(150,10){\line(-1,1){10}}

\put(167,23){\tiny $g$} \put(150,53){\tiny $c$} \put(150,5){\tiny $a$}
\put(125,30){\tiny $d$}
%\multiput(159,15)(15,0){2}{\tiny $c_1$}
%\multiput(162,39)(10,0){2}{\tiny $g_2$}
%\put(131,41){\tiny $c_2$} \put(134,15){\tiny $g_1$}
%\put(201,41){\tiny $c_2'$} \put(200,15){\tiny $g_1'$}
\put(147,27){$u$}

\put(170,30){\circle*{3}}

\put(190,50){\vector(1,-1){10}} \put(190,50){\vector(-1,-1){10}}
\put(170,30){\line(1,1){10}}    \put(210,30){\line(-1,1){10}}

\put(170,30){\vector(1,-1){10}} \put(210,30){\vector(-1,-1){10}}
\put(190,10){\line(1,1){10}}    \put(190,10){\line(-1,1){10}}
\put(190,53){\tiny $c$} \put(190,5){\tiny $a$} \put(215,30){\tiny $d'$}
\put(180,27){$-u$}

\put(233,30){$= \rho(u)\rho(-u)\ \delta_{d,d'}$}
\end{picture}
\label{unit1}
\ee
\smallskip
%%%%%%%%%%%%%%%%%%%%%%%%%%%%%%%%%%%%%%%%%

\be
\begin{picture}(280,25)(50,30)

\put(100,30){$\disp{\sum_b}$}
\put(150,50){\vector(1,-1){10}} \put(150,50){\line(-1,-1){10}}
\put(130,30){\vector(1,1){10}}  \put(170,30){\line(-1,1){10}}

\put(130,30){\vector(1,-1){10}} \put(170,30){\line(-1,-1){10}}
\put(150,10){\vector(1,1){10}}  \put(150,10){\line(-1,1){10}}

\put(167,23){\tiny $b$} \put(150,53){\tiny $c$} \put(150,5){\tiny $a$}
\put(125,30){\tiny $d$}
%\multiput(159,15)(15,0){2}{\tiny $c_1$}
%\multiput(162,39)(10,0){2}{\tiny $g_2$}
%\put(131,41){\tiny $c_2$} \put(134,15){\tiny $g_1$}
%\put(201,41){\tiny $c_2'$} \put(200,15){\tiny $g_1'$}
\put(136,27){$\lambda-u$}

\put(170,30){\circle*{3}}

\put(190,50){\line(1,-1){10}} \put(190,50){\vector(-1,-1){10}}
\put(170,30){\line(1,1){10}}  \put(210,30){\vector(-1,1){10}}

\put(170,30){\line(1,-1){10}} \put(210,30){\vector(-1,-1){10}}
\put(190,10){\line(1,1){10}}  \put(190,10){\vector(-1,1){10}}
\put(190,53){\tiny $c$} \put(190,5){\tiny $a$} \put(213,30){\tiny $d'$}
\put(176,27){$\lambda+u$}

\put(223,30){$\disp{S_b S_d\over S_c S_a}\;\;=\rho(u)\rho(-u)\
              \delta_{d,d'}$}
\end{picture}
\label{unit2}
\ee

\vspace{.4in}

\noindent
where $I$ is the identity matrix, $\rho(u)={\sin(\lambda-u)/\sin\lambda}$
and the second unitarity condition is obtained from the first by using the
crossing symmetry. Taking the braid limits we obtain the braid matrices
\be
b_j^{\pm 1}=\lim_{u\to\mp i\infty}{X_j(u)\over \rho(u)}
\ee
which satisfy
\be
b_{j+1}b_jb_{j+1}=b_jb_{j+1}b_j.
\ee

Consider two \ade models with intertwining adjacency matrices $A$ and $G$
and face weights $W^A$ and $W^G$. Intertwiners between the
face weights, or equivalently the two face algebras, can be obtained by
constructing cells. This leads to the Ocneanu cell calculus
\cite{Ocneanu,Roch:90}. A cell is a face transfer operator $C_j$. The
elements of the cell $C_j$ are given by
\be
\bra a C_j(u)\ket{a'}=\weight
C{a_{j}}{a_{j+1}}{a'_{j}}{a_{j-1}}
    \prod_{k\ne j}\delta(a_{k},a'_{k})
=
\setlength{\unitlength}{.0943in}
\diagface{a_j}{a_{j+1}}{a'_j}{a_{j-1}}{}{0}{0}
\setlength{\unitlength}{.1in}
\label{cell}
\ee
where
\be
\setlength{\unitlength}{.1in}
\weight Cabcd = \punit2
\sqedges {c_1}{g_2}{c_2}{g_1}\sqface abcd{}{0}{.3}
\ee
are a set of complex numbers which, in general, can depend on bond variables
$c_1,c_2,g_1,g_2$ in addition to the spins $a,b,c,d$. Here the cells vanish
unless the spins $d,a$ are adjacent sites of $A$, the spins $c,b$ are
adjacent sites of $G$ and the spins $a,b$ and $d,c$ are adjacent sites of
the intertwining graph $C$. We adopt the convention that the value of a cell
is changed into its complex conjugate after a reflection
\be
\setlength{\unitlength}{.1in}
\sqedges {c_1}{g_2}{c_2}{g_1}\sqface abcd{}{0}{.3}\punit2=\punit1
\left(\quad\sqedges {c_1}{g_1}{c_2}{g_2}\sqface badc{}{0}{.3}\quad
\right)^*
\ee

We say that a cell system $C_j$ intertwines the face weights $W^A$ and $W^G$
if
\be
X_{j+1}^A C_jC_{j+1}=C_jC_{j+1}X_j^G\label{eq:faceinter}
\ee
or graphically
\be
\setlength{\unitlength}{.1in}
\rule[-8\unitlength]{0in}{16\unitlength}
\raisebox{1.5\unitlength}{$
\diagface{a_j}{a_{j+1}\ \ \ }{}{a_{j-1}}{W^A}{0}{0}\punit{-9}
\diagface{}{a_{j+2}}{}{}{}{3}{-3}\punit{-9}
\diagface{}{\ \ \ a'_{j+1}}{a'_j}{a_{j-1}}{}{6}{0}
$}\punit7=\punit2
\raisebox{-1.5\unitlength}{$
\diagface{a_{j+1}}{a_{j+2}}{}{a_j\ \ \ }{}{0}{0}\punit{-9}
\diagface{}{}{}{a_{j-1}}{}{3}{3}\punit{-9}
\diagface{}{a_{j+2}}{a'_{j+1}}{\ \ \ a'_j}{W^G}{6}{0}
$}
\ee
where for simplicity we have omitted the bond variables. We also require that
the cells satisfy the unitarity conditions
\be
\begin{picture}(350,25)(50,30)

\put(80,30){$\disp{\sum_{(b,c_1,g_2)}}$}
\put(150,50){\vector(1,-1){10}} \put(150,50){\vector(-1,-1){10}}
\put(130,30){\line(1,1){10}}    \put(170,30){\line(-1,1){10}}

\put(130,30){\vector(1,-1){10}} \put(170,30){\vector(-1,-1){10}}
\put(150,10){\line(1,1){10}}    \put(150,10){\line(-1,1){10}}

\put(167,23){\tiny $b$} \put(150,53){\tiny $c$} \put(150,5){\tiny $a$}
\put(125,30){\tiny $d$}\multiput(159,15)(15,0){2}{\tiny $c_1$}
\multiput(162,39)(10,0){2}{\tiny $g_2$}
\put(131,41){\tiny $c_2$} \put(134,15){\tiny $g_1$}
\put(201,41){\tiny $c_2'$} \put(200,15){\tiny $g_1'$}

\put(170,30){\circle*{2}}

\put(190,50){\vector(1,-1){10}} \put(190,50){\vector(-1,-1){10}}
\put(170,30){\line(1,1){10}}    \put(210,30){\line(-1,1){10}}

\put(170,30){\vector(1,-1){10}} \put(210,30){\vector(-1,-1){10}}
\put(190,10){\line(1,1){10}}    \put(190,10){\line(-1,1){10}}
\put(190,53){\tiny $c$} \put(190,5){\tiny $a$} \put(215,30){\tiny $d'$}

\put(233,30){$= \delta_{d,d'} \delta_{g_1,g_1'} \delta_{c_2,c_2'}\; $}
%\put(360,30){(I)}
\end{picture}
\label{cellunit1}
\ee

%%%%%%%%%%%%%%%%%%%%%%%%%%%%%%%%%%%%%%%%%

\be
\begin{picture}(350,25)(50,30)

\put(80,30){$\disp{\sum_{(b,c_1,g_2)}}$}
\put(150,50){\vector(1,-1){10}} \put(150,50){\line(-1,-1){10}}
\put(130,30){\vector(1,1){10}}  \put(170,30){\line(-1,1){10}}

\put(130,30){\vector(1,-1){10}} \put(170,30){\line(-1,-1){10}}
\put(150,10){\vector(1,1){10}}  \put(150,10){\line(-1,1){10}}

\put(167,23){\tiny $b$} \put(150,53){\tiny $c$} \put(150,5){\tiny $a$}
\put(125,30){\tiny $d$}\multiput(159,15)(15,0){2}{\tiny $c_1$}
\multiput(162,39)(10,0){2}{\tiny $g_2$}
\put(131,41){\tiny $c_2$} \put(134,15){\tiny $g_1$}
\put(201,41){\tiny $c_2'$} \put(200,15){\tiny $g_1'$}

\put(170,30){\circle*{2}}

\put(190,50){\line(1,-1){10}} \put(190,50){\vector(-1,-1){10}}
\put(170,30){\line(1,1){10}}  \put(210,30){\vector(-1,1){10}}

\put(170,30){\line(1,-1){10}} \put(210,30){\vector(-1,-1){10}}
\put(190,10){\line(1,1){10}}  \put(190,10){\vector(-1,1){10}}
\put(190,53){\tiny $c$} \put(190,5){\tiny $a$} \put(213,30){\tiny $d'$}

\put(223,30){$\disp{S_b S_d\over S_c S_a}\;\;=\delta_{d,d'}
  \delta_{g_1,g_1'} \delta_{c_2,c_2'}\; $}

%\put(360,30){(II)}
\end{picture}
\label{cellunit2}
\ee

\vspace{.3in}
The cell intertwiner relation (\ref{eq:faceinter}) and the two cell unitarity
conditions (\ref{cellunit1}) and (\ref{cellunit2}) are analogues of the
Yang-Baxter equation (\ref{YBE}) and the usual unitarity conditions
(\ref{unit1}) and (\ref{unit2}).

Once the intertwining relation (\ref{eq:adjinter}) for the adjacency matrices
is found, we can then construct the cell system satisfying the two unitary
conditions and the intertwining relation for the face weights. There is a
certain freedom amounting to a gauge freedom in the construction of the
cells. For example, the phase angles of the cells can be changed such that
the two unitary conditions and the intertwining relation are still satisfied.
Solving the two unitary conditions almost completely determines the cells
but  both the unitary conditions and the intertwining relation are needed to
complete the determination. If the cells are found, the intertwining relation
(\ref{eq:faceinter}) tells us that the two models have the same integrability
properties. In the following subsections it is shown that the cell systems
for the classical \ade models are closely related to their corresponding
adjacency matrices. In particular, the cells depend directly on the
Perron-Frobenius eigenvectors of the graphs. We also give a systematic way
to derive all the cells for intertwining relations of the classical and
affine \ade models.

\subsection{Classical $A$--$D$ Cells}
Let us first consider the cell calculus for the classical $A$--$D$ models.
Let $A$ and $D$ be the adjacency matrices of the $A_{L}$ and
$D_{L+3 \over 2}$ models with odd $L$ and suppose that $AC=CD$. We build the
cells starting from the adjacency matrix $C$ of the intertwiner. Consider the
directed bonds $(a,d)$ in $C$ with $a\in A$ and $d\in D$. Two such edges
$(a,d)$ and $(a',d')$ in $C$ are considered to be adjacent if they form an
allowed cell, that is, if $a$ and $a'$ are adjacent in $A$ and $d$ and $d'$
are adjacent in $D$. In this way we represent a cell pictorially by
projecting it onto two adjacent edges of $C$ while maintaining the
orientation
\be
\setlength{\unitlength}{.1in}
\weight Cad{d'}{a'} = \punit2
\sqface ad{d'}{a'}{}{0}{.3}
\punit2=\punit1
\setlength{\unitlength}{1pt}
\begin{picture}(40,30)(0,0)
\put(10,23){\scriptsize $(a',d')$}
\put(20,15){\circle*{3}}
\put(20,-10){\line(0,1){25}}
\put(20,15){\vector(0,-1){13}}
\put(20,-10){\circle*{3}}
\put(10,-23){\scriptsize $(a,d)$}
\end{picture}
\ee
The allowed cells are thus specified by an adjacency matrix or graph of
adjacent edges of $C$.

The adjacency matrix intertwiner $C$ is an
$L$ by ${L+3 \over 2}$ rectanglar matrix with the elements
$0$ or $1$. The nonzero elements represent allowed edges $(a,d)$ of $C$. Thus
two nonzero elements are
adjacent or neighbours if the row numbers of the elements are  allowed
nearest-neighbour pairs of the
$A$-model and column numbers of the elements are allowed  nearest-neighbour
pairs of the $D$-model. We
connect all of the neighbour elements with oriented bonds such that all
arrows around one nonzero element go
into or out of the element. Therefore we obtain two graphs from the adjacency
matrix $C$ of the intertwiner
as pictured in Figure~4.

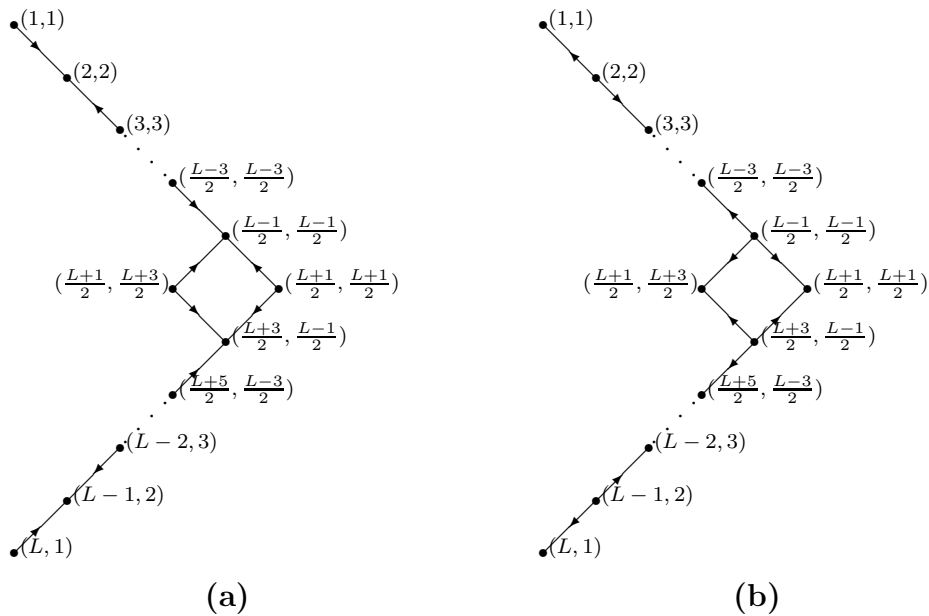
\begin{figure}[htb]
\begin{center}
\begin{picture}(300,240)(0,0)

\put(5,220){\vector(1,-1){10}}
\put(15,210){\line(1,-1){10}}
\put(25,200){\line(1,-1){10}}
\put(45,180){\vector(-1,1){10}}
\put(65,160){\vector(1,-1){10}}
\put(75,150){\line(1,-1){10}}
\put(85,140){\line(1,-1){10}}
\put(105,120){\vector(-1,1){10}}

\put(65,120){\vector(1,-1){10}}
\put(75,110){\line(1,-1){10}}

\put(85,140){\line(-1,-1){10}}
\put(65,120){\vector(1,1){10}}

\put(105,120){\vector(-1,-1){10}}
\put(95,110){\line(-1,-1){10}}
\put(85,100){\line(-1,-1){10}}
\put(65,80){\vector(1,1){10}}
\put(45,60){\vector(-1,-1){10}}
\put(35,50){\line(-1,-1){10}}
\put(25,40){\line(-1,-1){10}}
\put(5,20){\vector(1,1){10}}

\multiput(47,178)(5,-5){4}{\circle*{1}}
\multiput(47,62)(5,5){4}{\circle*{1}}

\put(7,220){\scriptsize(1,1)}
\put(27,200){\scriptsize(2,2)}
\put(47,180){\scriptsize(3,3)}
\put(67,160){\tiny $({L-3 \over 2},{L-3 \over 2})$}
\put(87,140){\tiny $({L-1 \over 2},{L-1 \over 2})$}
\put(107,120){\tiny $({L+1 \over 2},{L+1 \over 2})$}
\put(87,100){\tiny $({L+3 \over 2},{L-1 \over 2})$}
\put(20,120){\tiny $({L+1 \over 2},{L+3 \over 2})$}
\put(7,20){\tiny $(L,1)$}
\put(27,40){\tiny $(L-1,2)$}
\put(47,60){\tiny $(L-2,3)$}
\put(67,80){\tiny $({L+5 \over 2},{L-3 \over 2})$}

\multiput(5,220)(20,-20){6}{\circle*{3}}
\multiput(5,20)(20,20){5}{\circle*{3}}
\put(65,120){\circle*{3}}

\put(77,0){\bf (a)}

\put(205,220){\line(1,-1){10}}
\put(225,200){\vector(-1,1){10}}
\put(225,200){\vector(1,-1){10}}
\put(245,180){\line(-1,1){10}}
\put(265,160){\line(1,-1){10}}
\put(285,140){\vector(-1,1){10}}
\put(285,140){\vector(1,-1){10}}
\put(305,120){\line(-1,1){10}}

\put(265,120){\line(1,-1){10}}
\put(285,100){\vector(-1,1){10}}

\put(285,140){\vector(-1,-1){10}}
\put(265,120){\line(1,1){10}}

\put(305,120){\line(-1,-1){10}}
\put(285,100){\vector(1,1){10}}
\put(285,100){\vector(-1,-1){10}}
\put(265,80){\line(1,1){10}}
\put(245,60){\line(-1,-1){10}}
\put(225,40){\vector(1,1){10}}
\put(225,40){\vector(-1,-1){10}}
\put(205,20){\line(1,1){10}}

\multiput(247,178)(5,-5){4}{\circle*{1}}
\multiput(247,62)(5,5){4}{\circle*{1}}

\put(207,220){\scriptsize(1,1)}
\put(227,200){\scriptsize(2,2)}
\put(247,180){\scriptsize(3,3)}
\put(267,160){\tiny $({L-3 \over 2},{L-3 \over 2})$}
\put(287,140){\tiny $({L-1 \over 2},{L-1 \over 2})$}
\put(307,120){\tiny $({L+1 \over 2},{L+1 \over 2})$}
\put(287,100){\tiny $({L+3 \over 2},{L-1 \over 2})$}
\put(220,120){\tiny $({L+1 \over 2},{L+3 \over 2})$}
\put(207,20){\tiny $(L,1)$}
\put(227,40){\tiny $(L-1,2)$}
\put(247,60){\tiny $(L-2,3)$}
\put(267,80){\tiny $({L+5 \over 2},{L-3 \over 2})$}

\multiput(205,220)(20,-20){6}{\circle*{3}}
\multiput(205,20)(20,20){5}{\circle*{3}}
\put(265,120){\circle*{3}}

\put(277,0){\bf (b)}

\end{picture}
\end{center}
\caption{The cell graphs showing all the allowed cells intertwining the
classical $A$--$D$ models.
The number of vertices in the graph is given by the number of nonzero
elements of the adjacency intertwiner
$C$.}  \end{figure}

The oriented bonds in the two graphs of Figure~4 represent all of the
allowed cells intertwining the classical
$A$--$D$ models. Each bond occurs twice with arrows in the opposite
direction. So each bond corresponds to two
different nonzero cells. We refer to the two graphs of Figure~4 as cell
graphs. The cells in the two graphs of
Figure~4 are independent of each other so we can consider these two graphs
separately.  There are only two
classes of bonds in these graphs. We distinguish them as single bonds and
face bonds. The latter ones form
quadrilateral faces. Using the two unitarity conditions we find that the
values of cells corresponding to the
single bonds are independent and given by unimodular phase factors. In fact
we can choose them to be $1$.  On
the other hand, the cells corresponding to face bonds are involved in a set
of dependent equations through the
unitarity conditions. So to obtain the cells for face bonds, we initially
satisfy just the first unitarity
condition. The four cells on the face of Figure~4a, for example, are related
to each other by the first
unitarity condition and we can fix their values as
\begin{equation}
\begin{picture}(300,40)(0,20)

\put(110,50){\circle*{3}}
\put(110,50){\line(1,-1){10}}
\put(130,30){\vector(-1,1){10}}
\put(130,30){\circle*{3}}

\put(90,30){\circle*{3}}
\put(90,30){\vector(1,-1){10}}
\put(110,10){\line(-1,1){10}}
\put(110,10){\circle*{3}}

\put(110,50){\circle*{3}}
\put(110,50){\line(-1,-1){10}}
\put(90,30){\vector(1,1){10}}
\put(90,30){\circle*{3}}

\put(130,30){\circle*{3}}
\put(130,30){\vector(-1,-1){10}}
\put(110,10){\line(1,1){10}}
\put(110,10){\circle*{3}}

\put(90,55){\tiny $({L-1 \over 2},{L-1 \over 2})$}
\put(130,30){\tiny $({L+1 \over 2},{L+1 \over 2})$}
\put(90,5){\tiny $({L+3 \over 2},{L-1 \over 2})$}
\put(45,30){\tiny $({L+1 \over 2},{L+3 \over 2})$}

\put(260,50){\line(1,-1){20}}
\put(260,10){\line(-1,1){20}}
\put(260,50){\line(-1,-1){20}}
\put(260,10){\line(1,1){20}}

\put(185,30){=}
\put(210,40){\scriptsize $-\sin m$}
\put(275,40){\scriptsize $\cos m$}
\put(215,20){\scriptsize $\cos m$}
\put(275,20){\scriptsize $\sin m$}
\end{picture}
\label{eq:choice1}
\end{equation}

\bigskip\noindent
Obviously the values of these four cells satisfy the first unitarity
condition for any angle $m$. They
should also satisfy the equations from the second unitarity condition.
So we can take one of these equations to fix the parameter $m$.  For
instance, from
the normalization
\be
\sin^2 m\ {{S^A_{(L+1)/2} S^D_{(L-1)/2} \over
S^A_{(L+3)/2} S^D_{(L+1)/2}}}=1.
\ee
it follows that $m={\pi/4}$.

In (\ref{eq:choice1}) we have chosen the cell
\be
\weight C{(L-1)/2}{(L-1)/2}{(L+3)/2}{(L+1)/2}=-\sin m
\ee
to be negative and all the others
positive. We can change this to take any one of the four cells to be
negative with the other three cells
positive. Such choices also satisfy the unitarity conditions. The cells
around the face in Figure~4a
need only satisfy the first unitarity condition and we can set
them to $1$ as we did for single bonds. But the four cells around the face
of Figure~4b
are related by the second unitarity condition. This is satisfied by taking
only one of
the four cells to be negative, say
\begin{equation}
\begin{picture}(80,15)(60,0)
\put(0,-5){\tiny $({L+1 \over 2},{L+3 \over 2})$}
\put(40,4){\circle*{3}}
\put(40,4){\line(1,0){20}}
\put(80,4){\vector(-1,0){20}} \put(80,4){\circle*{3}}
\put(80,-5){\tiny $({L-1 \over 2},{L-1 \over 2})$}
\put(120,4){ $=-1.$}
\end{picture}
\label{eq:choicetwo}
\end{equation}

\medskip\noindent
Alternatively, this cell could be taken as $1$ and the other three to be
$-1$. So we have derived all
cells for the $A$--$D$ models with odd $L$. These results are the same as
those in \cite{Roch:90}.
The cells for the $A$--$D$ models  with even $L$ can be taken as $1$ because
there are only single bonds in
the cell graphs.

Lastly, we need to check that all of the cells satisfy the intertwining
relations for the face weights of the  $A$ and $D$ models.
Notice that the faces in the cell graphs are of two types:
\be
\mbox{Type A:}\qquad
\begin{picture}(54,24)(-3,12)

   \put(20,40){\line(1,-1){10}}   \put(20,40){\circle*{3}}
   \put(40,20){\vector(-1,1){10}} \put(40,20){\circle*{3}}

   \put(0,20){\vector(1,-1){10}} \put(0,20){\circle*{3}}
   \put(20,0){\line(-1,1){10}}  \put(20,0){\circle*{3}}

    \put(20,40){\line(-1,-1){10}}
   \put(0,20){\vector(1,1){10}}

    \put(40,20){\vector(-1,-1){10}}
    \put(20,0){\line(1,1){10}}
    \end{picture}
\qquad\mbox{or}\qquad\mbox{Type B:}\qquad
\begin{picture}(54,24)(-3,12)

   \put(20,40){\vector(1,-1){10}}   \put(20,40){\circle*{3}}
   \put(40,20){\line(-1,1){10}} \put(40,20){\circle*{3}}

   \put(0,20){\line(1,-1){10}} \put(0,20){\circle*{3}}
   \put(20,0){\vector(-1,1){10}}  \put(20,0){\circle*{3}}

    \put(20,40){\vector(-1,-1){10}}
   \put(0,20){\line(1,1){10}}

    \put(40,20){\line(-1,-1){10}}
    \put(20,0){\vector(1,1){10}}
    \end{picture}
\ee

\bigskip\noindent
Using these we can summarize the rules to
obtain the cells as follows:

\begin{enumerate}
\item The cells on single bonds are either $1$ or $-1$.
\item The absolute values of the cells on a face of type~A
are given by
\be
      \begin{picture}(100,30)(-3,15)
      \put(50,40){\line(1,-1){20}}
      \put(50,0){\line(-1,1){20}}
      \put(50,40){\line(-1,-1){20}}
      \put(50,0){\line(1,1){20}}

       \put(7,30){\scriptsize $\sin m$}
       \put(65,30){\scriptsize $\cos m$}
       \put(7,10){\scriptsize $\cos m$}
       \put(65,10){\scriptsize $\sin m$}
       \end{picture}
\ee

\medskip\noindent
One of these four cells
should be negative and the other three positive. The angle
$m$ is $\pi/4$ if only one such face exists in the graph.

\item The absolute values of cells on isolated faces of type~B
are $1$ and either one cell is negative and three are positive or one cell
is positive and three are negative.
\end{enumerate}
These rules determine the cells on single bonds and the absolute values of
the cells on faces.  The
intertwining relation (\ref{eq:faceinter}) can be used to determine which
cells on faces are negative.
In particular, it can be verified directly that the choice of the  minus
sign in (\ref{eq:choice1}) and
(\ref{eq:choicetwo}) ensures that the intertwining  relation of the $A$--$D$
models is satisfied.

\subsection{Classical $A_{11}$--$E_6$ and $A_{17}$--$E_7$ Cells}

Now consider the $A_{11}$--$E_6$ intertwiner. In this case the adjacency
matrix $C$ given in Appendix~A is a
$12$ by $6$ matrix with the entries $0$ or $1$ \cite{FrZu:89p}. We can
derive the cell graphs of all allowed
cells as we did for the $A$--$D$ intertwiner. The result is shown in
Figure~5.

\begin{figure}[htb]
\begin{center}
\begin{picture}(310,300)(-10,-10)

\put(10,260){\vector(1,-1){10}} \put(10,260){ \tiny{(1,1)}}
\put(10,260){\circle*{3}}
\put(30,240){\line(-1,1){10}}
\put(30,240){\circle*{3}}
\put(30,240){\line(1,-1){10}}  \put(30,240){\tiny{(2,2)}}
\put(30,240){\circle*{3}}
\put(50,220){\vector(-1,1){10}}
\put(50,220){\circle*{3}}
\put(50,220){\vector(1,-1){10}} \put(50,220){\tiny{(3,3)}}
\put(50,220){\circle*{3}}
\put(70,200){\line(-1,1){10}}
\put(70,200){\circle*{3}}
\put(70,200){\line(1,-1){10}}  \put(70,200){\tiny{(4,4)}}
\put(70,200){\circle*{3}}
\put(90,180){\vector(-1,1){10}} \put(90,180){\tiny{(5,5)}}
\put(90,180){\circle*{3}}

\put(30,200){\line(1,-1){10}} \put(10,200){ \tiny{(4,6)}}
\put(30,200){\circle*{3}}
\put(50,180){\vector(-1,1){10}}
\put(50,180){\circle*{3}}
\put(50,180){\vector(1,-1){10}} \put(30,180){ \tiny{(5,3)}}
\put(50,180){\circle*{3}}
\put(70,160){\line(-1,1){10}}
\put(70,160){\circle*{3}}

\put(30,160){\line(1,-1){10}} \put(10,160){ \tiny{(6,2)}}
\put(30,160){\circle*{3}}
\put(50,140){\vector(-1,1){10}}
\put(50,140){\circle*{3}}
\put(50,140){\vector(1,-1){10}}
\put(50,140){\circle*{3}}
\put(70,120){\line(-1,1){10}}
\put(70,120){\circle*{3}}

\put(10,140){\vector(1,-1){10}} \put(-10,140){ \tiny{(7,1)}}
\put(10,140){\circle*{3}}
\put(30,120){\line(-1,1){10}}
\put(30,120){\circle*{3}}
\put(30,120){\line(1,-1){10}}  \put(10,120){ \tiny{(8,2)}}
\put(30,120){\circle*{3}}
\put(50,100){\vector(-1,1){10}}
\put(50,100){\circle*{3}}
\put(50,100){\vector(1,-1){10}} \put(30,100){ \tiny{(9,3)}}
\put(50,100){\circle*{3}}
\put(70,80){\line(-1,1){10}}
\put(70,80){\circle*{3}}
\put(70,80){\line(1,-1){10}} \put(70,80){\tiny{(10,4)}}
\put(70,80){\circle*{3}}
\put(90,60){\vector(-1,1){10}} \put(90,60){\tiny{(11,5)}}
\put(90,60){\circle*{3}}

\put(50,220){\vector(-1,-1){10}}  \put(50,220){\circle*{3}}
\put(30,200){\line(1,1){10}}      \put(30,200){\circle*{3}}

\put(70,200){\line(-1,-1){10}}    \put(70,200){\circle*{3}}
\put(50,180){\vector(1,1){10}}    \put(50,180){\circle*{3}}
\put(50,180){\vector(-1,-1){10}}  \put(50,180){\circle*{3}}
\put(30,160){\line(1,1){10}}      \put(30,160){\circle*{3}}
\put(30,160){\line(-1,-1){10}}    \put(30,160){\circle*{3}}
\put(10,140){\vector(1,1){10}}

\put(90,180){\vector(-1,-1){10}}  \put(90,180){\circle*{3}}
\put(70,160){\line(1,1){10}}      \put(70,160){\circle*{3}}
\put(70,160){\line(-1,-1){10}} \put(70,160){\tiny{(6,4)}}
\put(70,160){\circle*{3}}
\put(50,140){\vector(1,1){10}}
\put(50,140){\circle*{3}}
\put(50,140){\vector(-1,-1){10}} \put(55,140){\tiny{(7,3)}}
\put(50,140){\circle*{3}}
\put(30,120){\line(1,1){10}}
\put(30,120){\circle*{3}}

\put(70,120){\line(-1,-1){10}} \put(70,120){\tiny{(8,6)}}
\put(70,120){\circle*{3}}
\put(50,100){\vector(1,1){10}}
\put(50,100){\circle*{3}}

\put(47,197){\small 1}
\put(67,177){\small 2}
\put(47,157){\small 3}
\put(27,137){\small 4}
\put(47,117){\small 5}

\put(50,0){\bf (a)}

\put(210,260){\line(1,-1){10}} \put(210,260){ \tiny{(1,1)}}
\put(210,260){\circle*{3}}
\put(230,240){\vector(-1,1){10}}
\put(230,240){\circle*{3}}
\put(230,240){\vector(1,-1){10}}  \put(230,240){\tiny{(2,2)}}
\put(230,240){\circle*{3}}
\put(250,220){\line(-1,1){10}}
\put(250,220){\circle*{3}}
\put(250,220){\line(1,-1){10}} \put(250,220){\tiny{(3,3)}}
 \put(250,220){\circle*{3}}
\put(270,200){\vector(-1,1){10}}
\put(270,200){\circle*{3}}
\put(270,200){\vector(1,-1){10}}  \put(270,200){\tiny{(4,4)}}
\put(270,200){\circle*{3}}
\put(290,180){\line(-1,1){10}} \put(290,180){\tiny{(5,5)}}
\put(290,180){\circle*{3}}

\put(230,200){\vector(1,-1){10}} \put(210,200){ \tiny{(4,6)}}
\put(230,200){\circle*{3}}
\put(250,180){\line(-1,1){10}}
\put(250,180){\circle*{3}}
\put(250,180){\line(1,-1){10}} \put(230,180){ \tiny{(5,3)}}
\put(250,180){\circle*{3}}
\put(270,160){\vector(-1,1){10}}
 \put(270,160){\circle*{3}}

\put(230,160){\vector(1,-1){10}} \put(210,160){ \tiny{(6,2)}}
\put(230,160){\circle*{3}}
\put(250,140){\line(-1,1){10}}
\put(250,140){\circle*{3}}
\put(250,140){\line(1,-1){10}}
\put(250,140){\circle*{3}}
\put(270,120){\vector(-1,1){10}}
\put(270,120){\circle*{3}}

\put(210,140){\line(1,-1){10}} \put(190,140){ \tiny{(7,1)}}
\put(210,140){\circle*{3}}
\put(230,120){\vector(-1,1){10}}
\put(230,120){\circle*{3}}
\put(230,120){\vector(1,-1){10}}  \put(210,120){ \tiny{(8,2)}}
 \put(230,120){\circle*{3}}
\put(250,100){\line(-1,1){10}}
\put(250,100){\circle*{3}}
\put(250,100){\line(1,-1){10}} \put(230,100){ \tiny{(9,3)}}
\put(250,100){\circle*{3}}
\put(270,80){\vector(-1,1){10}}
\put(270,80){\circle*{3}}
\put(270,80){\vector(1,-1){10}} \put(270,80){\tiny{(10,4)}}
\put(270,80){\circle*{3}}
\put(290,60){\line(-1,1){10}} \put(290,60){\tiny{(11,5)}}
\put(290,60){\circle*{3}}

\put(250,220){\line(-1,-1){10}}
\put(250,220){\circle*{3}}
\put(230,200){\vector(1,1){10}}
\put(230,200){\circle*{3}}

\put(270,200){\vector(-1,-1){10}}
\put(270,200){\circle*{3}}
\put(250,180){\line(1,1){10}}
\put(250,180){\circle*{3}}
\put(250,180){\line(-1,-1){10}}
\put(250,180){\circle*{3}}
\put(230,160){\vector(1,1){10}}
\put(230,160){\circle*{3}}
\put(230,160){\vector(-1,-1){10}}
\put(230,160){\circle*{3}}
\put(210,140){\line(1,1){10}}
\put(210,140){\circle*{3}}

\put(290,180){\line(-1,-1){10}}
\put(290,180){\circle*{3}}
\put(270,160){\vector(1,1){10}}
\put(270,160){\circle*{3}}
\put(270,160){\vector(-1,-1){10}} \put(270,160){\tiny{(6,4)}}
\put(270,160){\circle*{3}}
\put(250,140){\line(1,1){10}}
\put(250,140){\circle*{3}}
\put(250,140){\line(-1,-1){10}} \put(255,140){\tiny{(7,3)}}
\put(250,140){\circle*{3}}
\put(230,120){\vector(1,1){10}}
\put(230,120){\circle*{3}}

\put(270,120){\vector(-1,-1){10}} \put(270,120){\tiny{(8,6)}}
\put(270,120){\circle*{3}}
\put(250,100){\line(1,1){10}}
\put(250,100){\circle*{3}}

\put(247,197){\small 1}
\put(267,177){\small 2}
\put(247,157){\small 3}
\put(227,137){\small 4}
\put(247,117){\small 5}

\put(250,0){\bf (b)}
\end{picture}
\end{center}
\caption{The cell graphs for the $A_{11}$--$E_6$ intertwiner.}
\end{figure}
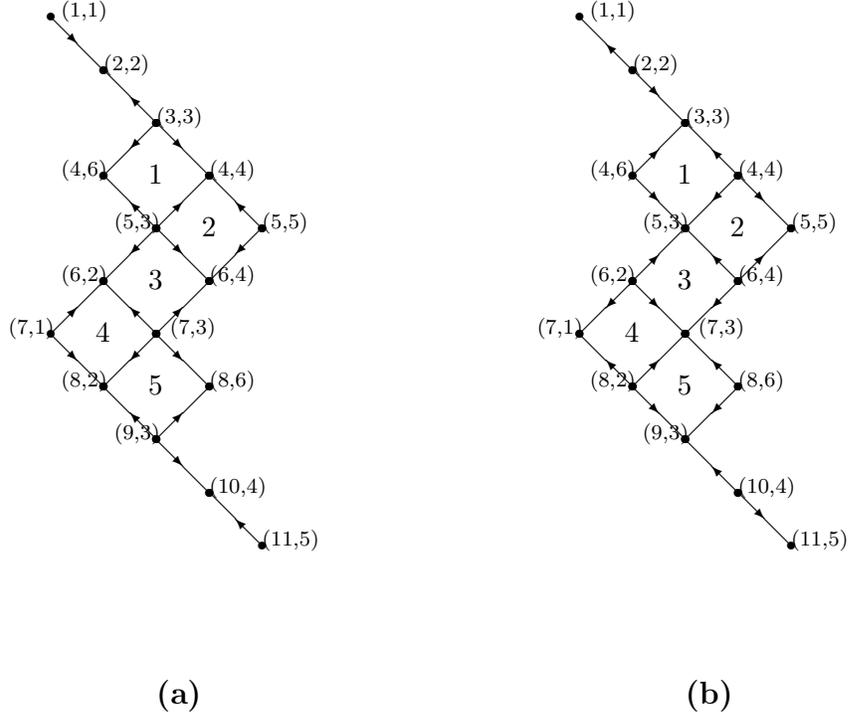

This time there are five faces in each graph which we number from $1$ to $5$.
 Unlike the $A$--$D$ models
some face bonds are common to two faces. We divide the face bonds into two
classes,
shared face bonds and unshared face bonds. We need to augment the above
rules to obtain cells on shared
bonds. By studying the cell graphs in Figure~5 we see that the shared bonds
always occur between a face of type $A$ and a face of type $B$ (see rules 2
and 3). The cells on a
face of type $A$ should be given by the rule 2 up to an angle $m$ and a
sign. To fix the angle $m$ we
introduce a new rule:

\begin{enumerate}
\setcounter{enumi}{3}
 \item
The absolute value of the cell of a shared bond
\begin{picture}(110,15)(3,0)
\put(10,4){\scriptsize $(a,b)$}
\put(40,4){\circle*{3}}
\put(40,4){\line(1,0){20}}
\put(80,4){\vector(-1,0){20}}
\put(80,4){\circle*{3}}
\put(90,4){\scriptsize $(c,d)$}
\end{picture}
in a face of type $B$ is related to the absolute value of the cell of the
opposite bond in the face.
Explicitly, if the opposite bond is \vspace{2 mm}
\begin{picture}(110,15)(6,0)
\put(10,4){\scriptsize $(a,b')$}
\put(40,4){\circle*{3}}
\put(40,4){\line(1,0){20}}
\put(80,4){\vector(-1,0){20}}
\put(80,4){\circle*{3}}
\put(90,4){\scriptsize $(c',d)$ }
\end{picture}
then
\be
\left|\weight Cabdc\right|=
{\sqrt{S^A_{c'} S^E_{b'}\over S^A_c S^E_b}}\  \left|\weight Ca{b'}d{c'}
\right|
\ee
The absolute values of cells of any unshared face bonds on a face of
type~B are $1$. The absolute value
of a cell of an unshared face bond
\begin{picture}(110,15)(3,0)
\put(10,4){\scriptsize $(a,b)$}
\put(40,4){\circle*{3}}
\put(40,4){\line(1,0){20}}
\put(80,4){\vector(-1,0){20}}
\put(80,4){\circle*{3}}
\put(90,4){\scriptsize $(c,d)$}
\end{picture}
on a face of type A is given by
\be
\left|\weight Cabcd\right|=
{\sqrt{S^A_a S^E_c\over S^A_d S^E_b}}
\ee
\end{enumerate}

The values of the nonzero cells for the $A$--$E$ intertwiner are determined
by these four rules
up to some sign factors which are determined by the intertwining relation
(\ref{eq:faceinter}). It  may happen that a face has two neighbouring faces
and so there are two ways to
find the absolute value of the cell on a shared bond. In such cases the two
results always agree. For example,
applying rule 2 to face 4 in Figure~5a gives
\be
\left|\weight C6237\right|=\cos m_4.
\ee
Then rule~4 tells us that
\be
\cos m_4={\sqrt{S^A_5 S^E_4\over S^A_7 S^E_2}}\ \left|\weight C6435\right|=
{\sqrt{S^A_5 S^E_4\over S^A_7 S^E_2}}\ \cos m_2
\ee
from face 3 where, from face 1, we see that $m_2$ is given by
\be
\left|\weight C4435\right|=\sin m_2={\sqrt{S^A_3 S^E_6\over S^A_5 S^E_4}}.
\ee
Also we get the same value of $|C(6,2,3,7)|$  from face 5 by using
\be
\sin m_4={\sqrt{S^A_9 S^E_6\over S^A_7 S^E_2}}
\ee
and
\be
\cos m_4=\sqrt{1-\sin^2 m_4}
\ee
We obtain the
unshared bonds in face 4 and also
\be
\left|\weight C6237\right|=\left|\weight C8217\right|=\cos m_4=
\sqrt{S^A_8 S^E_1\over S^A_7 S^E_2}
\ee
which gives us the same value.
Obviously it is simple to find the cell value $|C(6,2,3,7)|$ from
face $5$ or from the unshared bond in face 4. Then we can show that this
cell is negative, or
$C(6,2,3,7)=-|C(6,2,3,7)|$, using
the intertwining relation. Finally, because $\sin m_4=\sin m_2$ we have
$C(6,2,3,7)=-\cos m_2$. In this way we can determine all of the cells
completely for the
$A_{11}$--$E_6$ intertwiner. The results are tabulated in Appendix~C2.

The $A_{17}-E_7$ intertwiner can be obtained in a similar manner. The cell
graphs for
this intertwiner are shown in Figure~6. The derivation of the cells proceeds
by applying the four rules as for
the $A_{11}$--$E_6$ case. The results for the cells are tabulated in
Appendix~C3.

To summarize the procedure is (i) to draw the cell graphs from
the adjacency matrix of the intertwiner, (ii) to obtain the absolute
values of the cells by applying the four
rules and then (iii) to fix the remaining sign factors using the full
intertwining relations.

\begin{figure}[htb]
\begin{center}
\begin{picture}(370,350)(-15,0)

\multiput(10,350)(40,-40){3}{\vector(1,-1){10}}
\multiput(10,350)(40,-40){3}{\circle*{3}}
\multiput(30,330)(40,-40){3}{\line(-1,1){10}}
\multiput(30,330)(40,-40){3}{\circle*{3}}
\multiput(30,330)(40,-40){2}{\line(1,-1){10}}
\multiput(30,330)(40,-40){2}{\circle*{3}}
\multiput(50,310)(40,-40){2}{\vector(-1,1){10}}
\multiput(50,310)(40,-40){2}{\circle*{3}}

\put(50,270){\vector(1,-1){10}}  \put(50,270){\circle*{3}}
\put(70,250){\line(-1,1){10}}    \put(70,250){\circle*{3}}
\put(70,250){\line(1,-1){10}}    \put(70,250){\circle*{3}}
\put(90,230){\vector(-1,1){10}}  \put(90,230){\circle*{3}}

\put(50,230){\vector(1,-1){10}}  \put(50,230){\circle*{3}}
\put(70,210){\line(-1,1){10}}     \put(70,210){\circle*{3}}
\put(70,210){\line(1,-1){10}}     \put(70,210){\circle*{3}}
\put(90,190){\vector(-1,1){10}}  \put(90,190){\circle*{3}}

\multiput(30,210)(40,-40){2}{\line(1,-1){10}}
\multiput(30,210)(40,-40){2}{\circle*{3}}
\multiput(50,190)(40,-40){2}{\vector(-1,1){10}}
\multiput(50,190)(40,-40){2}{\circle*{3}}
\multiput(50,190)(40,-40){2}{\vector(1,-1){10}}
\multiput(50,190)(40,-40){2}{\circle*{3}}
\multiput(70,170)(40,-40){2}{\line(-1,1){10}}
\multiput(70,170)(40,-40){2}{\circle*{3}}

\multiput(10,190)(40,-40){2}{\vector(1,-1){10}}
\multiput(10,190)(40,-40){2}{\circle*{3}}
\multiput(30,170)(40,-40){2}{\line(-1,1){10}}
\multiput(30,170)(40,-40){2}{\circle*{3}}
\multiput(30,170)(40,-40){2}{\line(1,-1){10}}
\multiput(30,170)(40,-40){2}{\circle*{3}}
\multiput(50,150)(40,-40){2}{\vector(-1,1){10}}
\multiput(50,150)(40,-40){2}{\circle*{3}}

\put(50,110){\vector(1,-1){10}}    \put(50,110){\circle*{3}}
\put(70,90){\line(-1,1){10}}       \put(70,90){\circle*{3}}
%%%%%%%%%%%%
\put(70,290){\line(-1,-1){10}}     \put(70,290){\circle*{3}}
\put(50,270){\vector(1,1){10}}     \put(50,270){\circle*{3}}

\multiput(90,270)(-40,-40){2}{\vector(-1,-1){10}}
\multiput(90,270)(-40,-40){2}{\circle*{3}}
\multiput(70,250)(-40,-40){2}{\line(1,1){10}}
\multiput(70,250)(-40,-40){2}{\circle*{3}}
\multiput(70,250)(-40,-40){2}{\line(-1,-1){10}}
\multiput(70,250)(-40,-40){2}{\circle*{3}}
\multiput(50,230)(-40,-40){2}{\vector(1,1){10}}
\multiput(50,230)(-40,-40){2}{\circle*{3}}

\multiput(110,250)(-40,-40){2}{\line(-1,-1){10}}
\multiput(110,250)(-40,-40){2}{\circle*{3}}
\multiput(90,230)(-40,-40){2}{\vector(1,1){10}}
\multiput(90,230)(-40,-40){2}{\circle*{3}}
\multiput(90,230)(-40,-40){2}{\vector(-1,-1){10}}
\multiput(90,230)(-40,-40){2}{\circle*{3}}
\multiput(70,210)(-40,-40){2}{\line(1,1){10}}
\multiput(70,210)(-40,-40){2}{\circle*{3}}

\multiput(90,190)(0,-40){2}{\vector(-1,-1){10}}
\multiput(90,190)(0,-40){2}{\circle*{3}}
\multiput(70,170)(0,-40){2}{\line(1,1){10}}
\multiput(70,170)(0,-40){2}{\circle*{3}}
\multiput(70,170)(0,-40){2}{\line(-1,-1){10}}
 \multiput(70,170)(0,-40){2}{\circle*{3}}
\multiput(50,150)(0,-40){2}{\vector(1,1){10}}
\multiput(50,150)(0,-40){2}{\circle*{3}}

\multiput(110,130)(-40,-40){3}{\line(-1,-1){10}}
\multiput(110,130)(-40,-40){3}{\circle*{3}}
\multiput(90,110)(-40,-40){3}{\vector(1,1){10}}
\multiput(90,110)(-40,-40){3}{\circle*{3}}
\multiput(90,110)(-40,-40){2}{\vector(-1,-1){10}}
\multiput(90,110)(-40,-40){2}{\circle*{3}}
\multiput(70,90)(-40,-40){2}{\line(1,1){10}}
\multiput(70,90)(-40,-40){2}{\circle*{3}}
%%%%%%%%%%%%%%%%%%%%%
\put(13,350){\scriptsize(1,1)}
\put(33,330){\scriptsize(2,2)}
\put(53,310){\scriptsize(3,3)}
\put(73,290){\scriptsize(4,4)}
\put(93,270){\scriptsize(5,5)}
\put(113,250){\scriptsize(6,6)}
\put(93,230){\scriptsize(7,5)}
\put(73,210){\scriptsize(8,4)}
\put(93,190){\scriptsize(9,7)}
\put(73,170){\scriptsize(10,4)}
\put(93,150){\scriptsize(11,5)}
\put(113,130){\scriptsize(12,6)}
\put(93,110){\scriptsize(13,5)}
\put(73,90){\scriptsize(14,4)}
\put(53,70){\scriptsize(15,3)}
\put(33,50){\scriptsize(16,2)}
\put(13,30){\scriptsize(17,1)}

\put(27,270){\scriptsize(5,7)}
\put(47,250){\scriptsize(6,4)}
\put(27,230){\scriptsize(7,3)}
\put(7,210){\scriptsize(8,2)}
\put(-13,190){\scriptsize(9,1)}
\put(7,170){\scriptsize(10,2)}
\put(27,150){\scriptsize(11,3)}
\put(47,130){\scriptsize(12,4)}
\put(27,110){\scriptsize(13,7)}

\put(40,190){\scriptsize(9,3)}

%%%%%%%%%%%%%%%%%%%%%%%
\put(67,267){\small 1 }
\put(87,247){\small 2}
\put(67,227){\small 3}
\put(47,207){\small 4}
\put(67,187){\small 5}
\put(27,187){\small 6}
\put(47,167){\small 7}
\put(67,147){\small 8}
\put(87,127){\small 9}
\put(65,107){\small 10}

\put(53,0){\bf (a)}
%%%%%%%%%%%%%%%%%%%%%
\multiput(210,350)(40,-40){3}{\line(1,-1){10}}
\multiput(210,350)(40,-40){3}{\circle*{3}}
\multiput(230,330)(40,-40){3}{\vector(-1,1){10}}
\multiput(230,330)(40,-40){3}{\circle*{3}}
\multiput(230,330)(40,-40){2}{\vector(1,-1){10}}
\multiput(230,330)(40,-40){2}{\circle*{3}}
\multiput(250,310)(40,-40){2}{\line(-1,1){10}}
\multiput(250,310)(40,-40){2}{\circle*{3}}

\put(250,270){\line(1,-1){10}}        \put(250,270){\circle*{3}}
\put(270,250){\vector(-1,1){10}}      \put(270,250){\circle*{3}}
\put(270,250){\vector(1,-1){10}}      \put(270,250){\circle*{3}}
\put(290,230){\line(-1,1){10}}        \put(290,230){\circle*{3}}

\put(250,230){\line(1,-1){10}}       \put(250,230){\circle*{3}}
\put(270,210){\vector(-1,1){10}}     \put(270,210){\circle*{3}}
\put(270,210){\vector(1,-1){10}}     \put(270,210){\circle*{3}}
\put(290,190){\line(-1,1){10}}       \put(290,190){\circle*{3}}

\multiput(230,210)(40,-40){2}{\vector(1,-1){10}}
\multiput(230,210)(40,-40){2}{\circle*{3}}
\multiput(250,190)(40,-40){2}{\line(-1,1){10}}
\multiput(250,190)(40,-40){2}{\circle*{3}}
\multiput(250,190)(40,-40){2}{\line(1,-1){10}}
\multiput(250,190)(40,-40){2}{\circle*{3}}
\multiput(270,170)(40,-40){2}{\vector(-1,1){10}}
\multiput(270,170)(40,-40){2}{\circle*{3}}

\multiput(210,190)(40,-40){2}{\line(1,-1){10}}
\multiput(210,190)(40,-40){2}{\circle*{3}}
\multiput(230,170)(40,-40){2}{\vector(-1,1){10}}
\multiput(230,170)(40,-40){2}{\circle*{3}}
\multiput(230,170)(40,-40){2}{\vector(1,-1){10}}
\multiput(230,170)(40,-40){2}{\circle*{3}}
\multiput(250,150)(40,-40){2}{\line(-1,1){10}}
\multiput(250,150)(40,-40){2}{\circle*{3}}

\put(250,110){\line(1,-1){10}}       \put(250,110){\circle*{3}}
\put(270,90){\vector(-1,1){10}}      \put(270,90){\circle*{3}}
%%%%%%%%%%%%
\put(270,290){\vector(-1,-1){10}}    \put(270,290){\circle*{3}}
\put(250,270){\line(1,1){10}}        \put(250,270){\circle*{3}}

\multiput(290,270)(-40,-40){2}{\line(-1,-1){10}}
\multiput(290,270)(-40,-40){2}{\circle*{3}}
\multiput(270,250)(-40,-40){2}{\vector(1,1){10}}
\multiput(270,250)(-40,-40){2}{\circle*{3}}
\multiput(270,250)(-40,-40){2}{\vector(-1,-1){10}}
\multiput(270,250)(-40,-40){2}{\circle*{3}}
\multiput(250,230)(-40,-40){2}{\line(1,1){10}}
\multiput(250,230)(-40,-40){2}{\circle*{3}}

\multiput(310,250)(-40,-40){2}{\vector(-1,-1){10}}
\multiput(310,250)(-40,-40){2}{\circle*{3}}
\multiput(290,230)(-40,-40){2}{\line(1,1){10}}
\multiput(290,230)(-40,-40){2}{\circle*{3}}
\multiput(290,230)(-40,-40){2}{\line(-1,-1){10}}
\multiput(290,230)(-40,-40){2}{\circle*{3}}
\multiput(270,210)(-40,-40){2}{\vector(1,1){10}}
\multiput(270,210)(-40,-40){2}{\circle*{3}}

\multiput(290,190)(0,-40){2}{\line(-1,-1){10}}
\multiput(290,190)(0,-40){2}{\circle*{3}}
\multiput(270,170)(0,-40){2}{\vector(1,1){10}}
\multiput(270,170)(0,-40){2}{\circle*{3}}
\multiput(270,170)(0,-40){2}{\vector(-1,-1){10}}
\multiput(270,170)(0,-40){2}{\circle*{3}}
\multiput(250,150)(0,-40){2}{\line(1,1){10}}
\multiput(250,150)(0,-40){2}{\circle*{3}}

\multiput(310,130)(-40,-40){3}{\vector(-1,-1){10}}
\multiput(310,130)(-40,-40){3}{\circle*{3}}
\multiput(290,110)(-40,-40){3}{\line(1,1){10}}
\multiput(290,110)(-40,-40){3}{\circle*{3}}
\multiput(290,110)(-40,-40){2}{\line(-1,-1){10}}
\multiput(290,110)(-40,-40){2}{\circle*{3}}
\multiput(270,90)(-40,-40){2}{\vector(1,1){10}}
\multiput(270,90)(-40,-40){2}{\circle*{3}}

%%%%%%%%%%%%%%%%%%%%%%%%
\put(213,350){\scriptsize(1,1)}
\put(233,330){\scriptsize(2,2)}
\put(253,310){\scriptsize(3,3)}
\put(273,290){\scriptsize(4,4)}
\put(293,270){\scriptsize(5,5)}
\put(313,250){\scriptsize(6,6)}
\put(293,230){\scriptsize(7,5)}
\put(273,210){\scriptsize(8,4)}
\put(293,190){\scriptsize(9,7)}
\put(273,170){\scriptsize(10,4)}
\put(293,150){\scriptsize(11,5)}
\put(313,130){\scriptsize(12,6)}
\put(293,110){\scriptsize(13,5)}
\put(273,90){\scriptsize(14,4)}
\put(253,70){\scriptsize(15,3)}
\put(233,50){\scriptsize(16,2)}
\put(213,30){\scriptsize(17,1)}

\put(227,270){\scriptsize(5,7)}
\put(247,250){\scriptsize(6,4)}
\put(227,230){\scriptsize(7,3)}
\put(207,210){\scriptsize(8,2)}
\put(187,190){\scriptsize(9,1)}
\put(207,170){\scriptsize(10,2)}
\put(227,150){\scriptsize(11,3)}
\put(247,130){\scriptsize(12,4)}
\put(227,110){\scriptsize(13,7)}

\put(240,190){\scriptsize(9,3)}

%%%%%%%%%%%%%%%%%%%%%%%

\put(267,267){\small 1 }
\put(287,247){\small 2}
\put(267,227){\small 3}
\put(247,207){\small 4}
\put(267,187){\small 5}
\put(227,187){\small 6}
\put(247,167){\small 7}
\put(267,147){\small 8}
\put(287,127){\small 9}
\put(265,107){\small 10}
%%%%%%%%%%%%%%%%%%%

\put(283,280){--}
\put(281,240){--}
\put(283,200){--}
\put(238,200){--}
\put(252,160){--}
\put(278,120){--}
\put(83,280){--}
\put(81,240){--}
\put(83,200){--}
\put(38,200){--}
\put(52,160){--}
\put(78,120){--}
%%%%%%%%%%%%%%%%%%%%%%%%
\put(253,0){\bf (b)}
\end{picture}
\end{center}
\caption{The cell graphs for the $A_{17}$--$E_7$ intertwiner. The negative
signs indicate that the value of
the cell associated with that bond is negative, for example,
$C(4,4,5,5)=-\cos m_1$.}
\end{figure}
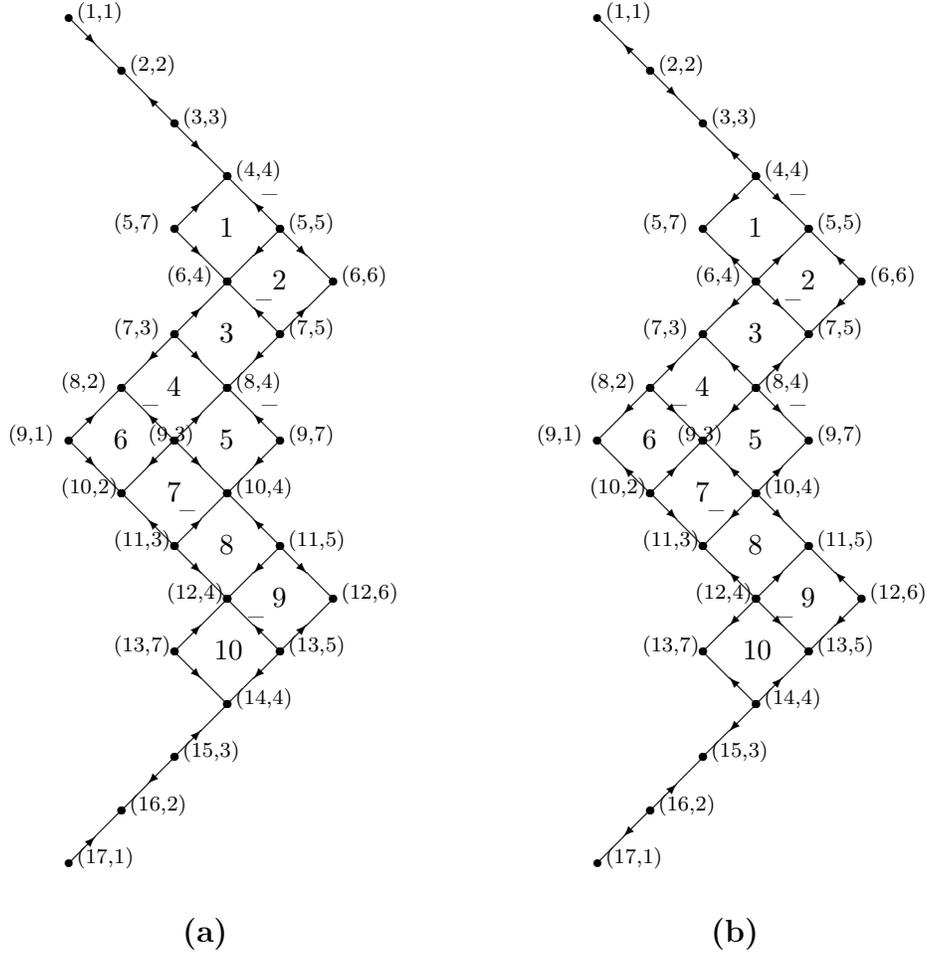

\subsection{Classical $A_{29}$--$E_8$ Cells}

Let us now consider the $A_{29}-E_8$ intertwiner.  The cell graphs for the
$A_{29}$--$E_8$ intertwiner shown
in Figure~7 are much bigger than those of $A_{17}$--$E_7$. Moreover, since
the element (15,5) of the adjacency
matrix is $2$, we treat it as two nonzero elements $(15,1,5)$ and $(15,2,5)$
distinguished by a bond variable
taking the values 1 or 2.  The cell graphs need this extra bond variable to
label the two different bonds
between the given nodes.  If there is only one bond state we label it by 1.
In the cell graphs we mark the
negative cells by minus signs in the upper parts of the two graphs. The
cells in the lower parts that occur in
faces 1 through to 12 are related to the cells in the upper parts by the
height reversal symmetry of the $A$ model
\be

\ee
A similar equation holds with the direction of the arrows reversed.

%\clearpage
\begin{figure}[p]
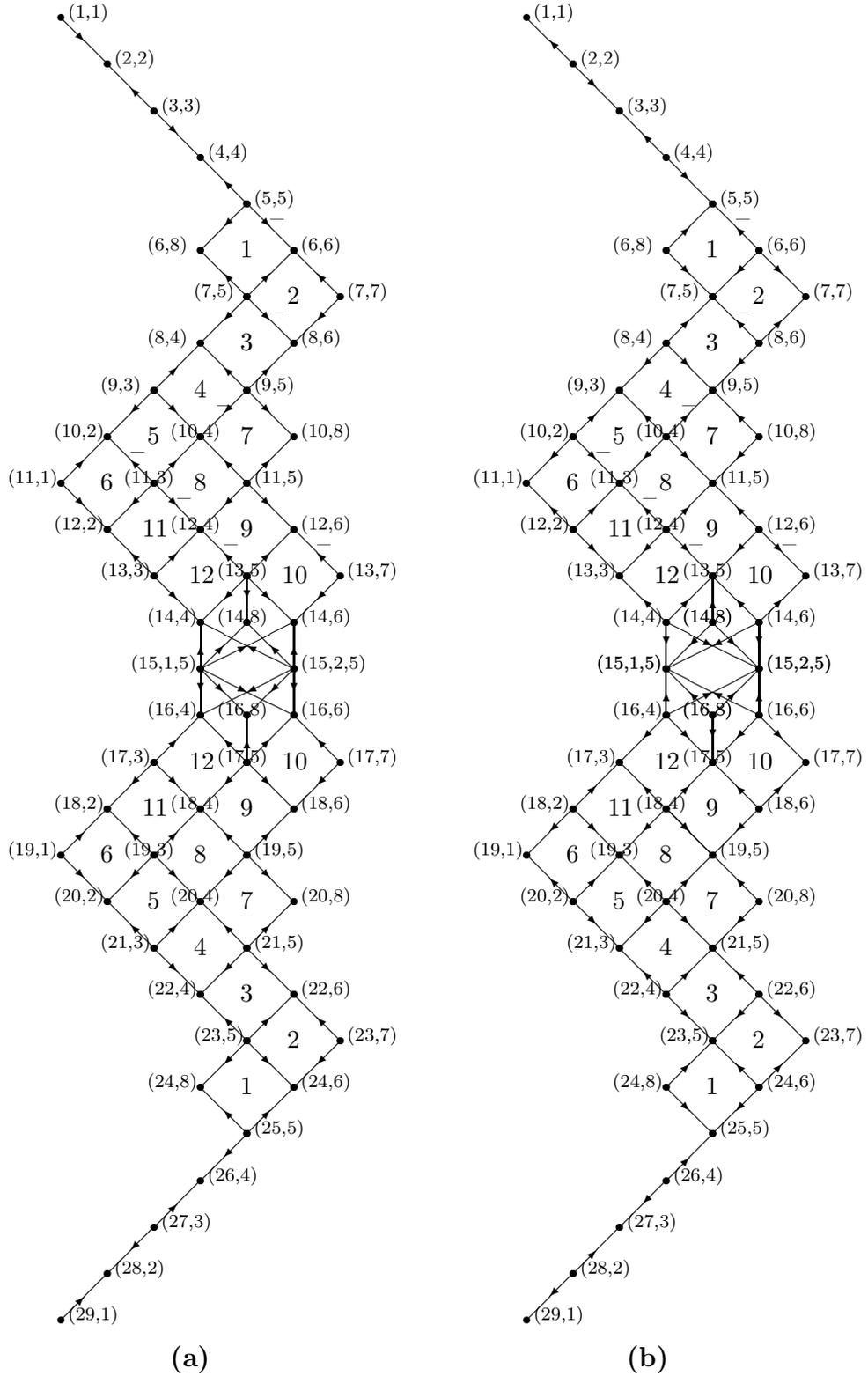

\begin{center}
%
\end{center}
\vspace{-.75in}
\caption{The cell graphs of the $A_{29}$--$E_8$ intertwiner. The negative
cells are indicated with a
minus sign in the upper parts of the graphs. The cells in the lower part are
related to the cells in the
upper part by the height reversal symmetry of the $A$ model.}
\end{figure}
%\clearpage

Using the four rules given above we can find all of the cells on the twelve
numbered faces. The
cells on the faces containing the special elements $(15,1,5)$ and $(15,2,5)$
have to be studied seperately.
The parts of the cell graphs around the special elements $(15,1,5)$ and
$(15,2,5)$ are drawn magnified in Figure~8.

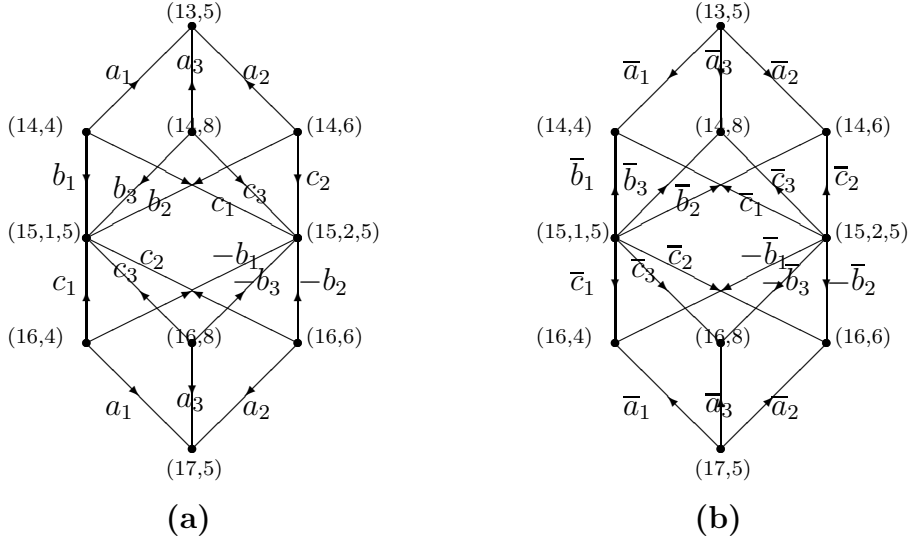
\begin{figure}[htb]
\begin{center}
\begin{picture}(320,230)(-10,10)
\put(70,200){\line(0,-1){20}} \put(70,160){\vector(0,1){20}}
\put(70,200){\line(-1,-1){20}} \put(30,160){\vector(1,1){20}}
\put(70,200){\line(1,-1){20}} \put(110,160){\vector(-1,1){20}}

\put(30,120){\line(0,1){20}}   \put(30,160){\vector(0,-1){20}}
\put(30,120){\line(1,1){20}}   \put(70,160){\vector(-1,-1){20}}
\put(30,120){\line(2,1){40}}   \put(110,160){\vector(-2,-1){40}}
\put(30,120){\line(0,-1){20}}  \put(30,80){\vector(0,1){20}}
\put(30,120){\line(1,-1){20}}  \put(70,80){\vector(-1,1){20}}
\put(30,120){\line(2,-1){40}}  \put(110,80){\vector(-2,1){40}}

\put(110,120){\line(0,1){20}}   \put(110,160){\vector(0,-1){20}}
\put(110,120){\line(-1,1){20}}  \put(70,160){\vector(1,-1){20}}
\put(110,120){\line(-2,1){40}}  \put(30,160){\vector(2,-1){40}}
\put(110,120){\line(0,-1){20}}  \put(110,80){\vector(0,1){20}}
\put(110,120){\line(-1,-1){20}} \put(70,80){\vector(1,1){20}}
\put(110,120){\line(-2,-1){40}} \put(30,80){\vector(2,1){40}}

\put(70,40){\line(1,1){20}}     \put(110,80){\vector(-1,-1){20}}
\put(70,40){\line(-1,1){20}}    \put(30,80){\vector(1,-1){20}}
\put(70,40){\line(0,1){20}}     \put(70,80){\vector(0,-1){20}}
%%%%%%%%%%%%%%%%%%%%%%%%%%%%%%%%%
\put(70,200){\circle*{3}} \put(70,160){\circle*{3}}
\put(70,200){\circle*{3}} \put(30,160){\circle*{3}}
\put(70,200){\circle*{3}} \put(110,160){\circle*{3}}

\put(30,120){\circle*{3}}   \put(30,160){\circle*{3}}
\put(30,120){\circle*{3}}   \put(70,160){\circle*{3}}
\put(30,120){\circle*{3}}   \put(110,160){\circle*{3}}
\put(30,120){\circle*{3}}  \put(30,80){\circle*{3}}
\put(30,120){\circle*{3}}  \put(70,80){\circle*{3}}
\put(30,120){\circle*{3}}  \put(110,80){\circle*{3}}

\put(110,120){\circle*{3}}   \put(110,160){\circle*{3}}
\put(110,120){\circle*{3}}  \put(70,160){\circle*{3}}
\put(110,120){\circle*{3}}  \put(30,160){\circle*{3}}
\put(110,120){\circle*{3}}  \put(110,80){\circle*{3}}
\put(110,120){\circle*{3}} \put(70,80){\circle*{3}}
\put(110,120){\circle*{3}} \put(30,80){\circle*{3}}

\put(70,40){\circle*{3}}     \put(110,80){\circle*{3}}
\put(70,40){\circle*{3}}    \put(30,80){\circle*{3}}
\put(70,40){\circle*{3}}     \put(70,80){\circle*{3}}

\put(60,203){\scriptsize(13,5)}
\put(60,160){\scriptsize(14,8)}
\put(0,160){\scriptsize(14,4)}
\put(113,160){\scriptsize(14,6)}
\put(0,120){\scriptsize(15,1,5)}
\put(113,120){\scriptsize(15,2,5)}
\put(0,80){\scriptsize(16,4)}
\put(113,80){\scriptsize(16,6)}
\put(60,80){\scriptsize(16,8)}
\put(60,30){\scriptsize(17,5)}

\multiput(37,180)(0,-127){2}{$a_1$}
\multiput(89,180)(0,-127){2}{$a_2$}
\multiput(64,184)(0,-128){2}{$a_3$}
\put(17,140){$b_1$} \put(77,110){$-b_1$}
\put(17,100){$c_1$} \put(77,130){$c_1$}
\put(110,100){$-b_2$} \put(53,130){$b_2$}
\put(85,100){$-b_3$} \put(40,135){$b_3$}
\put(113,140){$c_2$} \put(50,110){$c_2$}
\put(40,105){$c_3$} \put(89,135){$c_3$}
\put(60,10){\bf (a)}

%%%%%%%%%%%%%%%%%%%%%%%%%%%%%%%%%%%%%
\put(270,200){\vector(0,-1){20}} \put(270,160){\line(0,1){20}}
\put(270,200){\vector(-1,-1){20}} \put(230,160){\line(1,1){20}}
\put(270,200){\vector(1,-1){20}} \put(310,160){\line(-1,1){20}}

\put(230,120){\vector(0,1){20}}   \put(230,160){\line(0,-1){20}}
\put(230,120){\vector(1,1){20}}   \put(270,160){\line(-1,-1){20}}
\put(230,120){\vector(2,1){40}}   \put(310,160){\line(-2,-1){40}}
\put(230,120){\vector(0,-1){20}}  \put(230,80){\line(0,1){20}}
\put(230,120){\vector(1,-1){20}}  \put(270,80){\line(-1,1){20}}
\put(230,120){\vector(2,-1){40}}  \put(310,80){\line(-2,1){40}}

\put(310,120){\vector(0,1){20}}   \put(310,160){\line(0,-1){20}}
\put(310,120){\vector(-1,1){20}}  \put(270,160){\line(1,-1){20}}
\put(310,120){\vector(-2,1){40}}  \put(230,160){\line(2,-1){40}}
\put(310,120){\vector(0,-1){20}}  \put(310,80){\line(0,1){20}}
\put(310,120){\vector(-1,-1){20}} \put(270,80){\line(1,1){20}}
\put(310,120){\vector(-2,-1){40}} \put(230,80){\line(2,1){40}}

\put(270,40){\vector(1,1){20}}     \put(310,80){\line(-1,-1){20}}
\put(270,40){\vector(-1,1){20}}    \put(230,80){\line(1,-1){20}}
\put(270,40){\vector(0,1){20}}     \put(270,80){\line(0,-1){20}}

%%%%%%%%%%%%%%%%%%%%%%%%%%%%%%%%%%%%%
\put(270,200){\circle*{3}} \put(270,160){\circle*{3}}
\put(270,200){\circle*{3}} \put(230,160){\circle*{3}}
\put(270,200){\circle*{3}} \put(310,160){\circle*{3}}

\put(230,120){\circle*{3}}   \put(230,160){\circle*{3}}
\put(230,120){\circle*{3}}   \put(270,160){\circle*{3}}
\put(230,120){\circle*{3}}   \put(310,160){\circle*{3}}
\put(230,120){\circle*{3}}  \put(230,80){\circle*{3}}
\put(230,120){\circle*{3}}  \put(270,80){\circle*{3}}
\put(230,120){\circle*{3}}  \put(310,80){\circle*{3}}

\put(310,120){\circle*{3}}   \put(310,160){\circle*{3}}
\put(310,120){\circle*{3}}  \put(270,160){\circle*{3}}
\put(310,120){\circle*{3}}  \put(230,160){\circle*{3}}
\put(310,120){\circle*{3}}  \put(310,80){\circle*{3}}
\put(310,120){\circle*{3}} \put(270,80){\circle*{3}}
\put(310,120){\circle*{3}} \put(230,80){\circle*{3}}

\put(270,40){\circle*{3}}     \put(310,80){\circle*{3}}
\put(270,40){\circle*{3}}    \put(230,80){\circle*{3}}
\put(270,40){\circle*{3}}     \put(270,80){\circle*{3}}

\put(260,203){\scriptsize(13,5)}
\put(260,160){\scriptsize(14,8)}
\put(200,160){\scriptsize(14,4)}
\put(313,160){\scriptsize(14,6)}
\put(200,120){\scriptsize(15,1,5)}
\put(313,120){\scriptsize(15,2,5)}
\put(200,80){\scriptsize(16,4)}
\put(313,80){\scriptsize(16,6)}
\put(260,80){\scriptsize(16,8)}
\put(260,30){\scriptsize(17,5)}

\multiput(233,180)(0,-127){2}{$\aa_1$}
\multiput(289,180)(0,-127){2}{$\aa_2$}
\multiput(264,184)(0,-130){2}{$\aa_3$}
\put(213,140){$\bb_1$} \put(277,110){$-\bb_1$}
\put(213,100){$\cc_1$} \put(277,130){$\cc_1$}
\put(310,100){$-\bb_2$} \put(253,130){$\bb_2$}
\put(285,100){$-\bb_3$} \put(233,138){$\bb_3$}
\put(313,140){$\cc_2$} \put(250,110){$\cc_2$}
\put(236,105){$\cc_3$} \put(289,139){$\cc_3$}
\put(260,10){\bf (b)}
\end{picture}
\end{center}
\caption{Magnified view of the $A_{29}$--$E_8$ cell graphs in the vicinity
of the special elements $(15,1,5)$
and $(15,2,5)$ showing the symbols used to denote the values of the cells.}
\end{figure}

We have used symbols to represent the values of the cells on the bonds in
the vicinity of the
special elements $(15,1,5)$
and $(15,2,5)$. The same symbol has been used to denote cells with the same
values. Let us define a
$3\times 3$ matrix using the cells in the upper part of Figure~8a
\be
\begin{picture}(260,120)(-40,-50)
\multiput(30,60)(40,0){3}{\vector(1,0){10}}
\multiput(50,60)(40,0){3}{\line(-1,0){10}}
\multiput(30,40)(40,0){3}{\vector(1,0){10}}
\multiput(50,40)(40,0){3}{\line(-1,0){10}}
\multiput(30,60)(40,0){3}{\vector(0,-1){10}}
\multiput(30,40)(40,0){3}{\line(0,1){10}}
\multiput(50,60)(40,0){3}{\vector(0,-1){10}}
\multiput(50,40)(40,0){3}{\line(0,1){10}}

\multiput(30,20)(40,0){3}{\vector(1,0){10}}
\multiput(50,20)(40,0){3}{\line(-1,0){10}}
\multiput(50,20)(40,0){3}{\vector(0,-1){10}}
\multiput(50,0)(40,0){3}{\line(0,1){10}}
\multiput(30,20)(40,0){3}{\vector(0,-1){10}}
\multiput(30,0)(40,0){3}{\line(0,1){10}}
\multiput(30,0)(40,0){3}{\vector(1,0){10}}
\multiput(50,0)(40,0){3}{\line(-1,0){10}}

\multiput(30,-20)(40,0){3}{\vector(1,0){10}}
\multiput(50,-20)(40,0){3}{\line(-1,0){10}}
\multiput(30,-40)(40,0){3}{\vector(1,0){10}}
\multiput(50,-40)(40,0){3}{\line(-1,0){10}}
\multiput(30,-20)(40,0){3}{\vector(0,-1){10}}
\multiput(30,-40)(40,0){3}{\line(0,1){10}}
\multiput(50,-20)(40,0){3}{\vector(0,-1){10}}
\multiput(50,-40)(40,0){3}{\line(0,1){10}}

\put(13,10){$\left( \begin{array}{c} \\  \\   \\   \\   \\
\end{array} \right.$}
\put(126,10){$ \left. \begin{array}{c} \\ \\  \\   \\    \\
\end{array} \right) \;\;$ = }
\multiput(39,61)(40,0){3}{\scriptsize 1}
\multiput(39,35)(40,0){3}{\scriptsize 1}
\multiput(39,21)(40,0){3}{\scriptsize 1}
\multiput(39,-5)(40,0){3}{\scriptsize 1}
\multiput(39,-19)(40,0){3}{\scriptsize 1}
\multiput(39,-45)(40,0){3}{\scriptsize 2}
\put(165,10){$\left( \begin{array}{ccc}      a_3 & a_2 & a_1 \\  \\
                                             b_3 & b_2 & b_1   \\  \\
                                             c_3 & c_2 & c_1   \end{array}
\right)  \;  $}
%%%%%%%%%%%%%%

\multiput(21,60)(40,0){3}{\scriptsize 14}
\multiput(50,60)(0,-40){3}{\scriptsize 8}
\multiput(21,20)(40,0){3}{\scriptsize 14}
\multiput(90,60)(0,-40){3}{\scriptsize 6}
\multiput(21,-20)(40,0){3}{\scriptsize 14}
\multiput(130,60)(0,-40){3}{\scriptsize 4}

\multiput(21,38)(40,0){3}{\scriptsize 13}
\multiput(50,38)(0,-40){3}{\scriptsize 5}
\multiput(21,-2)(40,0){3}{\scriptsize 15}
\multiput(90,38)(0,-40){3}{\scriptsize 5}
\multiput(21,-42)(40,0){3}{\scriptsize 15}
\multiput(130,38)(0,-40){3}{\scriptsize 5}
\end{picture}
\ee
It is easy to see that these cells satisfy the first unitarity condition if
the matrix is orthogonal. Since
the matrix can be taken as a coordinate transformation in three dimensional
space we can write its elements
in terms of the Euler angles $\alpha, \beta, \gamma$
\begin{eqnarray}
  a_1 & = & \sin \gamma                             \nonumber          \\
  a_2 & = & \sin \alpha \cos \gamma       \nonumber          \\
  a_3 & = & \cos \alpha \cos \gamma                    \nonumber \\
  b_1 & = & \sin \beta \cos\gamma                     \nonumber \\
  b_2 & = & \cos \alpha \cos \beta -\sin \alpha \sin \gamma \sin \beta
\nonumber \\
  b_3 & = & \mbox{} -\sin \alpha \cos \beta -\cos \alpha \sin \gamma \sin
\beta
                               \label{eq:abc}  \\
  c_1 & = & \cos \beta \cos \gamma
   \nonumber   \\
  c_2 & = & \mbox{} -\cos \alpha \sin \beta -\sin \alpha \sin \gamma \cos
\beta\nonumber \\
  c_3 & = & \sin \alpha \sin \beta -\cos \alpha \sin \gamma \cos \beta .
\nonumber
\end{eqnarray}
The Euler angles can be determined from the second unitarity condition.
Consider the three
quadrilateral faces formed by the elements $(14,2i)$, $(16,2i)$, $(15,1,5)$
and $(15,2,5)$ in Figure~8a
where $i=2,3,4$.  The cells on these faces should satisfy three sets of
equations required by the
second  unitarity condition. Setting
\begin{eqnarray}
 \cell {15}{1}{5}{2i}{}{16} & = &\ c_{i-1},\;\;\;\;\; i=2,3,4  \nonumber \\
 \cell {15}{2}{5}{2i}{}{16} & = & -b_{i-1},\;\;\;\;\; i=2,3,4
\end{eqnarray}
these three sets of equations read
\begin{eqnarray}
 b_i^2+c_i^2=S^A_{15} S^E_{2i}/(S^A_{14} S^E_{5}),\quad\;\;\;\;i=2,3,4.
\end{eqnarray}
Substituting $b$ and $c$ into the above equations, we have
\begin{eqnarray}
 \cos \gamma  & = & \sqrt{S^A_{15} S^E_{4}/(S^A_{14} S^E_5)}\;\;\;\;{\rm for}
\;\;i=2       \\
 \sin \alpha  & = & \sqrt{S^A_{14} S^E_{5}-S^A_{15} S^E_6\over S^A_{15}
S^E_4}\;\;\;\;
                                              {\rm for}\;\;i=3\;\;{\rm or}
\;\;4
\end{eqnarray}
The variable $\beta$ is not determined by the above equations and we simply
take
$\beta =\alpha$. It is easy to check that taking
\be
\weight C{17}{5}{2i}{16}=a_{i-1}\;\;\;\;\; i=2,3,4
\ee
satisfies the
unitarity conditions.

In Figure~8b the nine cells with values in the matrix
\[  \left( \begin{array}{ccc}
\aa_3 & \aa_2 & \aa_1  \\
\bb_3 & \bb_2 & \bb_1  \\
\cc_3 & \cc_2 & \cc_1
\end{array} \right)  \]
are required to satisfy the second unitarity condition. We easily see that
the second unitarity condition is
obtainable from the first unitarity  condition with
\begin{eqnarray}
 \aa_i & = & a_i \sqrt{S^A_{14} S^E_{5}/(S^A_{13} S^E_{2i+2})}\;\;\;
i=1,2,3 \nonumber \\
 \bb_i & = & b_i \sqrt{S^A_{14} S^E_{5}/(S^A_{15} S^E_{2i+2})}\;\;\;
i=1,2,3  \\
 \cc_i & = & c_i \sqrt{S^A_{14} S^E_{5}/(S^A_{15} S^E_{2i+2})}\;\;\;
i=1,2,3  \nonumber
\end{eqnarray}
The cells on the three faces formed by the elements $(14,2i)$, $(16,2i)$,
$(15,1,5)$ and $(15,2,5)$
in Figure~8b should satisfy the first unitarity condition. So we can simply
take
\be
\cell {16}{\mbox{}}{2i}{5}{1}{15}=\cc_{i-1},\qquad \cell {16}{\mbox{}}{2i}{5}
{2}{15}=-\bb_{i-1}
\ee
and
\be
\weight C{16}{2i}{5}{17}=\aa_{i-1},\quad i=2,3,4
\ee
On the other hand, the cells $C(14,4,5,13)$ and
$C(14,6,5,13)$ occur on the shared bonds of the faces 10 and 12. So of
course we can obtain
the values $\aa_1$ and $\aa_2$ respectively from the faces 12 and 10 using
the usual rules. This
gives the same results. The complete list of cells for the $A_{29}$--$E_8$
intertwiner are given in
Appendix~C4.

This concludes the derivation of the cells for the classical \ade
intertwiners.
All of the cells can be found in Appendix~C. They satisfy the two unitarity
conditions and the
cell intertwining relations.

\subsection{Affine \ade}
In this section we obtain the intertwining cells for the affine \ade models.
 We first draw the cell graphs and find the absolute values of the cells
using our previous four rules or, equivalently, the unitarity relations.
To determine the remaining signs we use the intertwining relation of the
affine models.

The adjacency matrix $C$ for the affine
$A_{L-1}^{(1)}$--$D_{{L \over 2}+2}^{(1)}$ intertwiner is given in
Appendix~B. Note that the spins $L$ and $1$ are neighbours for the affine
A model. We obtain the cell graph shown in Figure~9 for the affine
$A_{L-1}^{(1)}$--$D_{{L \over 2}+2}^{(1)}$ model. The values of the cells
can be obtained as before. The negative signs in the the cell graphs
indicate the cells that are negative. Therefore we can read off the values
of all the cells from the cell graphs. It is easily verified that these
cells intertwine the faces of the affine $A$--$D$ models even for arbitrary
 crossing parameter $\lambda $. The results for the cells are listed in
Appendix~D1.
\begin{figure}[htb]
\begin{center}
\begin{picture}(310,260)(-10,0)
\put(25,240){\line(-1,-1){10}}
\put(5,220){\vector(1,1){10}}
\put(25,240){\line(1,-1){10}}
\put(45,220){\vector(-1,1){10}}
\put(45,220){\vector(-1,-1){10}}
\put(25,200){\line(1,1){10}}
\put(5,220){\vector(1,-1){10}}
\put(15,210){\line(1,-1){10}}
\put(25,200){\line(1,-1){10}}
\put(45,180){\vector(-1,1){10}}
\put(65,160){\vector(1,-1){10}}
\put(75,150){\line(1,-1){10}}
\put(85,140){\line(1,-1){10}}
\put(105,120){\vector(-1,1){10}}
\put(65,120){\vector(1,-1){10}}
\put(75,110){\line(1,-1){10}}
\put(85,140){\line(-1,-1){10}}
\put(65,120){\vector(1,1){10}}
\put(105,120){\vector(-1,-1){10}}
\put(95,110){\line(-1,-1){10}}
\put(85,100){\line(-1,-1){10}}
\put(65,80){\vector(1,1){10}}
\put(45,60){\vector(-1,-1){10}}
\put(35,50){\line(-1,-1){10}}
\put(25,40){\line(-1,-1){10}}
\put(5,20){\vector(1,1){10}}
\multiput(47,178)(5,-5){4}{\circle*{1}}
\multiput(47,62)(5,5){4}{\circle*{1}}
\put(27,240){\tiny $(L,3)$}
\put(47,220){\tiny $(1,{2})$}
\put(-5,220){\scriptsize(1,1)}
\put(27,200){\scriptsize(2,3)}
\put(47,180){\scriptsize(3,4)}
\put(67,160){\tiny $({L-2 \over 2},{L \over 2})$}
\put(87,140){\tiny $({L \over 2},{L +2\over 2})$}
\put(107,120){\tiny $({L+2 \over 2},{L+6 \over 2})$}
\put(87,100){\tiny $({L+4 \over 2},{L +2\over 2})$}
\put(22,120){\tiny $({L+2 \over 2},{L+4 \over 2})$}
\put(7,20){\tiny $(L,3)$}
\put(27,40){\tiny $(L\!-\!1,4)$}
\put(47,60){\tiny $(L\!-\!2,5)$}
\put(67,80){\tiny $({L+6 \over 2},{L \over 2})$}
\put(25,240){\circle*{3}} \put(45,220){\circle*{3}}
\multiput(5,220)(20,-20){6}{\circle*{3}}
\multiput(5,20)(20,20){5}{\circle*{3}}
\put(65,120){\circle*{3}}
%%%%%%%%%%%%%%%%%%%
\multiput(37,190)(40,-40){2}{--}
\put(13,230){--}
\put(72,130){--}
\put(298,130){--}
\put(210,233){--}
\put(239,208){--}
\put(210,208){--}
\put(77,0){\bf (a)}
\put(225,240){\vector(-1,-1){10}} \put(205,220){\line(1,1){10}}
\put(225,240){\vector(1,-1){10}} \put(245,220){\line(-1,1){10}}
\put(245,220){\vector(-1,-1){10}} \put(225,200){\line(1,1){10}}
\put(205,220){\line(1,-1){10}} \put(225,200){\vector(-1,1){10}}
\put(225,200){\vector(1,-1){10}} \put(245,180){\line(-1,1){10}}
\put(265,160){\line(1,-1){10}} \put(285,140){\vector(-1,1){10}}
\put(285,140){\vector(1,-1){10}} \put(305,120){\line(-1,1){10}}
\put(265,120){\line(1,-1){10}} \put(285,100){\vector(-1,1){10}}
\put(285,140){\vector(-1,-1){10}} \put(265,120){\line(1,1){10}}
\put(305,120){\line(-1,-1){10}} \put(285,100){\vector(1,1){10}}
\put(285,100){\vector(-1,-1){10}} \put(265,80){\line(1,1){10}}
\put(245,60){\line(-1,-1){10}} \put(225,40){\vector(1,1){10}}
\put(225,40){\vector(-1,-1){10}} \put(205,20){\line(1,1){10}}
\multiput(247,178)(5,-5){4}{\circle*{1}}
\multiput(247,62)(5,5){4}{\circle*{1}}
\put(227,240){\tiny $(L,3)$}
\put(247,220){\tiny $(1,{2})$} \put(189,220){\scriptsize(1,1)}
\put(227,200){\scriptsize(2,3)} \put(247,180){\scriptsize(3,4)}
\put(267,160){\tiny $({L-2 \over 2},{L \over 2})$}
\put(287,140){\tiny $({L \over 2},{L+2 \over 2})$}
\put(307,120){\tiny $({L+2 \over 2},{L+6 \over 2})$}
\put(287,100){\tiny $({L+4 \over 2},{L+2 \over 2})$}
\put(222,120){\tiny $({L+2 \over 2},{L+4 \over 2})$}
\put(207,20){\tiny $(L,3)$}
\put(227,40){\tiny $(L\!-\!1,4)$}
\put(247,60){\tiny $(L\!-\!2,5)$}
\put(267,80){\tiny $({L+6 \over 2},{L \over 2})$}
\put(225,240){\circle*{3}}
\put(245,220){\circle*{3}}
\multiput(205,220)(20,-20){6}{\circle*{3}}
\multiput(205,20)(20,20){5}{\circle*{3}}
\put(265,120){\circle*{3}}
\put(277,0){\bf (b)}
\end{picture}
\end{center}
\caption{The cell graphs of the $A_{L-1}^{(1)}-D_{{L \over 2}+2}^{(1)}$
intertwiner. The negative signs indicate that the cell associated with
that bond is negative.}
\end{figure}
\begin{figure}[p]
\begin{center}
\begin{picture}(310,200)(-10,0)
\multiput(30,200)(40,-120){2}{\line(-1,-1){10}}
\multiput(10,180)(40,-120){2}{\vector(1,1){10}}
\put(30,160){\line(1,1){10}}
\put(50,180){\vector(-1,-1){10}}
\put(30,200){\line(1,-1){10}}
\put(50,180){\vector(-1,1){10}}
\put(70,160){\line(-1,1){10}}
\put(50,180){\vector(1,-1){10}}
\put(70,160){\line(1,-1){10}}
\put(90,140){\vector(-1,1){10}}
\put(10,180){\vector(1,-1){10}}
\put(30,160){\line(-1,1){10}}
\put(50,140){\vector(-1,1){10}}
\put(30,160){\line(1,-1){10}}
\put(50,140){\vector(1,-1){10}}
\put(70,120){\line(-1,1){10}}
\put(70,160){\line(-1,-1){10}}
\put(50,140){\vector(1,1){10}}
\put(30,120){\line(1,1){10}}
\put(50,140){\vector(-1,-1){10}}
\put(30,120){\line(-1,-1){10}}
\put(10,100){\vector(1,1){10}}
\put(90,140){\vector(-1,-1){10}}
\put(70,120){\line(1,1){10}}
\put(50,100){\vector(1,1){10}}
\put(70,120){\line(-1,-1){10}}
\put(50,100){\vector(-1,-1){10}}
\put(30,80){\line(1,1){10}}
\put(30,120){\line(1,-1){10}}
\put(50,100){\vector(-1,1){10}}
\put(70,80){\line(-1,1){10}}
\put(50,100){\vector(1,-1){10}}
\put(10,100){\vector(1,-1){10}}
\put(30,80){\line(-1,1){10}}
\put(50,60){\vector(-1,1){10}}
\put(30,80){\line(1,-1){10}}
%%%%%%%%%%%%%%%%%%%%%%%
\multiput(30,200)(20,-20){4}{\circle*{3}}
\multiput(10,180)(20,-20){4}{\circle*{3}}
\multiput(30,120)(20,-20){3}{\circle*{3}}
\multiput(10,100)(20,-20){3}{\circle*{3}}
\put(20,205){\scriptsize(1,2)}
\put(52,180){\scriptsize(2,3)}
\put(72,160){\scriptsize(3,4)}
\put(92,140){\scriptsize(4,5)}
\put(72,120){\scriptsize(5,4)}
\put(54,98){\scriptsize(6,3)}
\put(72,80){\scriptsize(1,2)}
\put(40,50){\scriptsize(2,3)}
\put(-8,180){\scriptsize(2,1)}
\put(12,160){\scriptsize(3,2)}
\put(32,140){\scriptsize(4,3)}
\put(12,120){\scriptsize(5,6)}
\put(-8,100){\scriptsize(6,7)}
\put(12,80){\scriptsize(1,6)}
\put(40,15){\bf (a)}
%%%%%%%%%%%%%%%%%%
\multiput(230,200)(40,-120){2}{\vector(-1,-1){10}}
\multiput(210,180)(40,-120){2}{\line(1,1){10}}
\put(230,160){\vector(1,1){10}}
\put(250,180){\line(-1,-1){10}}
\put(230,200){\vector(1,-1){10}}
\put(250,180){\line(-1,1){10}}
\put(270,160){\vector(-1,1){10}}
\put(250,180){\line(1,-1){10}}
\put(270,160){\vector(1,-1){10}}
\put(290,140){\line(-1,1){10}}
\put(210,180){\line(1,-1){10}}
\put(230,160){\vector(-1,1){10}}
\put(250,140){\line(-1,1){10}}
\put(230,160){\vector(1,-1){10}}
\put(250,140){\line(1,-1){10}}
\put(270,120){\vector(-1,1){10}}
\put(270,160){\vector(-1,-1){10}}
\put(250,140){\line(1,1){10}}
\put(230,120){\vector(1,1){10}}
\put(250,140){\line(-1,-1){10}}
\put(230,120){\vector(-1,-1){10}}
\put(210,100){\line(1,1){10}}
\put(290,140){\line(-1,-1){10}}
\put(270,120){\vector(1,1){10}}
\put(250,100){\line(1,1){10}}
\put(270,120){\vector(-1,-1){10}}
\put(250,100){\line(-1,-1){10}}
\put(230,80){\vector(1,1){10}}
\put(230,120){\vector(1,-1){10}}
\put(250,100){\line(-1,1){10}}
\put(270,80){\vector(-1,1){10}}
\put(250,100){\line(1,-1){10}}
\put(210,100){\line(1,-1){10}}
\put(230,80){\vector(-1,1){10}}
\put(250,60){\line(-1,1){10}}
\put(230,80){\vector(1,-1){10}}
%%%%%%%%%%%%%%%%%%%%%%%
\multiput(230,200)(20,-20){4}{\circle*{3}}
\multiput(210,180)(20,-20){4}{\circle*{3}}
\multiput(230,120)(20,-20){3}{\circle*{3}}
\multiput(210,100)(20,-20){3}{\circle*{3}}
\put(220,205){\scriptsize(1,2)}
\put(252,180){\scriptsize(2,3)}
\put(272,160){\scriptsize(3,4)}
\put(292,140){\scriptsize(4,5)}
\put(272,120){\scriptsize(5,4)}
\put(254,98){\scriptsize(6,3)}
\put(272,80){\scriptsize(1,2)}
\put(240,50){\scriptsize(2,3)}
\put(192,180){\scriptsize(2,1)}
\put(212,160){\scriptsize(3,2)}
\put(232,140){\scriptsize(4,3)}
\put(212,120){\scriptsize(5,6)}
\put(192,100){\scriptsize(6,7)}
\put(212,80){\scriptsize(1,6)}
\put(240,15){\bf (b)}
%%%%%%%%%%%%%%%%%%%%%%%%%%%%%%%%%%%%%%%%%%%%%%
\put(239,170){--}
\put(260,130){--}
\put(243,88){--}
\put(38,170){--}
\put(60,130){--}
\put(43,88){--}
\end{picture}
\end{center}
\caption{The cell graphs of the $A_5^{(1)}$--$E_6^{(1)}$ intertwiner.
The negative signs indicate that the cell associated with that bond is
negative.}
\end{figure}
\begin{figure}[p]
\begin{center}
\begin{picture}(310,310)(200,0)
\multiput(230,280)(40,-40){3}{\vector(1,-1){10}}
\multiput(250,260)(40,-40){3}{\line(-1,1){10}}
\multiput(250,260)(40,-40){2}{\line(1,-1){10}}
\multiput(270,240)(40,-40){2}{\vector(-1,1){10}}

\multiput(210,260)(40,-40){3}{\line(1,-1){10}}
\multiput(230,240)(40,-40){3}{\vector(-1,1){10}}
\multiput(230,240)(40,-40){2}{\vector(1,-1){10}}
\multiput(250,220)(40,-40){2}{\line(-1,1){10}}
\multiput(310,200)(-40,-40){2}{\vector(-1,-1){10}}
\multiput(290,180)(-40,-40){2}{\line(1,1){10}}
\multiput(290,180)(-40,-40){2}{\line(-1,-1){10}}
\multiput(270,160)(-40,-40){2}{\vector(1,1){10}}
\multiput(330,180)(-40,-40){2}{\line(-1,-1){10}}
\multiput(310,160)(-40,-40){2}{\vector(1,1){10}}
\multiput(310,160)(-40,-40){2}{\vector(-1,-1){10}}
\multiput(290,140)(-40,-40){2}{\line(1,1){10}}
\multiput(230,280)(40,-40){3}{\vector(-1,-1){10}}
\multiput(210,260)(40,-40){3}{\line(1,1){10}}
\multiput(230,240)(40,-40){2}{\vector(1,1){10}}
\multiput(250,260)(40,-40){2}{\line(-1,-1){10}}
\multiput(250,180)(0,-40){3}{\line(1,-1){10}}
\multiput(270,160)(0,-40){3}{\vector(-1,1){10}}
\put(250,180){\line(1,1){10}}
\put(270,200){\vector(-1,-1){10}}
\multiput(270,160)(-40,-40){2}{\vector(1,-1){10}}
\multiput(290,140)(-40,-40){2}{\line(-1,1){10}}
\put(270,120){\vector(1,-1){10}}
\put(290,100){\line(-1,1){10}}
\put(290,100){\line(-1,-1){10}}
\put(270,80){\vector(1,1){10}}
%%%%%%%%%%%%%%%%%%%%%%%%%%%%%%%%%%%%%%%%%%%%%%%%%%%%%%%%%%%%%%%%%%%%
\put(228,258){\small 8}
\put(248,238){\small 1}
\put(268,218){\small 2}
\put(288,198){\small 3}
\put(308,178){\small 4}
\put(288,158){\small 5}
\put(268,138){\small 6}
\put(248,118){\small 7}
\put(268,178){\small 9}
\put(268,98){\small 0}
%%%%%%%%%%%%%%%%%%%%%%%%%%%%%%%%%%%%%%%%%%%%%%%%%%%%%%%%%%%%%%%%%%%%%
%%%%%%%%%%%%%%%%%%%%%%%
\put(231,280){\scriptsize(8,2)}
\put(251,260){\scriptsize(1,3)}
\put(271,240){\scriptsize(2,4)}
\put(291,220){\scriptsize(3,5)}
\put(311,200){\scriptsize(4,6)}
\put(331,180){\scriptsize(5,7)}
\put(311,160){\scriptsize(6,6)}
\put(291,140){\scriptsize(7,5)}
\put(271,120){\scriptsize(8,4)}
\put(232,97){\scriptsize(1,3)}
\put(291,100){\scriptsize(1,8)}
\put(265,75){\scriptsize(2,4)}
\put(192,260){\scriptsize(1,1)}
\put(212,237){\scriptsize(2,2)}
\put(232,217){\scriptsize(3,3)}
\put(252,200){\scriptsize(4,4)}
\put(280,180){\scriptsize(5,5)}
\put(232,180){\scriptsize(5,8)}
\put(252,160){\scriptsize(6,4)}
\put(232,140){\scriptsize(7,3)}
\put(212,120){\scriptsize(8,2)}
%%%%%%%%%%%%%%%%%%%
\multiput(210,260)(20,-20){6}{\circle*{3}}
\multiput(250,180)(20,-20){3}{\circle*{3}}
\multiput(250,140)(20,-20){3}{\circle*{3}}
\multiput(230,120)(20,-20){3}{\circle*{3}}
\put(218,270){--}
\put(238,270){--}
\put(238,250){--}
\put(298,190){--}
\put(278,170){--}
\put(258,130){--}
\put(270,30){\bf (a)}
%%%%%%%%%%%%%%%%%%%%%%%%%%%%%%%%%%%%%%%%%%%%%%%%%%%%%%%%%%%%%
%%%%%%%%%%%%%%%%%% %%%%%%%%%%%%%%%%%%%%%%%%%%%%%%%%%%%%%%%%%%%%%%%%%%%
%%%%%%%%%%%%%%%%%%%%%%%%%%%%%%%
\multiput(430,280)(40,-40){3}{\line(1,-1){10}}
\multiput(450,260)(40,-40){3}{\vector(-1,1){10}}
\multiput(450,260)(40,-40){2}{\vector(1,-1){10}}
\multiput(470,240)(40,-40){2}{\line(-1,1){10}}
\multiput(410,260)(40,-40){3}{\vector(1,-1){10}}
\multiput(430,240)(40,-40){3}{\line(-1,1){10}}
\multiput(430,240)(40,-40){2}{\line(1,-1){10}}
\multiput(450,220)(40,-40){2}{\vector(-1,1){10}}
\multiput(510,200)(-40,-40){2}{\line(-1,-1){10}}
\multiput(490,180)(-40,-40){2}{\vector(1,1){10}}
\multiput(490,180)(-40,-40){2}{\vector(-1,-1){10}}
\multiput(470,160)(-40,-40){2}{\line(1,1){10}}
\multiput(530,180)(-40,-40){2}{\vector(-1,-1){10}}
\multiput(510,160)(-40,-40){2}{\line(1,1){10}}
\multiput(510,160)(-40,-40){2}{\line(-1,-1){10}}
\multiput(490,140)(-40,-40){2}{\vector(1,1){10}}
\multiput(430,280)(40,-40){3}{\line(-1,-1){10}}
\multiput(410,260)(40,-40){3}{\vector(1,1){10}}
\multiput(430,240)(40,-40){2}{\line(1,1){10}}
\multiput(450,260)(40,-40){2}{\vector(-1,-1){10}}
\multiput(450,180)(0,-40){3}{\vector(1,-1){10}}
\multiput(470,160)(0,-40){3}{\line(-1,1){10}}
\put(450,180){\vector(1,1){10}}
\put(470,200){\line(-1,-1){10}}
\multiput(470,160)(-40,-40){2}{\line(1,-1){10}}
\multiput(490,140)(-40,-40){2}{\vector(-1,1){10}}
\put(470,120){\line(1,-1){10}}
\put(490,100){\vector(-1,1){10}}
\put(490,100){\vector(-1,-1){10}}
\put(470,80){\line(1,1){10}}
%%%%%%%%%%%%%%%%%%%%%%%%%%%%%%%%%%%%%%%%%%%%%%%%%
%%%%%%%%%%%%%%%%%%%
\put(428,258){\small 8}
\put(448,238){\small 1}
\put(468,218){\small 2}
\put(488,198){\small 3}
\put(508,178){\small 4}
\put(488,158){\small 5}
\put(468,138){\small 6}
\put(448,118){\small 7}
\put(468,178){\small 9}
\put(468,98){\small 0}
%%%%%%%%%%%%%%%%%%%%%%%%%%%%%%%%%%%%%%%%%%%%%%%%%%%%%%%%%%%%%%%%%%%%%%%
%%%%%%%%%%%%%%%%%%%%%
\put(431,280){\scriptsize(8,2)}
\put(451,260){\scriptsize(1,3)}
\put(471,240){\scriptsize(2,4)}
\put(491,220){\scriptsize(3,5)}
\put(511,200){\scriptsize(4,6)}
\put(531,180){\scriptsize(5,7)}
\put(511,160){\scriptsize(6,6)}
\put(491,140){\scriptsize(7,5)}
\put(471,120){\scriptsize(8,4)}
\put(432,97){\scriptsize(1,3)}
\put(491,100){\scriptsize(1,8)}
\put(465,75){\scriptsize(2,4)}
\put(392,260){\scriptsize(1,1)}
\put(412,237){\scriptsize(2,2)}
\put(432,217){\scriptsize(3,3)}
\put(452,200){\scriptsize(4,4)}
\put(480,180){\scriptsize(5,5)}
\put(432,180){\scriptsize(5,8)}
\put(452,160){\scriptsize(6,4)}
\put(432,140){\scriptsize(7,3)}
\put(412,120){\scriptsize(8,2)}
%%%%%%%%%%%%%%%%%%%
\multiput(410,260)(20,-20){6}{\circle*{3}}
\multiput(450,180)(20,-20){3}{\circle*{3}}
\multiput(450,140)(20,-20){3}{\circle*{3}}
\multiput(430,120)(20,-20){3}{\circle*{3}}
\put(418,270){--}
\put(438,270){--}
\put(438,250){--}
\put(498,190){--}
\put(478,170){--}
\put(458,130){--}
\put(470,30){\bf (b)}
\end{picture}
\end{center}
\caption{The cell graphs of the $A_7^{(1)}$--$E_7^{(1)}$ intertwiner.
The negative signs indicate that the cell associated with that bond is
negative.}
\end{figure}
\begin{figure}[htb]
\begin{center}
\begin{picture}(450,270)(-15,200)
\put(90,460){\vector(1,-1){10}}
\put(90,460){\circle*{3}}
\put(110,440){\line(-1,1){10}}
\multiput(50,460)(40,-40){2}{\vector(1,-1){10}}
\multiput(50,460)(20,-20){5}{\circle*{3}}
\multiput(70,440)(40,-40){2}{\line(-1,1){10}}
\multiput(70,440)(40,-40){2}{\line(1,-1){10}}
\multiput(90,420)(40,-40){2}{\vector(-1,1){10}}
\multiput(10,460)(40,-40){3}{\vector(1,-1){10}}
\multiput(10,460)(20,-20){6}{\circle*{3}}
\multiput(30,440)(40,-40){3}{\line(-1,1){10}}
\multiput(30,440)(40,-40){2}{\line(1,-1){10}}
\multiput(50,420)(40,-40){2}{\vector(-1,1){10}}
\multiput(-10,440)(40,-40){2}{\line(1,-1){10}}
\multiput(-10,440)(20,-20){5}{\circle*{3}}
\multiput(10,420)(40,-40){2}{\vector(-1,1){10}}
\multiput(10,420)(40,-40){2}{\vector(1,-1){10}}
\multiput(30,400)(40,-40){2}{\line(-1,1){10}}
%%%%%%%%%%%%%%%%%%%%%%%%%%%%%%
\multiput(10,460)(120,-80){2}{\vector(-1,-1){10}}
\multiput(-10,440)(120,-80){2}{\line(1,1){10}}
\put(50,460){\vector(-1,-1){10}}
\put(30,440){\line(1,1){10}}
\multiput(30,440)(40,0){2}{\line(-1,-1){10}}
\multiput(10,420)(40,0){2}{\vector(1,1){10}}
\multiput(90,460)(-40,-40){2}{\vector(-1,-1){10}}
\multiput(70,440)(-40,-40){2}{\line(1,1){10}}
\multiput(110,440)(-40,-40){2}{\line(-1,-1){10}}
\multiput(110,440)(-40,-40){2}{\circle*{3}}
\multiput(90,420)(-40,-40){2}{\vector(1,1){10}}
\multiput(90,420)(-40,-40){2}{\circle*{3}}
\put(90,420){\vector(-1,-1){10}}
\put(90,420){\circle*{3}}
\put(70,400){\line(1,1){10}}
\put(70,400){\circle*{3}}
\put(110,400){\line(-1,-1){10}}
\put(110,400){\circle*{3}}
\put(90,380){\vector(1,1){10}}
\put(90,380){\circle*{3}}
\put(90,380){\vector(-1,-1){10}}
\put(90,380){\circle*{3}}
\put(70,360){\line(1,1){10}}
\put(70,360){\circle*{3}}
%%%%%%%%%%%%%%%%%%%%%%%%%%%%%%%%%%%%%%%%%%%%%%%%%%%%%%%%%%%%
\put(81,462){\scriptsize(10,6)}
\put(36,462){\scriptsize(10,4)}
\put(-3,462){\scriptsize(10,2)}
\put(113,440){\scriptsize(1,9)}
\put(93,420){\scriptsize(2,6)}
\put(113,400){\scriptsize(3,7)}
\put(133,380){\scriptsize(4,8)}
\put(113,360){\scriptsize(5,7)}
\put(-33,440){\scriptsize(1,1)}
\put(-13,420){\scriptsize(2,2)}
\put(7,400){\scriptsize(3,3)}
\put(27,380){\scriptsize(4,4)}
\put(47,360){\scriptsize(5,5)}
\put(22,440){\scriptsize(1,3)}
\put(42,420){\scriptsize(2,4)}
\put(62,400){\scriptsize(3,5)}
\put(82,380){\scriptsize(4,6)}
\put(61,440){\scriptsize(1,5)}
%%%%%%%%%%%%%%%%%%%%%%%%%%%%%%%%%%%%%%%%%%%%%%%%%%%%%%%%%%%%
\put(7,437){\small 7}
\put(27,417){\small 1}
\put(47,397){\small 2}
\put(67,377){\small 3}
\put(47,437){\small 8}
\put(87,437){\small 9}
\put(67,417){\small 4}
\put(87,397){\small 5}
\put(107,377){\small 6}
\multiput(-5,427)(20,20){2}{--}
\multiput(15,427)(60,0){2}{--}
\multiput(55,387)(40,0){2}{--}
%%%%%%%%%%%%%%%%%%%%%%%%%%%%%%%%%%%%%%%%%%%%%%%%%%%%%%%%%%%%%%%%%%%%%%%%%
%%%%%%%
\put(90,380){\vector(0,-1){10}}
\put(90,380){\circle*{3}}
\put(90,360){\line(0,1){10}}
\put(90,360){\circle*{3}}
\put(70,340){\vector(0,1){10}}
\put(70,360){\line(0,-1){10}}
\put(70,340){\vector(1,1){10}}
\put(90,360){\line(-1,-1){10}}
\put(70,340){\vector(2,1){20}}
\put(110,360){\line(-2,-1){20}}
\put(70,340){\vector(0,-1){10}}
\put(70,320){\line(0,1){10}}
\put(70,340){\vector(1,-1){10}}
\put(90,320){\line(-1,1){10}}
\put(70,340){\vector(2,-1){20}}
\put(110,320){\line(-2,1){20}}
\put(110,340){\vector(0,1){10}}
\put(110,360){\line(0,-1){10}}
\put(110,340){\vector(-1,1){10}}
\put(90,360){\line(1,-1){10}}
\put(110,340){\vector(-2,1){20}}
\put(70,360){\line(2,-1){20}}
\put(110,340){\vector(0,-1){10}}
\put(110,320){\line(0,1){10}}
\put(110,340){\vector(-1,-1){10}}
\put(90,320){\line(1,1){10}}
\put(110,340){\vector(-2,-1){20}}
\put(70,320){\line(2,1){20}}
\put(90,300){\vector(0,1){10}}
\put(90,300){\circle*{3}}
\put(90,320){\line(0,-1){10}}
\put(90,320){\circle*{3}}
\put(82,360){\scriptsize(5,9)}
\put(40,340){\scriptsize(6,1,6)}
\put(113,340){\scriptsize(6,2,6)}
\put(82,320){\scriptsize(7,9)}
\put(70,340){\circle*{3}}
\put(70,360){\circle*{3}}
\put(70,340){\circle*{3}}
\put(90,360){\circle*{3}}
\put(70,340){\circle*{3}}
\put(110,360){\circle*{3}}
\put(70,340){\circle*{3}}
\put(70,320){\circle*{3}}
\put(70,340){\circle*{3}}
\put(90,320){\circle*{3}}
\put(70,340){\circle*{3}}
\put(110,320){\circle*{3}}
\put(110,340){\circle*{3}}
\put(110,360){\circle*{3}}
\put(110,340){\circle*{3}}
\put(90,360){\circle*{3}}
\put(110,340){\circle*{3}}
\put(70,360){\circle*{3}}
\put(110,340){\circle*{3}}
\put(110,320){\circle*{3}}
\put(110,340){\circle*{3}}
\put(90,320){\circle*{3}}
\put(110,340){\circle*{3}}
\put(70,320){\circle*{3}}
%%%%%%%%%%%%%%%%%%%%%%%%%%%
\multiput(70,320)(-40,-40){2}{\line(-1,-1){10}}
\multiput(70,320)(-20,-20){4}{\circle*{3}}
\multiput(50,300)(-40,-40){2}{\vector(1,1){10}}
\put(50,300){\vector(-1,-1){10}}
\put(30,280){\line(1,1){10}}
\multiput(110,320)(-40,-40){2}{\line(-1,-1){10}}
\multiput(110,320)(-20,-20){5}{\circle*{3}}
\multiput(90,300)(-40,-40){2}{\vector(1,1){10}}
\multiput(90,300)(-40,-40){2}{\vector(-1,-1){10}}
\multiput(70,280)(-40,-40){2}{\line(1,1){10}}
\multiput(130,300)(-40,-40){2}{\vector(-1,-1){10}}
\multiput(130,300)(-20,-20){4}{\circle*{3}}
\multiput(110,280)(-40,-40){2}{\line(1,1){10}}
\put(110,280){\line(-1,-1){10}}
\put(90,260){\vector(1,1){10}}
%%%%%%%%%%%%%%%%%%%%%%%%%%%%%%%%%%%%%%%%%%
\multiput(110,320)(-40,-40){2}{\line(1,-1){10}}
\multiput(130,300)(-40,-40){2}{\vector(-1,1){10}}
\multiput(70,320)(-40,-40){2}{\line(1,-1){10}}
\multiput(90,300)(-40,-40){2}{\vector(-1,1){10}}
\multiput(90,300)(-40,-40){2}{\vector(1,-1){10}}
\multiput(110,280)(-40,-40){2}{\line(-1,1){10}}
\multiput(50,300)(-40,-40){2}{\vector(1,-1){10}}
\multiput(70,280)(-40,-40){2}{\line(-1,1){10}}
%%%%%%%%%%%%%%%%%%%%%%%%%%%%%%%%%%%%%%%%%%%%%%%%%%%%%%%%%%%%%%%%%%%%%%%%%%
\put(93,260){\scriptsize(10,6)}
\put(113,280){\scriptsize(9,7)}
\put(133,300){\scriptsize(8,8)}
\put(113,320){\scriptsize(7,7)}
\put(20,234){\scriptsize(1,3)}
\put(-13,260){\scriptsize(10,2)}
\put(7,280){\scriptsize(9,3)}
\put(27,300){\scriptsize(8,4)}
\put(47,320){\scriptsize(7,5)}
\put(61,234){\scriptsize(1,5)}
\put(37,260){\scriptsize(10,4)}
\put(61,280){\scriptsize(9,5)}
\put(82,300){\scriptsize(8,6)}
%%%%%%%%%%%%%%%%%%%%%%%%%%%%%%%%%%%%%%%%%%%%%%%%%%
%%%%%%%%%%%%%%%%%%%%%%%%%%%%%%
\put(25,257){\small 1}
\put(45,277){\small 2}
\put(65,297){\small 3}
\put(67,257){\small 4}
\put(87,277){\small 5}
\put(105,297){\small 6}
\multiput(35,287)(40,0){3}{--}
\put(15,247){--}
\put(47,200){ \bf (a)}
%%%%%%%%%%%%%%%%%%%%%%%%%%%%%%%%%%%%%
%%%%%%%%%%%%%%%%%%%%%%%%%%%%%%%%%%%%%%%%%%%%
%%%%%%%%%%%%%%%%%%%%%%%%%%%%%%%%%%%%%%%%%%%%%%%%%%%%%%%%%%%%%%%%%%%%%%%%
\put(290,460){\line(1,-1){10}}
\put(290,460){\circle*{3}}
\put(310,440){\vector(-1,1){10}}
\multiput(250,460)(40,-40){2}{\line(1,-1){10}}
\multiput(250,460)(20,-20){5}{\circle*{3}}
\multiput(270,440)(40,-40){2}{\vector(-1,1){10}}
\multiput(270,440)(40,-40){2}{\vector(1,-1){10}}
\multiput(290,420)(40,-40){2}{\line(-1,1){10}}
\multiput(210,460)(40,-40){3}{\line(1,-1){10}}
\multiput(210,460)(20,-20){6}{\circle*{3}}
\multiput(230,440)(40,-40){3}{\vector(-1,1){10}}
\multiput(230,440)(40,-40){2}{\vector(1,-1){10}}
\multiput(250,420)(40,-40){2}{\line(-1,1){10}}
\multiput(190,440)(40,-40){2}{\vector(1,-1){10}}
\multiput(190,440)(20,-20){5}{\circle*{3}}
\multiput(210,420)(40,-40){2}{\line(-1,1){10}}
\multiput(210,420)(40,-40){2}{\line(1,-1){10}}
\multiput(230,400)(40,-40){2}{\vector(-1,1){10}}
%%%%%%%%%%%%%%%%%%%%%%%%%%%%%%
\multiput(210,460)(120,-80){2}{\line(-1,-1){10}}
\multiput(190,440)(120,-80){2}{\vector(1,1){10}}
\put(250,460){\line(-1,-1){10}}
\put(230,440){\vector(1,1){10}}
\multiput(230,440)(40,0){2}{\vector(-1,-1){10}}
\multiput(210,420)(40,0){2}{\line(1,1){10}}
\multiput(290,460)(-40,-40){2}{\line(-1,-1){10}}
\multiput(270,440)(-40,-40){2}{\vector(1,1){10}}
\multiput(310,440)(-40,-40){2}{\vector(-1,-1){10}}
\multiput(310,440)(-40,-40){2}{\circle*{3}}
\multiput(290,420)(-40,-40){2}{\line(1,1){10}}
\multiput(290,420)(-40,-40){2}{\circle*{3}}
\put(290,420){\line(-1,-1){10}}
\put(290,420){\circle*{3}}
\put(270,400){\vector(1,1){10}}
\put(270,400){\circle*{3}}
\put(310,400){\vector(-1,-1){10}}
\put(310,400){\circle*{3}}
\put(290,380){\line(1,1){10}}
\put(290,380){\circle*{3}}
\put(290,380){\line(-1,-1){10}}
\put(290,380){\circle*{3}}
\put(270,360){\vector(1,1){10}}
\put(270,360){\circle*{3}}
%%%%%%%%%%%%%%%%%%%%%%%%%%%%%%%%%%%%%%%%%%%%%%%%%%%%%%%%%%%%
\put(281,462){\scriptsize(10,6)}
\put(236,462){\scriptsize(10,4)}
\put(187,462){\scriptsize(10,2)}
\put(313,440){\scriptsize(1,9)}
\put(293,420){\scriptsize(2,6)}
\put(313,400){\scriptsize(3,7)}
\put(333,380){\scriptsize(4,8)}
\put(313,360){\scriptsize(5,7)}
\put(167,440){\scriptsize(1,1)}
\put(187,420){\scriptsize(2,2)}
\put(207,400){\scriptsize(3,3)}
\put(227,380){\scriptsize(4,4)}
\put(247,360){\scriptsize(5,5)}
\put(222,440){\scriptsize(1,3)}
\put(242,420){\scriptsize(2,4)}
\put(262,400){\scriptsize(3,5)}
\put(282,380){\scriptsize(4,6)}
\put(261,440){\scriptsize(1,5)}
%%%%%%%%%%%%%%%%%%%%%%%%%%%%%%%%%%%%%
%%%%%%%%%%%%%%%%%%%%%%%
\put(207,437){\small 7}
\put(227,417){\small 1}
\put(247,397){\small 2}
\put(267,377){\small 3}
\put(247,437){\small 8}
\put(287,437){\small 9}
\put(267,417){\small 4}
\put(287,397){\small 5}
\put(307,377){\small 6}
\multiput(195,427)(20,20){2}{--}
\multiput(215,427)(60,0){2}{--}
\multiput(255,387)(40,0){2}{--}
%%%%%%%%%%%%%%%%%%%%%%%%%%%%%%%%%%%%%%%%%%%%%%%%%%%%%%%
%%%%%%%%%%%%%%%%%%%%%%%%%
\put(290,380){\line(0,-1){10}}
\put(290,380){\circle*{3}}
\put(290,360){\vector(0,1){10}}
\put(290,360){\circle*{3}}
\put(270,340){\line(0,1){10}}
\put(270,360){\vector(0,-1){10}}
\put(270,340){\line(1,1){10}}
\put(290,360){\vector(-1,-1){10}}
\put(270,340){\line(2,1){20}}
\put(310,360){\vector(-2,-1){20}}
\put(270,340){\line(0,-1){10}}
\put(270,320){\vector(0,1){10}}
\put(270,340){\line(1,-1){10}}
\put(290,320){\vector(-1,1){10}}
\put(270,340){\line(2,-1){20}}
\put(310,320){\vector(-2,1){20}}
\put(310,340){\line(0,1){10}}
\put(310,360){\vector(0,-1){10}}
\put(310,340){\line(-1,1){10}}
\put(290,360){\vector(1,-1){10}}
\put(310,340){\line(-2,1){20}}
\put(270,360){\vector(2,-1){20}}
\put(310,340){\line(0,-1){10}}
\put(310,320){\vector(0,1){10}}
\put(310,340){\line(-1,-1){10}}
\put(290,320){\vector(1,1){10}}
\put(310,340){\line(-2,-1){20}}
\put(270,320){\vector(2,1){20}}
\put(290,300){\line(0,1){10}}
\put(290,300){\circle*{3}}
\put(290,320){\vector(0,-1){10}}
\put(290,320){\circle*{3}}
\put(282,360){\scriptsize(5,9)}
\put(240,340){\scriptsize(6,1,6)}
\put(313,340){\scriptsize(6,2,6)}
\put(282,320){\scriptsize(7,9)}
\put(270,340){\circle*{3}}
\put(270,360){\circle*{3}}
\put(270,340){\circle*{3}}
\put(290,360){\circle*{3}}
\put(270,340){\circle*{3}}
\put(310,360){\circle*{3}}
\put(270,340){\circle*{3}}
\put(270,320){\circle*{3}}
\put(270,340){\circle*{3}}
\put(290,320){\circle*{3}}
\put(270,340){\circle*{3}}
\put(310,320){\circle*{3}}
\put(310,340){\circle*{3}}
\put(310,360){\circle*{3}}
\put(310,340){\circle*{3}}
\put(290,360){\circle*{3}}
\put(310,340){\circle*{3}}
\put(270,360){\circle*{3}}
\put(310,340){\circle*{3}}
\put(310,320){\circle*{3}}
\put(310,340){\circle*{3}}
\put(290,320){\circle*{3}}
\put(310,340){\circle*{3}}
\put(270,320){\circle*{3}}
%%%%%%%%%%%%%%%%%%%%%%%%%%%
\multiput(270,320)(-40,-40){2}{\vector(-1,-1){10}}
\multiput(270,320)(-20,-20){4}{\circle*{3}}
\multiput(250,300)(-40,-40){2}{\line(1,1){10}}
\put(250,300){\line(-1,-1){10}}
\put(230,280){\vector(1,1){10}}
\multiput(310,320)(-40,-40){2}{\vector(-1,-1){10}}
\multiput(310,320)(-20,-20){5}{\circle*{3}}
\multiput(290,300)(-40,-40){2}{\line(1,1){10}}
\multiput(290,300)(-40,-40){2}{\line(-1,-1){10}}
\multiput(270,280)(-40,-40){2}{\vector(1,1){10}}
\multiput(330,300)(-40,-40){2}{\line(-1,-1){10}}
\multiput(330,300)(-20,-20){4}{\circle*{3}}
\multiput(310,280)(-40,-40){2}{\vector(1,1){10}}
\put(310,280){\vector(-1,-1){10}}
\put(290,260){\line(1,1){10}}
%%%%%%%%%%%%%%%%%%%%%%%%%%%%%%%%%%%%%%%%%%
\multiput(310,320)(-40,-40){2}{\vector(1,-1){10}}
\multiput(330,300)(-40,-40){2}{\line(-1,1){10}}
\multiput(270,320)(-40,-40){2}{\vector(1,-1){10}}
\multiput(290,300)(-40,-40){2}{\line(-1,1){10}}
\multiput(290,300)(-40,-40){2}{\line(1,-1){10}}
\multiput(310,280)(-40,-40){2}{\vector(-1,1){10}}
\multiput(250,300)(-40,-40){2}{\line(1,-1){10}}
\multiput(270,280)(-40,-40){2}{\vector(-1,1){10}}
%%%%%%%%%%%%%%%%%%%%%%%%%%%%%%%%%%%%%%%%%%%%%%%%%%%%%%%%%%%%%%%%%%%%%%%%%%
\put(293,260){\scriptsize(10,6)}
\put(313,280){\scriptsize(9,7)}
\put(333,300){\scriptsize(8,8)}
\put(313,320){\scriptsize(7,7)}
\put(220,234){\scriptsize(1,3)}
\put(187,260){\scriptsize(10,2)}
\put(207,280){\scriptsize(9,3)}
\put(227,300){\scriptsize(8,4)}
\put(247,320){\scriptsize(7,5)}
\put(261,234){\scriptsize(1,5)}
\put(237,260){\scriptsize(10,4)}
\put(261,280){\scriptsize(9,5)}
\put(282,300){\scriptsize(8,6)}
%%%%%%%%%%%%%%%%%%%%%%%%%%%%%%%%%%%%%%%%%%%%%%%%%%%%%%%%%%%%%%%%%%%%
%%%%%%%%%%%%
\put(225,257){\small 1}
\put(245,277){\small 2}
\put(265,297){\small 3}
\put(267,257){\small 4}
\put(287,277){\small 5}
\put(305,297){\small 6}
\multiput(235,287)(40,0){3}{--}
\put(215,247){--}
\put(247,200){ \bf (b)}
\end{picture}
\end{center}
\caption{The cell graphs of the $A_9^{(1)}$--$E_8^{(1)}$ intertwiner.
The negative signs indicate that the cell associated with that bond is
 negative.}
\end{figure}
\begin{figure}[htb]
\begin{center}
\begin{picture}(410,230)(-10,10)
\put(70,200){\line(0,-1){20}}
\put(70,160){\vector(0,1){20}}
\put(70,200){\line(-1,-1){20}}
\put(30,160){\vector(1,1){20}}
\put(70,200){\line(1,-1){20}}
\put(110,160){\vector(-1,1){20}}
\put(30,120){\line(0,1){20}}
\put(30,160){\vector(0,-1){20}}
\put(30,120){\line(1,1){20}}
\put(70,160){\vector(-1,-1){20}}
\put(30,120){\line(2,1){40}}
\put(110,160){\vector(-2,-1){40}}
\put(30,120){\line(0,-1){20}}
\put(30,80){\vector(0,1){20}}
\put(30,120){\line(1,-1){20}}
\put(70,80){\vector(-1,1){20}}
\put(30,120){\line(2,-1){40}}
\put(110,80){\vector(-2,1){40}}
\put(110,120){\line(0,1){20}}	\put(110,160){\vector(0,-1){20}}
\put(110,120){\line(-1,1){20}}
\put(70,160){\vector(1,-1){20}}
\put(110,120){\line(-2,1){40}}
\put(30,160){\vector(2,-1){40}}
\put(110,120){\line(0,-1){20}}
\put(110,80){\vector(0,1){20}}
\put(110,120){\line(-1,-1){20}}
\put(70,80){\vector(1,1){20}}
\put(110,120){\line(-2,-1){40}}
\put(30,80){\vector(2,1){40}}
\put(70,40){\line(1,1){20}}		\put(110,80){\vector(-1,-1){20}}
\put(70,40){\line(-1,1){20}}	\put(30,80){\vector(1,-1){20}}
\put(70,40){\line(0,1){20}}		\put(70,80){\vector(0,-1){20}}
%%%%%%%%%%%%%%%%%%%%%%%%%%%%%%%%%
\put(70,200){\circle*{3}}
\put(70,160){\circle*{3}}
\put(70,200){\circle*{3}}
\put(30,160){\circle*{3}}
\put(70,200){\circle*{3}}
\put(110,160){\circle*{3}}
\put(30,120){\circle*{3}}	\put(30,160){\circle*{3}}
\put(30,120){\circle*{3}}	\put(70,160){\circle*{3}}
\put(30,120){\circle*{3}}	\put(110,160){\circle*{3}}
\put(30,120){\circle*{3}}
\put(30,80){\circle*{3}}
\put(30,120){\circle*{3}}
\put(70,80){\circle*{3}}
\put(30,120){\circle*{3}}
\put(110,80){\circle*{3}}
\put(110,120){\circle*{3}}
\put(110,160){\circle*{3}}
\put(110,120){\circle*{3}}
\put(70,160){\circle*{3}}
\put(110,120){\circle*{3}}
\put(30,160){\circle*{3}}
\put(110,120){\circle*{3}}
\put(110,80){\circle*{3}}
\put(110,120){\circle*{3}}
\put(70,80){\circle*{3}}
\put(110,120){\circle*{3}}
\put(30,80){\circle*{3}}
\put(70,40){\circle*{3}}
\put(110,80){\circle*{3}}
\put(70,40){\circle*{3}}	\put(30,80){\circle*{3}}
\put(70,40){\circle*{3}}
\put(70,80){\circle*{3}}
\put(60,203){\scriptsize(4,6)}
\put(60,160){\scriptsize(5,9)}
\put(0,160){\scriptsize(5,5)}
\put(113,160){\scriptsize(5,7)}
\put(0,120){\scriptsize(6,1,6)}
\put(113,120){\scriptsize(6,2,6)}
\put(0,80){\scriptsize(7,5)}
\put(113,80){\scriptsize(7,7)}
\put(60,80){\scriptsize(7,9)}
\put(60,30){\scriptsize(8,6)}
\multiput(37,180)(0,-127){2}{$a_1$}
\multiput(89,180)(0,-127){2}{$a_2$}
\multiput(64,184)(0,-128){2}{$a_3$}
\put(17,140){$b_1$}
\put(77,110){$-b_1$}
\put(17,100){$c_1$}
\put(77,130){$c_1$}
\put(110,100){$-b_2$}
\put(53,130){$b_2$}
\put(85,100){$-b_3$}
\put(40,135){$b_3$}
\put(113,140){$c_2$}
\put(50,110){$c_2$}
\put(40,105){$c_3$}
\put(89,135){$c_3$}
\put(60,10){\bf (a)}
%%%%%%%%%%%%%%%%%%%%%%%%%%%%%%%%%%%%%
\put(270,200){\vector(0,-1){20}}
\put(270,160){\line(0,1){20}}
\put(270,200){\vector(-1,-1){20}}
\put(230,160){\line(1,1){20}}
\put(270,200){\vector(1,-1){20}}
\put(310,160){\line(-1,1){20}}
\put(230,120){\vector(0,1){20}}	\put(230,160){\line(0,-1){20}}
\put(230,120){\vector(1,1){20}}	\put(270,160){\line(-1,-1){20}}
\put(230,120){\vector(2,1){40}}	\put(310,160){\line(-2,-1){40}}
\put(230,120){\vector(0,-1){20}}
\put(230,80){\line(0,1){20}}
\put(230,120){\vector(1,-1){20}}
\put(270,80){\line(-1,1){20}}
\put(230,120){\vector(2,-1){40}}
\put(310,80){\line(-2,1){40}}
\put(310,120){\vector(0,1){20}}	\put(310,160){\line(0,-1){20}}
\put(310,120){\vector(-1,1){20}}
\put(270,160){\line(1,-1){20}}
\put(310,120){\vector(-2,1){40}}
\put(230,160){\line(2,-1){40}}
\put(310,120){\vector(0,-1){20}}
\put(310,80){\line(0,1){20}}
\put(310,120){\vector(-1,-1){20}}
\put(270,80){\line(1,1){20}}
\put(310,120){\vector(-2,-1){40}}
\put(230,80){\line(2,1){40}}
\put(270,40){\vector(1,1){20}}
\put(310,80){\line(-1,-1){20}}
\put(270,40){\vector(-1,1){20}}
\put(230,80){\line(1,-1){20}}
\put(270,40){\vector(0,1){20}}
\put(270,80){\line(0,-1){20}}
%%%%%%%%%%%%%%%%%%%%%%%%%%%%%%%%%%%%%
\put(270,200){\circle*{3}}
\put(270,160){\circle*{3}}
\put(270,200){\circle*{3}}
\put(230,160){\circle*{3}}
\put(270,200){\circle*{3}}
\put(310,160){\circle*{3}}
\put(230,120){\circle*{3}}
\put(230,160){\circle*{3}}
\put(230,120){\circle*{3}}
\put(270,160){\circle*{3}}
\put(230,120){\circle*{3}}
\put(310,160){\circle*{3}}
\put(230,120){\circle*{3}}
\put(230,80){\circle*{3}}
\put(230,120){\circle*{3}}
\put(270,80){\circle*{3}}
\put(230,120){\circle*{3}}
\put(310,80){\circle*{3}}
\put(310,120){\circle*{3}}
\put(310,160){\circle*{3}}
\put(310,120){\circle*{3}}
\put(270,160){\circle*{3}}
\put(310,120){\circle*{3}}
\put(230,160){\circle*{3}}
\put(310,120){\circle*{3}}
\put(310,80){\circle*{3}}
\put(310,120){\circle*{3}}
\put(270,80){\circle*{3}}
\put(310,120){\circle*{3}}
\put(230,80){\circle*{3}}
\put(270,40){\circle*{3}}
\put(310,80){\circle*{3}}
\put(270,40){\circle*{3}}
\put(230,80){\circle*{3}}
\put(270,40){\circle*{3}}
\put(270,80){\circle*{3}}
\put(260,203){\scriptsize(4,6)}
\put(260,160){\scriptsize(5,9)}
\put(200,160){\scriptsize(5,5)}
\put(313,160){\scriptsize(5,7)}
\put(200,120){\scriptsize(6,1,6)}
\put(313,120){\scriptsize(6,2,6)}
\put(200,80){\scriptsize(7,5)}
\put(313,80){\scriptsize(7,7)}
\put(260,80){\scriptsize(7,9)}
\put(260,30){\scriptsize(8,6)}
\multiput(233,180)(0,-127){2}{$\aa_1$}
\multiput(289,180)(0,-127){2}{$\aa_2$}
\multiput(264,184)(0,-130){2}{$\aa_3$}
\put(213,140){$\bb_1$}
\put(277,110){$-\bb_1$}
\put(213,100){$\cc_1$}
\put(277,130){$\cc_1$}
\put(310,100){$-\bb_2$}
\put(253,130){$\bb_2$}
\put(285,100){$-\bb_3$}
\put(233,138){$\bb_3$}
\put(313,140){$\cc_2$}
\put(250,110){$\cc_2$}
\put(236,105){$\cc_3$}
\put(289,139){$\cc_3$}
\put(260,10){\bf (b)}
\end{picture}
\end{center}
\caption{Magnified view of the $A_9^{(1)}$--$E_8^{(1)}$ cell graphs in the
vicinity of the special elements $(6,1,6)$ and $(6,2,6)$.}
\end{figure}
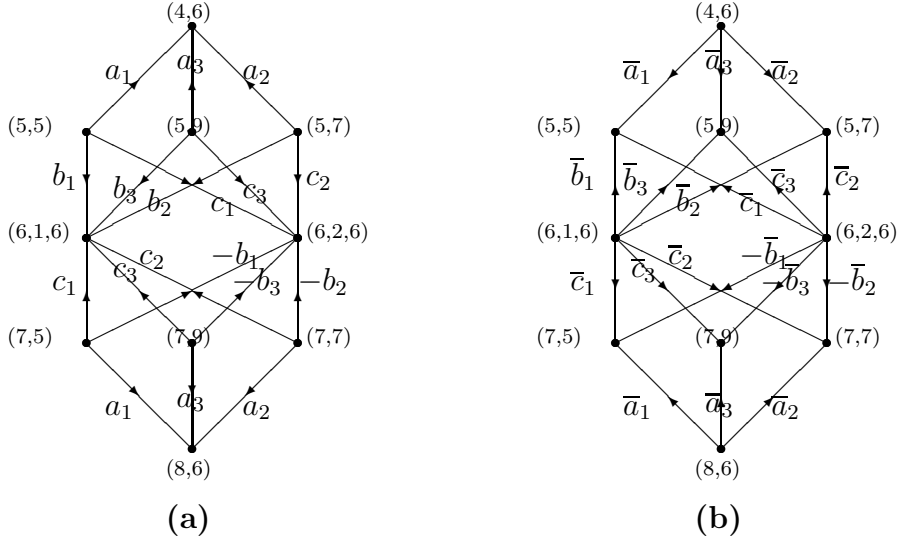
The adjacency matrices $C$ for the $A_5^{(1)}$--$E_6^{(1)}$,
$A_7^{(1)}$--$E_7^{(1)}$ and $A_9^{(1)}$--$E_8^{(1)}$ intertwiners are
also given in Appendix~B. The cell graphs for these intertwiners are shown
in Figures~10--12 respectively. We have indicated which cells are negative
in the graphs. We obtain  all of the cells for the $A_5^{(1)}$--$E_6^{(1)}$
 and $A_7^{(1)}$--$E_7^{(1)}$ intertwiners by applying our previous rules.
For the $A_9^{(1)}$--$E_8^{(1)}$ intertwiner we need to treat the doubly
degenerate element $(6,6)$ by introducing bond variables as we did for the
classical $A_{29}$--$E_8$
intertwiner. The cell graphs of the $A_9^{(1)}$--$E_8^{(1)}$ intertwiner
contains three parts: the upper part containing the faces 1--9, the middle
part around the elements $(6,1,6)$ and $(6,2,6)$ and the lower part
containing the faces 1--6. By our rules the absolute values of the cells
in the upper part are the same as those in the lower part, that is,
$|C(a,1,b,c,1,d)|=|C(12-a,1,b,c,1,12-d)|$. So we have marked the faces in
the lower part by the same number as the symmetric faces in the upper part.
However, the intertwining relation of the models does not give the same
sign for $C(a,1,b,c,1,d)$ and $C(12-a,1,b,c,1,12-d)$ generally. We can
treat the middle part in the same way as the $A_{29}$--$E_8$ case. The
cell graphs for this part are shown magnified in Figure~13.
Following the arguments for the $A_{29}$--$E_8$ intertwiner we obtain the
following results \begin{eqnarray}
 b_i^2+c_i^2= \sqrt{S^E_{3+2i}\over S^E_{6}}\;\;\;\;i=1,2,3.
\end{eqnarray}
Substituting $b$ and $c$ into the above equations, we have
\begin{eqnarray}
 \cos\gamma & = & \sqrt{S^E_{5}\over S^E_6}\;\;\;\;{\rm for}\;\;i=1;     \\
 \sin\alpha & = & \sqrt{S^E_{6}- S^E_7\over  S^E_5}\;\;\;\;
	{\rm for}\;\;i=2\;\;{\rm or}\;\;3.
\end{eqnarray}
We simply set $\beta =\alpha $. We also find
\begin{eqnarray}
 \aa_i & = & a_i \sqrt{ S^E_{6}\over S^E_{2i+3}}\;\;\;i=1,2,3;  \\
\bb_i & = & b_i \sqrt{ S^E_{6}\over  S^E_{2i+3}}\;\;\;i=1,2,3;  \\
 \cc_i & = & c_i \sqrt{ S^E_{6}\over  S^E_{2i+3}}\;\;\;i=1,2,3.
\end{eqnarray}
In this way we obtain all the cells for the affine $A$--$D$ and $A$--$E$
intertwiners. These cells satisfy the  unitarity conditions and the
intertwining relations. We list all of the
cells for the affine \ade intertwiners in Appendix~D.
%\clearpage
\subsection{Dilute \ade}
In this section consider the intertwining cells for the critical classical
dilute \ade models.
We use rules 1--4 and the cell graphs of the usual \ade models to obtain
the cells for the dilute models. The nonzero cells of the classical dilute
\ade models are exactly the same as those for the corresponding classical
Temperley-Lieb \ade models except that there are extra nonzero cells for
the dilute models. The number of the extra cells is just the sum of the
nonzero elements of the adjacency matrix of the adjacency intertwiner.
For example, we find $15$ extra nonzero cells for the dilute
$A_{11}$--$E_6$ intertwiner and $58$ extra nonzero cells for the dilute
$A_{29}$--$E_8$. These extra
cells all occur on the nodes (nonzero elements of the  adjacency matrix of
the intertwiner) and all take the value $1$ or $-1$. This means that the
extra nonzero cells are given by
$$
\weight Cabba =1;\;\;\mbox{\rm if the element\ } (a,b) \mbox{\rm \ of the
adjacency matrix\ } C \mbox{\rm \ is 1}
$$
for the dilute
$A$--$D$, $A_{11}$--$E_6$ and $A_{17}$--$E_7$ models and by
$$
\cell asbbsa =(-1)^{(s-1)};\;\;
\mbox{\rm if the element\ } (a,b) \mbox{\rm \  of the adjacency matrix\ } C
 \mbox{\rm \ is 1}
$$
for the dilute $A_{29}$--$E_8$ models where the bond variable $s=1,2$.
These cells satisfy the unitarity conditions and the intertwining relations.
%%%%%%%%%%%%%%%%%%%%%%%%%%%%%%%%%%%%%%%%%%%%%%%%

\section{Row Transfer Matrix Intertwiners}
\setcounter{equation}{0}

In this section we study intertwiners relating the row transfer matrices of
\ade models. This is the
third level of the intertwining relation. The first two levels relate the
adjacency matrices
and the face weights. These three levels are closely related. Suppose that
\mbox{\boldmath $a$} and \mbox{\boldmath $b$}  are allowed spin
configurations of two consecutive rows of $N$ (even) spins with periodic
boundary
conditions. The elements of the row transfer matrix ${\bf T}(u)$ are given
by
\begin{eqnarray}
\langle\mbox{\boldmath $a$}|{\bf T}(u)|\mbox{\boldmath $b$}\rangle &=&
\prod_{j=1}^N \wt W{a_j}{a_{j+1}}{b_{j+1}}{b_j}u  \nonumber \\
%&\put(80,30){=}&
&=&
\begin{picture}(240,30)(90,25)
\put(95,43){\tiny $b_1$} \put(118,43){\tiny $b_2$} \put(278,43){\tiny
$b_{N+1}$}
\put(95,16){\tiny $a_1$} \put(118,16){\tiny $a_2$} \put(278,16){\tiny
$a_{N+1}$}
\multiput(100,40)(20,0){9}{\vector(1,0){10}}
\multiput(120,40)(20,0){9}{\line(-1,0){10}}
\multiput(100,40)(20,0){5}{\line(0,-1){10}}
\multiput(100,20)(20,0){5}{\vector(0,1){10}}
\multiput(100,20)(20,0){9}{\vector(1,0){10}}
\multiput(120,20)(20,0){9}{\line(-1,0){10}}
\multiput(280,40)(-20,0){3}{\line(0,-1){10}}
\multiput(280,20)(-20,0){3}{\vector(0,1){10}}
\end{picture}
\end{eqnarray}
where $a_{N+1}=a_1$ and $b_{N+1}=b_1$. The Yang-Baxter equation
implies the commutation relations
\begin{equation}
[{\bf T}(u),{\bf T}(v)] = 0. \label{rowcommute}
\end{equation}
More precisely, the proof \cite{Baxter:82} uses the Yang-Baxter equation
and unitarity.

Let us consider the row transfer matrices in the braid limits and define
\be
{\cal B},\overline{\cal B}=\lim_{u\to\mp i\infty}
\left({\sin\lambda\over \sin(u-\lambda/2)}\right)^N{\bf T}(u)
\ee
then clearly
\be
[{\bf T}(u),{\cal B}] = [{\bf T}(u),\overline{\cal B}] = 0
\ee
so that ${\cal B}$ and $\overline{\cal B}$ can be simultaneously
diagonalized with ${\bf T}(u)$.
Moreover, in this braid limit, it can be shown \cite{Paul:91p} that the
eigenvalues of ${\cal B}$ and
$\overline{\cal B}$ are given by
\be
2\cos m_j,\qquad 2\cos \overline{m}_j
\ee
where $m_j$ and $\overline{m}_j$ are the Coxeter exponents of the relevant
\ade model. In this way, the Coxeter
exponents $(m_j,\overline{m}_j)$ are seen as quantum numbers labelling the
sectors of row
transfer matrix eigenvalues. It can also be shown \cite{KlPe:92} that for
the row transfer matrices of the $A_L$
models ${\cal B}_A=\overline{\cal B}_A$ and hence $m_j=\overline{m}_j$. The
braid row transfer matrices play
a similar role at the row transfer matrix level to the role played by the
adjacency matrices at the level of the
adjacency matrices.

Let us denote the row transfer matrices of two \ade models by ${\cal A}$
and
${\cal G}$. Invariably, if an intertwiner exists, the largest eigenvalue of
these matrices is in common. Since this
eigenvalue is exactly the same for a finite size system, the corresponding
critical models will have the same finite
size corrections and hence the same central charge $c$ in the thermodynamic
limit. For the affine \ade models this is
$c=1$. For the classical \ade models the central charge is
\be
c=1-{6\over h(h-1)}
\ee
and, similarly, the conformal weights
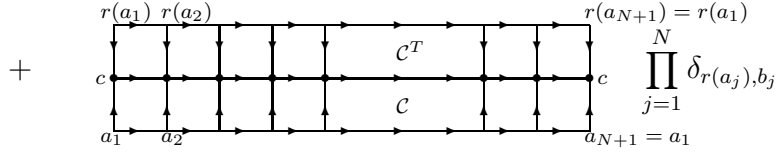
\be
(\Delta_{r,s},\overline{\Delta}_{r,s})
\ee
of the excited states corresponding to the other common
eigenvalues will be exactly the same \cite{KlPe:92,Paul:91p} where $r$ and
$s$ label the columns and rows of
the Kac table. Indeed, in this case, it turns out \cite{KlPe:92} that the
row label $s$ in the Kac Table is
given precisely by the Coxeter exponents
\be
s=m_j.
\ee

\subsection{Construction of Row Intertwiners}

We construct the intertwiners between \ade row transfer matrices from the
cells obtained in Section~3.
Let ${\cal A}$ and ${\cal G}$ be row transfer matrices of two \ade models.
Using the intertwining relation
(\ref{eq:faceinter}) for the face weights and the unitarity conditions for
the cells we can easily construct
the intertwiner for the row transfer matrices ${\cal A}$ and ${\cal G}$.
The intertwining relation
\be
{\cal A}(u){\cal C}={\cal C}{\cal G}(u)
\ee
can be pictured as follows
\begin{eqnarray}
& &\begin{picture}(300,40)(60,10)

\put(95,43){\tiny $a'_1$} \put(118,43){\tiny $a'_2$} \put(278,43){\tiny
$a'_{N+1}=a'_1$}
\put(95,-4){\tiny $a_1$} \put(118,-4){\tiny $a_2$} \put(278,-4){\tiny
$a_{N+1}=a_1$}
\multiput(93,18)(190,0){2}{\tiny $b$}
\multiput(100,40)(20,0){9}{\vector(1,0){10}}
\multiput(120,40)(20,0){9}{\line(-1,0){10}}
\multiput(100,40)(20,0){5}{\line(0,-1){10}}
\multiput(100,20)(20,0){5}{\vector(0,1){10}}
\multiput(100,20)(20,0){9}{\vector(1,0){10}}
\multiput(120,20)(20,0){9}{\line(-1,0){10}}
\multiput(280,40)(-20,0){3}{\line(0,-1){10}}
\multiput(280,20)(-20,0){3}{\vector(0,1){10}}
\multiput(100,20)(20,0){5}{\circle*{3}}
\multiput(280,20)(-20,0){3}{\circle*{3}}

\multiput(100,20)(20,0){5}{\vector(0,-1){10}}
\multiput(100,0)(20,0){5}{\line(0,1){10}}
\multiput(100,0)(20,0){9}{\vector(1,0){10}}
\multiput(120,0)(20,0){9}{\line(-1,0){10}}
\multiput(280,20)(-20,0){3}{\vector(0,-1){10}}
\multiput(280,0)(-20,0){3}{\line(0,1){10}}
\put(207,7){\tiny {${\cal A}$}} \put(207,27){\tiny {${\cal C}$}}
\end{picture}
  \nonumber \\  \label{eq:traninter} \\
& & \begin{picture}(300,40)(60,10)
\put(80,20){=}
\put(95,43){\tiny $a'_1$} \put(118,43){\tiny $a'_2$} \put(278,43){\tiny
$a'_{N+1}=a'_1$}
\put(95,-4){\tiny $a_1$} \put(118,-4){\tiny $a_2$} \put(278,-4){\tiny
$a_{N+1}=a_1$}
\multiput(93,18)(190,0){2}{\tiny $b$}
\multiput(100,40)(20,0){9}{\vector(1,0){10}}
\multiput(120,40)(20,0){9}{\line(-1,0){10}}
\multiput(100,40)(20,0){5}{\vector(0,-1){10}}
\multiput(100,20)(20,0){5}{\line(0,1){10}}
\multiput(100,20)(20,0){9}{\vector(1,0){10}}
\multiput(120,20)(20,0){9}{\line(-1,0){10}}
\multiput(280,40)(-20,0){3}{\vector(0,-1){10}}
\multiput(280,20)(-20,0){3}{\line(0,1){10}}
\multiput(100,20)(20,0){5}{\circle*{3}}
\multiput(280,20)(-20,0){3}{\circle*{3}}

\multiput(100,20)(20,0){5}{\line(0,-1){10}}
\multiput(100,0)(20,0){5}{\vector(0,1){10}}
\multiput(100,0)(20,0){9}{\vector(1,0){10}}
\multiput(120,0)(20,0){9}{\line(-1,0){10}}
\multiput(280,20)(-20,0){3}{\line(0,-1){10}}
\multiput(280,0)(-20,0){3}{\vector(0,1){10}}
\put(207,27){\tiny {${\cal G}$}} \put(207,7){\tiny {${\cal C}$}}
 \nonumber
\end{picture}
\end{eqnarray}

\bigskip\noindent
where the solid circles mean summation over the corresponding spins.
The row intertwiner ${\cal C}$ is constructed from $N$ cells with periodic
boundary
conditions. Specifically, if \mbox{\boldmath{$a$}} and
\mbox{\boldmath{$b$}} are allowed row configurations of
the graph $A$ and $G$ with periodic boundary conditions $a_{N+1}=a_1$ and
$b_{N+1}=b_1$, then the elements
of  the row intertwiner are given by
\begin{equation}
\begin{picture}(300,40)(20,26)
\put(20,25){$\langle\mbox{\boldmath $a$}|{\cal C}|\mbox{\boldmath
$b$}\rangle$ $\;$=}
\put(95,43){\tiny $b_1$} \put(118,43){\tiny $b_2$} \put(278,43){\tiny
$b_{N+1}$}
\put(95,16){\tiny $a_1$} \put(118,16){\tiny $a_2$} \put(278,16){\tiny
$a_{N+1}$}
\multiput(100,40)(20,0){9}{\vector(1,0){10}}
\multiput(120,40)(20,0){9}{\line(-1,0){10}}
\multiput(100,40)(20,0){5}{\line(0,-1){10}}
\multiput(100,20)(20,0){5}{\vector(0,1){10}}
\multiput(100,20)(20,0){9}{\vector(1,0){10}}
\multiput(120,20)(20,0){9}{\line(-1,0){10}}
\multiput(280,40)(-20,0){3}{\line(0,-1){10}}
\multiput(280,20)(-20,0){3}{\vector(0,1){10}}
\put(310,19){}
\end{picture}
\end{equation}

\bigskip\smallskip\noindent
Generally, the row intertwiner is a rectangular matrix.
The proof of the intertwining relation (\ref{eq:traninter}) between the row
transfer matrices proceeds in
exactly the same way as the proof of the commutation relation
(\ref{rowcommute}) with the cell intertwiner
relation playing the role of the Yang-Baxter relation.

\subsection{Properties of Row Intertwiners}

The intertwining relation between row transfer matrices is an equivalence
relation
even though the row transfer matrices are not in general symmetric. The
symmetry of the relation follows from
the crossing symmetry
\be
{\cal G}(u)^T={\cal G}(\lambda-u).
\ee
The row intertwiner projects the eigenvectors of the transfer matrix ${\cal
A}$ corresponding to common
eigenvalues onto those of ${\cal G}$ and annihilates all eigenvectors
corresponding to eigenvalues which are
not in common.

Let us consider the symmetry operators given by the square matrices ${\cal
C} {\cal C}^T$
and  ${\cal C}^T {\cal C}$. Just as for the adjacency matrix level, we can
show that
\be
[{\cal C}{\cal C}^T,{\cal A}(u)]=[{\cal C}^T{\cal C},{\cal G}(u)]=0
\ee
so that the row transfer matrix and its corresponding symmetry operator
have the same eigenvectors and can be
simultaneously diagonalized. The eigenvectors that are not annihilated by
the symmetry operators give the
eigenvalues that are intertwined or common to ${\cal A}$ and ${\cal G}$.
Moreover,
\be
{\cal C}{\cal C}^T\stackrel{\cal C}{\sim}{\cal C}^T{\cal C}
\ee
so that the nonzero eigenvalues of these symmetry operators are in common.
%Notice that since the positive
%Perron-Frobenius eigenvectors of ${\cal A}$ and ${\cal G}$ cannot be
annihilated by the positive
%symmetry operators, the largest eigenvalues of ${\cal A}$ and ${\cal G}$
are necessarily in common.

\subsection{An Example: The $A$--$D$ Row Intertwiner}

Let us now consider the $A_L$--$D_{(L+3)/2}$ models and define the height
reversal operators ${\cal R}_A$ and
${\cal R}_D$ for the $A$ and $D$ models by the elements
\be
\langle\mbox{\boldmath $a$}|{\cal R}_A|\mbox{\boldmath
$b$}\rangle=\prod_{j=1}^N \delta_{a_j,r(b_j)},
\qquad\qquad
\langle\mbox{\boldmath $a$}|{\cal R}_D|\mbox{\boldmath
$b$}\rangle=\prod_{j=1}^N \delta_{a_j,r(b_j)}
\ee
where for the $A$ model $r(b)=h-b$ and for the $D$ model $r(b)=b$ for
$b=1,2,\cdots ,(L-1)/2$,
$r({(L+1)/2})={(L+3)/2}$ and $r({(L+3)/2})={(L+1)/2}$. These matrix
operators reflect the $\Z_2$
symmetry of the models. We show that
\be
{\cal C}{\cal C}^T=I+{\cal R}_A,\qquad\qquad {\cal C}^T{\cal C}=I+{\cal
R}_D\label{symops}
\ee
An immediate consequence of this is that the eigenvalues of ${\cal A}$ and
${\cal D}$ are in common if and
only if the corresponding eigenvectors are even under the $\Z_2$ symmetry.
In particular, since the largest
eigenvalue has an even eigenvector, the largest eigenvalue is in common.

The proof of (\ref{symops}) is straightforward. Using the adjacency
conditions of the graphs $A_L$ and
$D_{(L+3)/2}$, we easily see that the operators ${\cal C}{\cal C}^T$ or
${\cal C}^T{\cal C}$ can be
decomposed into two nonzero terms pictured as follows

\begin{eqnarray}
& & \begin{picture}(400,50)(0,10)
\put(-5,20){$\langle\mbox{\boldmath $a$}|{\cal C}{\cal C}^T|\mbox{\boldmath
$b$}\rangle\ \ \ =\mbox{}\quad$}
\put(95,43){\tiny $b_1$} \put(118,43){\tiny $b_2$} \put(278,43){\tiny
$b_{N+1}=b_1$}
\put(95,-4){\tiny $a_1$} \put(118,-4){\tiny $a_2$} \put(278,-4){\tiny
$a_{N+1}=a_1$}
\multiput(93,18)(190,0){2}{\tiny $c$}
\multiput(100,40)(20,0){9}{\vector(1,0){10}}
\multiput(120,40)(20,0){9}{\line(-1,0){10}}
\multiput(100,40)(20,0){5}{\vector(0,-1){10}}
\multiput(100,20)(20,0){5}{\line(0,1){10}}
\multiput(100,20)(20,0){9}{\vector(1,0){10}}
\multiput(120,20)(20,0){9}{\line(-1,0){10}}
\multiput(280,40)(-20,0){3}{\vector(0,-1){10}}
\multiput(280,20)(-20,0){3}{\line(0,1){10}}
\multiput(100,20)(20,0){5}{\circle*{3}}
\multiput(280,20)(-20,0){3}{\circle*{3}}

\multiput(100,20)(20,0){5}{\line(0,-1){10}}
\multiput(100,0)(20,0){5}{\vector(0,1){10}}
\multiput(100,0)(20,0){9}{\vector(1,0){10}}
\multiput(120,0)(20,0){9}{\line(-1,0){10}}
\multiput(280,20)(-20,0){3}{\line(0,-1){10}}
\multiput(280,0)(-20,0){3}{\vector(0,1){10}}
\put(207,27){\tiny ${\cal C}^T$} \put(207,7){\tiny ${\cal C}$}
\end{picture}  \nonumber \\
& & \begin{picture}(400,60)(0,10)
\put(55,20){$\mbox{}=\mbox{}$}
\put(95,43){\tiny $a_1$} \put(118,43){\tiny $a_2$} \put(278,43){\tiny
$a_{N+1}=a_1$}
\put(95,-4){\tiny $a_1$} \put(118,-4){\tiny $a_2$} \put(278,-4){\tiny
$a_{N+1}=a_1$}
\multiput(93,18)(190,0){2}{\tiny $c$}
\multiput(100,40)(20,0){9}{\vector(1,0){10}}
\multiput(120,40)(20,0){9}{\line(-1,0){10}}
\multiput(100,40)(20,0){5}{\vector(0,-1){10}}
\multiput(100,20)(20,0){5}{\line(0,1){10}}
\multiput(100,20)(20,0){9}{\vector(1,0){10}}
\multiput(120,20)(20,0){9}{\line(-1,0){10}}
\multiput(280,40)(-20,0){3}{\vector(0,-1){10}}
\multiput(280,20)(-20,0){3}{\line(0,1){10}}
\multiput(100,20)(20,0){5}{\circle*{3}}
\multiput(280,20)(-20,0){3}{\circle*{3}}

\multiput(100,20)(20,0){5}{\line(0,-1){10}}
\multiput(100,0)(20,0){5}{\vector(0,1){10}}
\multiput(100,0)(20,0){9}{\vector(1,0){10}}
\multiput(120,0)(20,0){9}{\line(-1,0){10}}
\multiput(280,20)(-20,0){3}{\line(0,-1){10}}
\multiput(280,0)(-20,0){3}{\vector(0,1){10}}
\put(207,27){\tiny ${\cal C}^T$} \put(207,7){\tiny ${\cal C}$}
\put(300,20){$\disp{\prod_{j=1}^N \delta_{a_j,b_j}}$}
\end{picture}   \\
& & \begin{picture}(400,60)(0,10)
\put(60,20){$+$}
\put(95,43){\tiny $r(a_1)$} \put(118,43){\tiny $r(a_2)$} \put(278,43){\tiny
$r(a_{N+1})=r(a_1)$}
\put(95,-4){\tiny $a_1$} \put(118,-4){\tiny $a_2$} \put(278,-4){\tiny
$a_{N+1}=a_1$}
\multiput(93,18)(190,0){2}{\tiny $c$}
\multiput(100,40)(20,0){9}{\vector(1,0){10}}
\multiput(120,40)(20,0){9}{\line(-1,0){10}}
\multiput(100,40)(20,0){5}{\vector(0,-1){10}}
\multiput(100,20)(20,0){5}{\line(0,1){10}}
\multiput(100,20)(20,0){9}{\vector(1,0){10}}
\multiput(120,20)(20,0){9}{\line(-1,0){10}}
\multiput(280,40)(-20,0){3}{\vector(0,-1){10}}
\multiput(280,20)(-20,0){3}{\line(0,1){10}}
\multiput(100,20)(20,0){5}{\circle*{3}}
\multiput(280,20)(-20,0){3}{\circle*{3}}

\multiput(100,20)(20,0){5}{\line(0,-1){10}}
\multiput(100,0)(20,0){5}{\vector(0,1){10}}
\multiput(100,0)(20,0){9}{\vector(1,0){10}}
\multiput(120,0)(20,0){9}{\line(-1,0){10}}
\multiput(280,20)(-20,0){3}{\line(0,-1){10}}
\multiput(280,0)(-20,0){3}{\vector(0,1){10}}
\put(207,27){\tiny ${\cal C}^T$} \put(207,7){\tiny ${\cal C}$}
\put(300,20){$\disp{\prod_{j=1}^N \delta_{r(a_j),b_j}}$}
\end{picture}  \nonumber
\end{eqnarray}
%\clearpage

\bigskip\noindent
A similar picture can be drawn for
$\langle\mbox{\boldmath $a$}|{\cal C}^T{\cal C}|\mbox{\boldmath
$b$}\rangle$. The solid circles indicate
summation. Since the $A$ and $D$ models posess $\Z_2$ symmetry, the row
transfer matrices of the
models commute with the corresponding height reversal operators.

It is relatively simple to determine if an eigenvector is even or odd under
the given $\Z_2$ symmetry.
Indeed, for the row transfer matrix ${\cal A}$, the quantum number under
the $\Z_2$ symmetry
\be
{\cal R}_A={\cal U}^{(L-1)}({\cal B}_A)
\ee
is given by
\be
r_A={\cal U}^{(L-1)}(2\cos m_j)=\cases{1,& $m_j$ odd\cr -1,& $m_j$ even\cr}
\ee
where $2\cos m_j$ is the corresponding eigenvalue of the braid transfer
matrix ${\cal B}_A=\overline{\cal B}_A$.
For the $A$--$D$ models we therefore conclude that the overlap of the row
transfer matrix eigenvalues is
as shown pictorially in Figure~14.
\begin{figure}[htb]
%\hspace{.7in}\mbox{}\hfil\scaledpicture 5.11in by 2.92in (roweig scaled
%%500)\hfil\mbox{}
\[\epsfbox{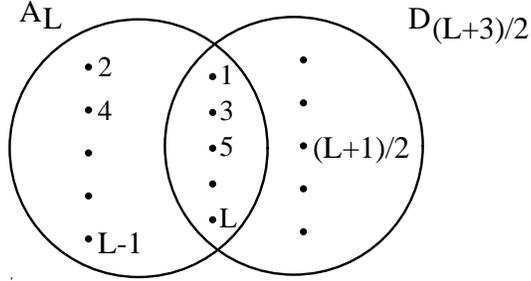}\]
\caption{Pictorial representation of the overlap of the $A_L$ and
$D_{(L+3)/2}$ row transfer matrix
eigenvalues. The common eigenvalues have quantum numbers given by the
Coxeter exponents
$m_j=\overline{m}_j=1,3,5,\ldots,L$ and are even under the $\Z_2$
symmetry.  The other eigenvalues are odd under the $\Z_2$ symmetry. The
eigenvalues lying in $A$ but not in
$D$ have quantum numbers $m_j=\overline{m}_j=2,4,6,\ldots,L-1$ whereas the
eigenvalues lying in $D$ but not
$A$ have quantum numbers $\overline{m}_j=L+1-m_j$ so that $m_j\ne
\overline{m}_j$ except for
$m_j=\overline{m}_j=(L+1)/2$.}
\end{figure}

This overlap can be interpreted in the thermodynamic limit in terms of
conformal weights, or better, in terms
of modular invariant partition functions.
Let us consider the $A_5$--$D_4$ intertwiner relating the
tetracritical Ising model to the critical 3-state Potts model. The
conformal weights in this case are shown in
Table~3.
{\tabcolsep 0pt
\begin{table}[htb]
\begin{center}
\begin{tabular}{p{.75in}p{.75in}p{.75in}p{.75in}p{.75in}p{.75in}}
\pb{$s=m_j$}&&&&&\\  \cline{2-5}
\pb{5}&\fb{3}&\fb{7/5}&\fb{2/5}&\fb{0}&\\  \cline{2-5}
\pb{4}&\fb{13/8}&\fb{21/40}&\fb{1/40}&\fb{1/8}&\\ \cline{2-5}
\pb{3}&\fb{2/3}&\fb{1/15}&\fb{1/15}&\fb{2/3}&\\ \cline{2-5}
\pb{2}&\fb{1/8}&\fb{1/40}&\fb{21/40}&\fb{13/8}&\\ \cline{2-5}
\pb{1}&\fb{0}&\fb{2/5}&\fb{7/5}&\fb{3}&\\ \cline{2-5}
\multicolumn{6}{c}{}\\
\pb{}&\pb{1}&\pb{2}&\pb{3}&\pb{4}&\pb{r}\\
\end{tabular}
\end{center}
\caption{Grid of conformal weights $\Delta_{r,s}$ for $h=6$. The row index
$s$ is identified with the Coxeter
exponent $m_j$. Only the left half of the table with $r=1, 2$ is needed.}
\end{table}
}
We find that the towers of eigenvalues in the two
modular invariant partition functions are divided as follows:
\bea
\mbox{In $A_5$ and $D_4$:}&&
|\chi_0|^2+|\chi_3|^2+|\chi_{2/5}|^2+|\chi_{7/5}|^2+|\chi_{1/15}|^2+|\chi_{2
3}|^2
\nonumber\\
\mbox{In $A_5$ not $D_4$:}&&
|\chi_{1/8}|^2+|\chi_{1/40}|^2+|\chi_{12/40}|^2+|\chi_{13/8}|^2\\
\mbox{In $D_4$ not $A_5$:}&& |\chi_{1/15}|^2+|\chi_{2/3}|^2+
\chi_0\overline{\chi}_3+\overline{\chi}_0\chi_3+\chi_{2/5}\overline{\chi}_{7
5}+\overline{\chi}_{2/5}\chi_{7/5}.
\nonumber
\eea
Here $\chi_\Delta(q)$ are the Virasoro characters, $q$ is the modular
parameter and the bars denote complex
conjugation.

%%%%%%%%%%%%%%%%%%%%%%%%%%%%%%%%%%%%%%%%%%%%%%%%

\section*{Acknowledgements} This research has been supported by the
Australian Research Council. PAP thanks
Prof. Jean Marie Maillard for the invitation to speak in Paris and for the
kind hospitality extended to him.
The authors also thank David O'Brien for discussions.

\clearpage
\appendix
\section{Appendix~A: Classical Adjacency Intertwiners}
\setcounter{equation}{0}
\renewcommand{\theequation}{A.\arabic{equation}}
The adjacency matrix intertwiners $C$ for the classical \ade models are as
follows:
\bigskip
\twocol
\hfil{$A_L$--$D_{L+3\over 2}$}\hfil&\hfil{$A_{11}$--$E_6$}\hfil\\
\smallskip
\twocol
$C=\smat{
1&0&\ldots&0&0&0\cr
0&1&\ldots&0&0&0\cr
\vdots&\vdots&\ddots&\vdots&\vdots&\vdots\cr
0&0&\ldots&1&0&0\cr
0&0&\ldots&0&1&1\cr
0&0&\ldots&1&0&0\cr
\vdots&\vdots&\dddots&\vdots&\vdots&\vdots\cr
0&1&\ldots&0&0&0\cr
1&0&\ldots&0&0&0}$&
$C=\smat{
1&0&0&0&0&0\cr
0&1&0&0&0&0\cr
0&0&1&0&0&0\cr
0&0&0&1&0&1\cr
0&0&1&0&1&0\cr
0&1&0&1&0&0\cr
1&0&1&0&0&0\cr
0&1&0&0&0&1\cr
0&0&1&0&0&0\cr
0&0&0&1&0&0\cr
0&0&0&0&1&0}$\\
\bigskip
\twocol
\hfil{$A_{17}$--$E_7$}\hfil&\hfil{$A_{29}$--$E_8$}\hfil\\
\smallskip
\twocol
$C=\smat{
1&0&0&0&0&0&0\cr
0&1&0&0&0&0&0\cr
0&0&1&0&0&0&0\cr
0&0&0&1&0&0&0\cr
0&0&0&0&1&0&1\cr
0&0&0&1&0&1&0\cr
0&0&1&0&1&0&0\cr
0&1&0&1&0&0&0\cr
1&0&1&0&0&0&1\cr
0&1&0&1&0&0&0\cr
0&0&1&0&1&0&0\cr
0&0&0&1&0&1&0\cr
0&0&0&0&1&0&1\cr
0&0&0&1&0&0&0\cr
0&0&1&0&0&0&0\cr
0&1&0&0&0&0&0\cr
1&0&0&0&0&0&0}$&
$C=\smat{
1&0&0&0&0&0&0&0\cr
0&1&0&0&0&0&0&0\cr
0&0&1&0&0&0&0&0\cr
0&0&0&1&0&0&0&0\cr
0&0&0&0&1&0&0&0\cr
0&0&0&0&0&1&0&1\cr
0&0&0&0&1&0&1&0\cr
0&0&0&1&0&1&0&0\cr
0&0&1&0&1&0&0&0\cr
0&1&0&1&0&0&0&1\cr
1&0&1&0&1&0&0&0\cr
0&1&0&1&0&1&0&0\cr
0&0&1&0&1&0&1&0\cr
0&0&0&1&0&1&0&1\cr
0&0&0&0&2&0&0&0\cr
0&0&0&1&0&1&0&1\cr
0&0&1&0&1&0&1&0\cr
0&1&0&1&0&1&0&0\cr
1&0&1&0&1&0&0&0\cr
0&1&0&1&0&0&0&1\cr
0&0&1&0&1&0&0&0\cr
0&0&0&1&0&1&0&0\cr
0&0&0&0&1&0&1&0\cr
0&0&0&0&0&1&0&1\cr
0&0&0&0&1&0&0&0\cr
0&0&0&1&0&0&0&0\cr
0&0&1&0&0&0&0&0\cr
0&1&0&0&0&0&0&0\cr
1&0&0&0&0&0&0&0}$\\
\clearpage
\section{Appendix~B: Affine Adjacency Intertwiners}
\setcounter{equation}{0}
\renewcommand{\theequation}{B.\arabic{equation}}
The adjacency matrix intertwiners $C$ for the affine \ade models are as
follows:
\bigskip
\twocol
\hfil{$A^{(1)}_{L-1}$--$D^{(1)}_{{L\over 2}+2}$}\hfil&\hfil{$
A^{(1)}_5$--$E^{(1)}_6$}\hfil\\ \smallskip
\twocol
$C=\smat{
1&1&0&\ldots&0&0&0\cr
0&0&1&\ldots&0&0&0\cr
\vdots&\vdots&\vdots&\ddots&\vdots&\vdots&\vdots\cr
0&0&0&\ldots&1&0&0\cr
0&0&0&\ldots&0&1&1\cr
0&0&0&\ldots&1&0&0\cr
\vdots&\vdots&\vdots&\dddots&\vdots&\vdots&\vdots\cr
0&0&1&\ldots&0&0&0}$&
$C=\smat{
0 & 1 & 0 & 0 & 0 & 1 & 0\cr
1 & 0 & 1 & 0 & 0 & 0 & 0\cr
0 & 1 & 0 & 1 & 0 & 0 & 0\cr
0 & 0 & 1 & 0 & 1 & 0 & 0\cr
0 & 0 & 0 & 1 & 0 & 1 & 0\cr
0 & 0 & 1 & 0 & 0 & 0 & 1}$\\
\bigskip
\twocol
\hfil{$A^{(1)}_{7}$--$E^{(1)}_7$}\hfil&\hfil{$A^{(1)}_{9}$--$E^{(1)}_8$}
\hfil\\ \smallskip
\twocol
$C=\smat{
	1 & 0 & 1 & 0 & 0 & 0 & 0 & 1  \cr
0 & 1 & 0 & 1 & 0 & 0 & 0 & 0  \cr 0 & 0 & 1 & 0 & 1 & 0 & 0 & 0  \cr
0 & 0 & 0 & 1 & 0 & 1 & 0 & 0  \cr  0 & 0 & 0 & 0 & 1 & 0 & 1 & 1
\cr  0 & 0 & 0 & 1 & 0 & 1 & 0 & 0  \cr  0 & 0 & 1 & 0 & 1 & 0 & 0 & 0  \cr
  0 & 1 & 0 & 1 & 0 & 0 & 0 & 0    }$&
$C=\smat{
1 & 0 & 1 & 0 & 1 & 0 & 0 & 0 & 1  \cr 0 & 1 & 0 & 1 & 0 & 1 & 0 & 0 & 0
\cr 0 & 0 & 1 & 0 & 1 & 0 & 1 & 0 & 0  \cr
0 & 0 & 0 & 1 & 0 & 1 & 0 & 1 & 0  \cr  0 & 0 & 0 & 0 & 1 & 0 & 1 & 0 & 1
\cr
0 & 0 & 0 & 0 & 0 & 2 & 0 & 0 & 0  \cr   0 & 0 & 0 & 0 & 1 & 0 & 1 & 0 & 1
 \cr  0 & 0 & 0 & 1 & 0 & 1 & 0 & 1 & 0  \cr  0 & 0 & 1 & 0 & 1 & 0 & 1 & 0
& 0  \cr 0 & 1 & 0 & 1 & 0 & 1 & 0 & 0 & 0
}$\\
\section{Appendix~C: Classical Cell Intertwiners}
\setcounter{equation}{0}
\renewcommand{\theequation}{C.\arabic{equation}}
\mbox{}\vspace{-.3in}
\subsection{Nonzero Cells for the $A_L$--$D_{L+3\over 2}$ Intertwiner}
\mbox{}\vspace{-.4in}
\begin{eqnarray*}
& &
\clearpage
\subsection{Nonzero Cells for the $A_{11}-E_6$ Intertwiner}
\begin{eqnarray*}
& & %
$$
\sin m_1 = {\sqrt{{{{ S^A_6}\,{ S^E_5}}\over
	{{ S^A_4}\,{ S^E_3}}}}};\;\;
\ sin m_2 = {\sqrt{{{{ S^A_3}\,{ S^E_6}}\over
	{{ S^A_5}\,{ S^E_4}}}}};\;\;
\cos m_3 = {\sqrt{{{{ S^A_4}\,{ S^E_5}}\over
	{{ S^A_6}\,{ S^E_3}}}}}.
$$
%%%%%%%%%%%%%%%%%%%%%%%%%%%%%%%%%%%%%%%%%%%%%%%%%%%%%%%%%%%%%%%%%%%%%%
%\clearpage
\subsection{Nonzero Cells for the $A_{17}-E_7$ Intertwiner}
\centerline{}
\begin{eqnarray*}
& & %
\begin{eqnarray*}
& & \sin m_1 = {\sqrt{{{{S^A_7}\,{S^E_6}}\over {{S^A_5}\,{S^E_4}}}}};\;\;
\;\;\;\;
 \sin m_2 = {\sqrt{{{{S^A_4}\,{S^E_7}}\over {{S^A_6}\,{S^E_5}}}}};\\
& & \cos m_3 = {\sqrt{{{{S^A_5}\,{S^E_6}}\over {{S^A_7}\,{S^E_4}}}}};\;\;
\;\;\;\;
 \cos m_4 = {\sqrt{{{{S^A_{10}}\,{S^E_1}} \over {{S^A_8}\,{S^E_3}}}}};\\
&	& \sin m_5 = {\sqrt{{{{S^A_7}\,{S^E_2}}\over {{S^A_9}\,{S^E_4}}}}};
\;\;\;\;\;\;
	\cos m_6 = {\sqrt{{{{S^A_5}\,{S^E_6}}\over {{S^A_9}\,{S^E_2}}}}};\\
& & \cos m_7 = {\sqrt{{{{S^A_8}\,{S^E_7}}\over {{S^A_{10}}\,{S^E_3}}}}}.
\end{eqnarray*}
%%%%%%%%%%%%%%%%%%%%%%%%%%%%%%%%%%%%%%%%%%%%%%%%%%%%%%%%%%%%%%%%%%%%%%
\clearpage
\subsection{Nonzero Cells for the $A_{29}-E_8$ Intertwiner}
\begin{eqnarray*}
&	&
%\clearpage
\begin{eqnarray*}
&	& a_1=\sin\gamma; \qquad a_2=\cos\gamma\,\sin\alpha; \\
&	& a_3=\cos\alpha\,\cos\gamma; \qquad b_1=\cos\gamma\,\sin\beta; \\
&	& b_2=\cos\alpha\,\cos\beta -
	\sin\alpha\,\sin\beta\,\sin\gamma; \\
&	& b_3= -\cos\beta\,\sin\alpha  -
\cos\alpha\,\sin\beta\,\sin\gamma; \\
&	& c_1=\cos\beta\,\cos\gamma; \\
&	& c_2=-\cos\alpha\,\sin\beta -
 \cos\beta\,\sin\alpha\,\sin\gamma; \\
&	& c_3=\sin\alpha\,\sin\beta -
	\cos\alpha\,\cos\beta\,\sin\gamma; \\
&	& \aa_1=\sin m_{12}=a_1\,\sqrt{{{ S^A_{14}}\,{ S^E_5}}\,\over
				{{ S^A_{13}}\,{ S^E_4}}}; \\
&	& \aa_3= a_3\,\sqrt{{{{ S^A_{14}}\,{ S^E_5}\,
			\over {{S^A_{13}}\,{S^E_8}}}}}; \;\;\;\;\;\;\;
\aa_2=\cos m_{10}=a_2\,\sqrt{{{S^A_{14}}\,{S^E_5}\,
		\over {{S^A_{13}}\,{S^E_6}}}}; \\
&	& \bb_1=b_1\,\sqrt{{{S^A_{14}}\,{S^E_5}\,\over
					{{S^A_{15}}\,{S^E_4}}}};
\;\;\;\;\;\;\;
\bb_2= b_2\,\sqrt{{{S^A_{14}}\,{S^E_5}\,
					\over {{S^A_{15}}\,{S^E_6}}}}; \\
&	& \bb_3= b_3\,\sqrt{{{S^A_{14}}\,{S^E_5}\,\over {{S^A_{15}}\,
{S^E_8}}}}; \;\;\;\;\;\;\; \cc_1=c_1\,\sqrt{{{S^A_{14}}\,{S^E_5}\,\over
{{S^A_{15}}\,{S^E_4}}}}; \\
&	& \cc_2= c_2\,\sqrt{{{S^A_{14}}\,{S^E_5}\,\over {{S^A_{15}}\,
{S^E_6}}}}; \;\;\;\;\;\;\; \cc_3=c_3\,\sqrt{{{S^A_{14}}\,{S^E_5}\,
				\over {{S^A_{15}}\,{S^E_8}}}}; \\
&	& \cos\gamma={\sqrt{{{{S^A_{15}}\,{S^E_4}}\over {{S^A_{14}}\,
{S^E_5}}}}}; \;\;\;\;\; \sin\alpha=\sin\alpha={\sqrt{{{{S^A_{14}}\,{S^E_5}
{S^A_{15}}\,{S^E_6}}\over {{S^A_{15}}\,{S^E_4}}}}}\\
&	& \sin m_1= {\sqrt{{{{S^A_8}\,{S^E_7}}\over
					{{S^A_6}\,{S^E_5}}}}}; \;\;\;\;
\sin m_2={\sqrt{{{{S^A_5}\,{S^E_8}}\over {{S^A_7}\,{S^E_6}}}}}; \\ &	&
 \cos m_3={\sqrt{{{{S^A_6}\,{S^E_7}}\over
				{{S^A_8}\,{S^E_5}}}}}; \;\;\;\;
\sin m_4={\sqrt{{{{S^A_{11}}\,{S^E_8}}\over {{S^A_9}\,{S^E_4}}}}}; \\ &	&
\cos m_5={\sqrt{{{{S^A_{12}}\,{S^E_1}}\over
					{{S^A_{10}}\,{S^E_3}}}}}; \;\;\;\;
\cos m_6= {\sqrt{{{{S^A_9}\,{S^E_4}}\over {{S^A_{11}}\,{S^E_2}}}}}\,
			{\sqrt{1 - {{{S^A_{11}}\,{S^E_8}}\over
						{{S^A_9}\,{S^E_4}}}}}; \\
&	& \sin m_7={\sqrt{{{{S^A_8}\,{S^E_3}}\over {{S^A_{10}}\,{S^E_5}}}}};
 \;\;\;\; \sin m_8= {\sqrt{{{{S^A_9}\,{S^E_2}}\over
{{S^A_{11}}\,{S^E_4}}}}}; \\
&	& \cos m_9= {\sqrt{{{{S^A_{14}}\,{S^E_3}}\over
					{{S^A_{12}}\,{S^E_5}}}}}; \;\;\;\;
\sin m_{10}= {\sqrt{{{{S^A_9}\,{S^E_2}}\over {{S^A_{11}}\,{S^E_4}}}}}\,
{\sqrt{{{{S^A_{11}}\,{S^E_4}}\over {{S^A_{13}}\,{S^E_6}}}}}; \\ &	&
 \sin m_{11}= {\sqrt{{{{S^A_{10}}\,{S^E_1}}\over
					{{S^A_{12}}\,{S^E_3}}}}}; \;\;\;\;
\sin m_{12}= {\sqrt{{{{S^A_{11}}\,{S^E_2}}\over {{S^A_{13}}\,{S^E_4}}}}}\,
{\sqrt{1 - {{{S^A_9}\,{S^E_4} }\over
{{S^A_{11}}\,{S^E_2}}}\left( 1 - {{{S^A_{11}}\,{S^E_8}}\over
{{S^A_9}\,{S^E_4}}} \right)}}.
\end{eqnarray*}
%\clearpage
\section{Appendix~D: Affine Cell Intertwiners}
\setcounter{equation}{0} \renewcommand{\theequation}{D.\arabic{equation}}
\subsection{Nonzero Cells for the $A_{L-1}^{(1)}-D_{{L\over 2}+2}^{(1)}$
Intertwiner}
\begin{eqnarray*}

\end{eqnarray*}
%\clearpage
%\mbox{}\vspace{-.3in}
\subsection{Nonzero Cells for the $A_5^{(1)}-E_6^{(1)}$ Intertwiner}
\vspace{-.1in}
\begin{eqnarray*}
& & %
\vspace{-.2in}
%%%%%%%%%%%%%%%%%%%%%%%%%%%%%%%%%%%%%%%%%%%%%%%%%%%%%%%%%%%
\begin{eqnarray*}
& & \sin m_1=\sqrt{S^E_1 \over S^E_2};\qquad \sin m_2=\sqrt{S^E_5 \over S^E_3};
\qquad \sin m_3=\sqrt{S^E_5 \over S^E_4};\\
& & \cos m_4=\sqrt{S^E_7 \over S^E_3}; \qquad
\sin m_5=\sqrt{S^E_7 \over S^E_6};\qquad \sin m_6=\sqrt{S^E_1 \over S^E_3};
\end{eqnarray*}
%\clearpage
\subsection{Nonzero Cells for the $A_7^{(1)}-E_7^{(1)}$ Intertwiner}
\centerline{}
\begin{eqnarray*}
& & %
%%%%%%%%%%%%%%%%%%%%%%%%%%%%%%%%%%%%%%%%%%%%%%%%%%%%%%%%%%
\begin{eqnarray*}
& & \cos m_0=\sqrt{S^E_2 \over S^E_4}; \qquad \cos m_1=\sqrt{S^E_2 \over
S^E_3}; \qquad \cos m_2=\sqrt{S^E_5 \over S^E_4};\\
& & \cos m_3=\sqrt{S^E_6 \over S^E_5}; \qquad
\cos m_4=\sqrt{S^E_7 \over S^E_6}; \qquad \sin m_5=\sqrt{S^E_6 \over S^E_5}; \\
& & \sin m_6=\sqrt{S^E_5 \over S^E_4}; \qquad \sin m_7=\sqrt{S^E_2 \over
S^E_3}; \qquad \cos m_8=\sqrt{S^E_1 \over S^E_2}; \\
& & \sin m_9=\sqrt{S^E_8 \over S^E_4};
\end{eqnarray*}
%\clearpage
\subsection{Nonzero Cells for the $A_9^{(1)}-E_8^{(1)}$ Intertwiner}
\centerline{}
\begin{eqnarray*}
& &
\vspace{-.1in}
\begin{eqnarray*}
&	& a_1=\sin\gamma; \\
&	& a_2=\cos\gamma\,\sin\alpha; \\
&	& a_3=\cos\alpha\,\cos\gamma; \\
&	& b_1=\cos\gamma\,\sin\beta; \\
&	& b_2=\cos\alpha\,\cos\beta -
\sin\alpha\,\sin\beta\,\sin\gamma; \\
&	& b_3= -\cos\beta\,\sin\alpha  -
\cos\alpha\,\sin\beta\,\sin\gamma; \\
&	& c_1=\cos\beta\,\cos\gamma; \\
&	& c_2=-\cos\alpha\,\sin\beta -
 \cos\beta\,\sin\alpha\,\sin\gamma; \\
&	& c_3=\sin\alpha\,\sin\beta -
	\cos\alpha\,\cos\beta\,\sin\gamma; \\
&	& \aa_1=\sin m_{3}=a_1\,\sqrt{{{S^E_6}}\over {{S^E_5}}}; \\
&	& \aa_3= a_3\,\sqrt{{{{S^E_6}\over {{S^E_9}}}}}; \;\;\;\;\;\;\;
\aa_2=\cos m_{6}=a_2\,\sqrt{{{S^E_6}   \over {{S^E_7}}}}; \\
&	& \bb_1=b_1\,\sqrt{{{S^E_6}\over  {{S^E_5}}}}; \;\;\;\;\;\;\;
\bb_2= b_2\,\sqrt{{{S^E_6}\,   \over {{S^E_7}}}}; \\
&	& \bb_3= b_3\,\sqrt{{{S^E_6} \over {{S^E_9}}}}; \;\;\;\;\;\;\;
\cc_1=c_1\,\sqrt{{{S^E_6}\over  {{S^E_5}}}}; \\
&	& \cc_2= c_2\,\sqrt{{{S^E_6}\,   \over {{S^E_7}}}}; \;\;\;\;\;\;\;
\cc_3=c_3\,\sqrt{{{S^E_6} \over {{S^E_9}}}}; \\
&	& \cos\gamma={\sqrt{{{{S^E_5}}\over   {{S^E_6}}}}}; \;\;\;\;\;\;\;
\sin\alpha=\sin\beta={\sqrt{{{{S^E_6}-{S^E_7}}\over
			{{S^E_5}}}}}; \\
&	& \sin m_1=\sqrt{S^E_1 \over S^E_3}; \qquad
\cos m_2=\sqrt{S^E_3 \over S^E_4}; \qquad
\cos m_3=\sqrt{S^E_4 \over S^E_5}; \\
&	&  \cos m_4=\sqrt{S^E_3 \over S^E_5}; \qquad
\sin m_5=\sqrt{S^E_8 \over S^E_6}; \qquad
\sin m_6=\sqrt{S^E_8 \over S^E_7}; \\
& & \cos m_7=\sqrt{S^E_1 \over S^E_2}; \qquad
\sin m_8=\sqrt{S^E_2 \over S^E_4}; \qquad \cos m_9=\sqrt{S^E_9 \over S^E_6};
\end{eqnarray*}
%%%%%%%%%%%%%%%%%%%%%%%%%%%%%%%%%%%%%%

%%%%%%%%%%%%%%%%%%%%%


\begin{thebibliography}{99}
\bibitem{Ocneanu} A. Ocneanu, ``Quantized Groups, String Algebras and
Galois Theory for Algebras in
Operator Algebras and Applications", Lond. Math. Soc. Lecture Note Series
{\bf 136}.
\bibitem{PasqEtiol} V. Pasquier, Commun. Math. Phys. {\bf 118} (1988) 335.
\bibitem{Roch:90} Ph. Roche, Commun. Math. Phys. {\bf 127} (1990) 395.
\bibitem{FrZu:89p} P. Di Francesco and J. B. Zuber, Nucl. Phys. {\bf B338}
(1990) 602.
\bibitem{Pasq:87} V. Pasquier, Nucl. Phys. {\bf B28} (1987) 162; J. Phys. A
{\bf 20} (1987) L1229, 5707.
\bibitem{OwBa:87} A. L. Owczarek and R. J. Baxter, J. Stat. Phys. {\bf 49}
(1987) 1093.
\bibitem{Ginsparg:89?} P. Ginsparg, Nucl. Phys. {\bf B295} [FS21] (1988)
153.
\bibitem{Paul:90} P. A. Pearce, Int. J. Mod. Phys. {\bf B4} (1990) 715.
\bibitem{WaNiSe:92} S. O. Warnaar, B. Nienhuis and K. A. Seaton, Phys. Rev.
Lett. {\bf 69} (1992) 710.
\bibitem{Roch:92} Ph. Roche, Phys. Lett. {\bf B4} (1992) 929.
\bibitem{Baxter:82} R. J. Baxter, ``Exactly Solved Models in Statistical
Mechanics", Academic Press,
                       London, 1982.
\bibitem{ABF:84} G. E. Andrews, R. J. Baxter and P. J. Forrester, J. Stat.
Phys. {\bf 35} (1984) 193.
\bibitem{Yang:67} C. N. Yang, Phys. Rev. Lett. {\bf 19} (1967) 1312.
\bibitem{PeSe:88} P. A. Pearce and K. A. Seaton, Phys. Rev. Lett. {\bf 60}
(1988) 1347;
Ann. Phys. (N.Y) {\bf 193} (1989) 326.
\bibitem{KuYa:88} A. Kuniba and T. Yajima, J. Stat. Phys. {\bf 52} (1988)
829.
\bibitem{DJKMO:87} E. Date, M. Jimbo, A. Kuniba, T. Miwa and M. Okado,
Nucl. Phys. {\bf B290} (1987) 231.
\bibitem{DJKMO:88} E. Date, M. Jimbo, A. Kuniba, T. Miwa and M. Okado, Adv.
Stud. Pure Math.,
                 {\bf 16} (1988) 17.
\bibitem{BaRe:89} V. V. Bazhanov and N. Yu Reshetikhin, Int. J. Mod. Phys.
B {\bf 4} (1989) 115.
\bibitem{KlPe:92} A. Kl{\"{u}}mper and P. A. Pearce, Physica A {\bf 183}
(1992) 304.
\bibitem{Paul:91p} P. A. Pearce, Int. J. Mod. Phys. {\bf A7} Suppl. 1B
(1992) 791.
\bibitem{FendGins}  P. Fendley and P. Ginsparg, Nucl. Phys. {\bf B324}
(1989) 549.

\end{thebibliography}
\end{document}